\documentclass[aps,prb,twocolumn,nofootinbib,showpacs,secnumarabic,superscriptaddress,floatfix]{revtex4}

\usepackage{amssymb,amsmath}
\usepackage[pdftex]{graphicx}
\usepackage{color}
\usepackage[usenames,dvipsnames]{xcolor}
\usepackage[colorlinks=true,linkcolor=RoyalBlue,citecolor=OliveGreen,linktoc=page]{hyperref}
\usepackage{enumerate}

\usepackage[english]{babel}

\usepackage[caption=false]{subfig}
\usepackage{balance}
\usepackage{float}

\DeclareMathOperator{\arcsinh}{arcsinh}

\newcommand{\rs}{\rm\scriptscriptstyle}

\newcommand{\rL}{{\rm\scriptscriptstyle L}}
\newcommand{\rR}{{\rm\scriptscriptstyle R}}
\newcommand{\rB}{{\rm\scriptscriptstyle B}}
\newcommand{\rLR}{{\rL{\scriptscriptstyle (}\rR{\scriptscriptstyle )}}}
\newcommand{\sLR}{{\!\rL{\scriptscriptstyle (}\rR{\scriptscriptstyle )}}}
\newcommand{\rio}{{\rm i{\scriptscriptstyle (}o{\scriptscriptstyle )}}}
\newcommand{\reh}{{\rm e{\scriptscriptstyle (}h{\scriptscriptstyle )}}}
\newcommand{\rS}{{\rm\scriptscriptstyle S}}
\newcommand{\rY}{{\rm\scriptscriptstyle Y}}
\newcommand{\rF}{{\rm\scriptscriptstyle F}}

\begin{document}

\title{Unitary limit in crossed Andreev transport}

\author{I.\,A.\,Sadovskyy}
\affiliation{Materials Science Division, Argonne National Laboratory, 9700 S. Cass Av., Argonne, IL 60637, USA}
\author{G.\,B.\,Lesovik}
\affiliation{Landau Institute for Theoretical Physics, RAS, Prosp. Akad. Semenova 1-A, 142432 Chernogolovka, Moscow region, Russia}
\author{V.\,M.\,Vinokur}
\affiliation{Materials Science Division, Argonne National Laboratory, 9700 S. Cass Av., Argonne, IL 60637, USA}

\date{\today}

\begin{abstract}
One of the most promising approaches of generating spin- and energy-entangled electron pairs is splitting a Cooper pair into the metal through spatially separated terminals. Utilizing hybrid systems with the energy-dependent barriers at the superconductor-normal metal interfaces, one can achieve practically 100\% efficiency outcome of entangled electrons. We investigate minimalistic one-dimensional model comprising a superconductor and two metallic leads and derive an expression for an electron-to-hole transmission probability as a measure of splitting efficiency. We find the conditions for achieving 100\% efficiency and present analytical results for the differential conductance and differential noise.
\end{abstract}


\maketitle

\section{Introduction}

\begin{figure}[b]
	\begin{center}
		\includegraphics[width=8.0cm]{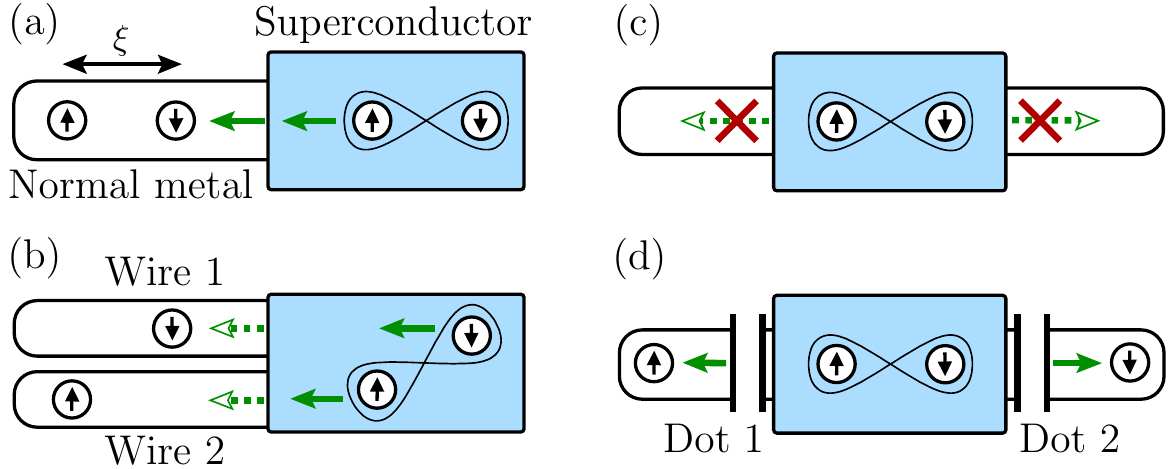}
		\subfloat{\label{fig:Andreev_reflection_regular}}
		\subfloat{\label{fig:Andreev_reflection_crossed}}
		\subfloat{\label{fig:Andreev_reflection_crossed1D}}
		\subfloat{\label{fig:Andreev_reflection_dots}}
	\end{center} \vspace{-3mm}
	\caption{
		Different scenarios of Andreev reflection.
		(a)~Regular Andreev reflection at the ideal NS interface. 
		In a one-dimensional case a Cooper pair converts into two 
		entangled electrons in normal metal wire with the unity probability.
		(b)~Three-dimensional crossed Andreev reflection (CAR) with two 
		different exits to the normal metal. Cooper pair converts into two 
		electrons in different normal wires with probability suppressed by 
		the distance between normal contacts.
		(c)~One-dimensional CAR to the opposite sides of NSN system is 
		completely suppressed for the ideal NS boundaries.
		(d)~Additional energy-dependent normal reflection at NS boundaries 
		may increase the probability of the CAR up to unity.
	}
	\label{fig:Andreev_reflection}
\end{figure}

Quantum entanglement dating back to Einsten's and Schr{\"o}dinger's seminal papers~\cite{Einstein:1935,Schrodinger:1935} has emerged as one of the most active research areas of contemporary condensed matter physics. The interest is motivated both, by the important promise of utilizing entanglement effects in communication and computation technologies,~\cite{Bennett:1993,Steane:1998,Nielsen:2011} and by the intellectual appeal of dealing with the most fundamental issues of quantum mechanics.~\cite{Bohm:2001} One of the major experimental tasks is the creation of and subsequent manipulation by the entangled quantum states. A Cooper pair comprising two electrons endowed with a unique inseparable quantum state is a natural source of electrons with states inextricably linked to each other and remaining entangled with respect to spin and/or energy despite having become spatially separated.~\cite{Lesovik:2001,Recher:2001} The initial stage of splitting can occur via an Andreev reflection (AR) phenomenon,~\cite{Andreev:1964} see Fig.~\subref{fig:Andreev_reflection_regular}, where the Cooper pair crosses the ideal normal metal-superconductor (NS) interface and enters normal metal as two electrons entangled with respect to energy and spin, more precisely, the electron-like quasiparticle enters and hole-like quasiparticle leaves normal metal, with the probability of unity. Applying the external magnetic field, one can take electrons and holes further apart.~\cite{Bozhko:1982,Benistant:1983}

A different recipe for the Cooper pair splitting (CPS) and spatial separation of electrons and holes~\cite{Lesovik:2001} proposed the normal-metal fork with leads linked to a superconductor by the NS barriers with the different resonance energies. This idea evolved into a crossed Andreev reflection (CAR) approach for CPS, utilizing two-terminal configuration in which a Cooper pair generates electrons escaping through two {\it separated} normal terminals~\cite{Byers:1995,Falci:2001,Golubev:2009,Deutscher:2000,Beckmann:2004,Russo:2005} as shown in Fig.~\subref{fig:Andreev_reflection_crossed}. Yet at first sight this approach seemed to be problematic: since the initial separation of exiting electron is the size of the Cooper pair, the escape probability was to remain appreciable only for terminal separation not much exceeding the coherence length. Furthermore, the CAR probability is exactly zero in a one-dimensional geometry with the ideal NS boundary, see Fig.~\subref{fig:Andreev_reflection_crossed1D}. In higher dimensions, the amplitude of the CAR decays exponentially with the distance~$L$ between the terminals. What more, in clean $D$-dimensional superconductors it acquires a small prefactor $\propto 1/(k_\rF L)^{D-1}$.~\cite{Choi:2000,Falci:2001,Recher:2001,Leijnse:2013} In three-dimensional disordered superconductors the amplitude drops by the factor $\propto 1/k_\rF \sqrt{l L}$ with~$k_\rF$ being the Fermi wave vector and $l$ being the mean-free path.~\cite{Feinberg:2003} 

This suppression of the CAR efficiency was mended by making the normal scattering amplitudes at the NS interface the energy dependent ones over the scale of order of the superconducting gap $\Delta$, in such a way that they became essentially different for electrons and holes. This was implemented via setting up quantum dots endowed with the different resonance levels at the NS interfaces. Varying the gates potentials one could appropriately change the positions of these resonances with respect to the Fermi level, see Fig.~\subref{fig:Andreev_reflection_crossed1D}. Thus the efficiency of the two-terminal configuration for the CPS was enhanced by plugging a quantum dot into the each lead,~\cite{Recher:2001} and utilizing Coulomb blockade for manipulating the electrons. The similar idea of utilizing asymmetric quantum dots was adopted for studying the dynamic conductance and noise in dot-superconductor-dot systems.\cite{Melin:2008,Chevallier:2011,Floser:2013} Reference~\onlinecite{Melin:2008} calculated noise in the framework of the tight-binding model for a setup comprising a superconductor and and two metal leads attached at the same side. Reference~\onlinecite{Chevallier:2011} addressed the noise in a N-dot-S-dot-N system. The analytical result although somewhat cumbersome for differential conductance and noise was derived in Ref.~\onlinecite{Floser:2013}. In the meantime, it was shown experimentally that using nanowires,~\cite{Hofstetter:2009,Hofstetter:2011} carbon nanotubes,~\cite{Herrmann:2010,Schindele:2012} and graphene~\cite{Tan:2015} to split Cooper pairs one can achieve from a few percents to near-unity efficiency. This stimulated a parallel theoretical development. In particular, a scheme realizing asymmetric dots via attaching n- and p-type semiconductor quantum dots at either sides of a superconductor was proposed in Ref.~\onlinecite{Veldhorst:2010}. It was demonstrated that choosing specific relations between the valence bands and bias voltage one can achieve the 100\% efficiency. Reference~\onlinecite{Burset:2011} dealt with the effects of the Coulomb and spin-orbit interactions on the CPS efficiency also showing the feasibility of the 100\% efficiency even in the presence of interactions.

In the present work we investigate a minimalistic model allowing for a 100\% efficiency. We consider a one-dimensional CAR-based splitter with the initially zero CAR amplitude. Adding an energy-dependent double barrier to each terminal of the setup shown in Fig.~\subref{fig:Andreev_reflection_crossed1D}, we demonstrate that in an equilibrium noninteracting system the proper tuning resonance levels in separately biased output terminals can provide the hundred percent outcome. This simple model utilizes non-equal transmission amplitudes of the electron and hole-like Bogoliubov quasiparticles (below the superconducting gap) across the energy dependent scatterer and, remarkably, allow for a full analytical treatment. We further consider the splitter comprising the Y-geometry junctions, each connected to the respective infinite superconductors, capturing the main features of experimental setups, and show that the 100\% outcome holds in this geometry as well. We derive a simple analytical result for a stationary Josephson effect situation where Fermi levels of both terminals are set equal by using a grounded superconductor. The simple model of the proposed setup allows for a detailed quantitative analysis revealing the underlying subtleties and conditions necessary for achieving a 100\% outcome.

\begin{figure}[tb]
	\begin{center}
		\includegraphics[width=8.7cm]{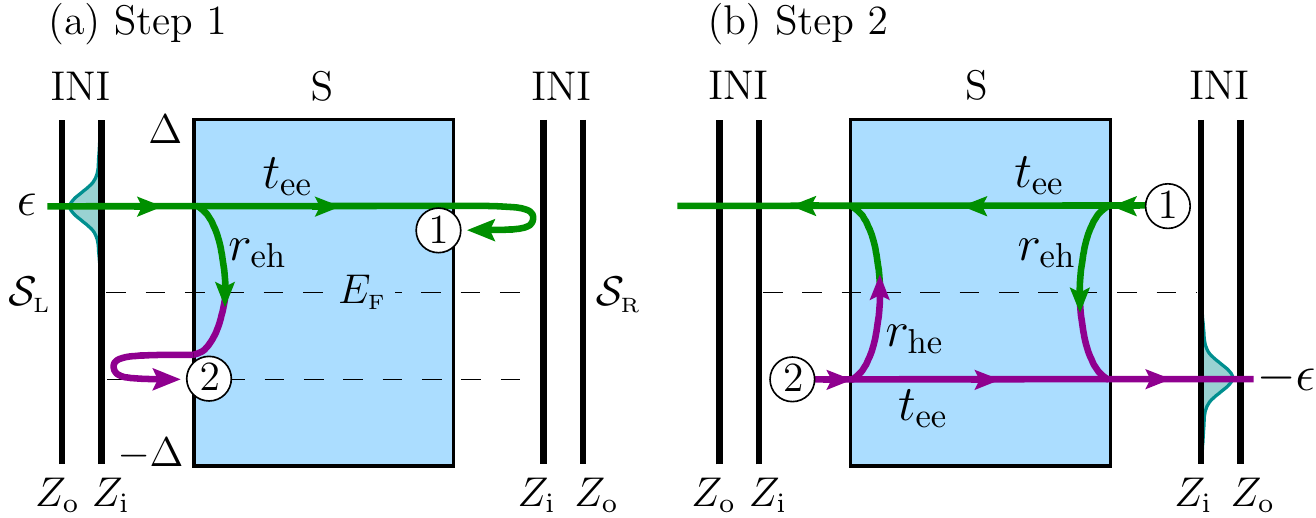}
		\subfloat{\label{fig:setup_nsn_stage1}}
		\subfloat{\label{fig:setup_nsn_stage2}}
		\subfloat{\label{fig:setup_nsn_ideal}}
		\subfloat{\label{fig:setup_nsn_interferometer}}
	\end{center} \vspace{-6mm}
	\caption{
		One-dimensional INI-S-INI system. A single sharp resonance 
		of the left symmetric INI double barrier lies at $\varepsilon$ such 
		that $|t_\rL(\varepsilon)| = 1$, $r_\rL(\varepsilon) = 0$, 
		$t_\rL(-\varepsilon) = 0$, and $|r_\rL(-\varepsilon)| = 1$.
		For the right INI double barrier a single resonance is located 
		at $-\varepsilon$, so that $t_\rR(\varepsilon) = 0$, 
		$|r_\rR(\varepsilon)| = 1$, $|t_\rR(-\varepsilon)| = 1$, 
		and \mbox{$r_\rR(-\varepsilon) = 0$}.
		(a)~An incident electron `splits' into the electron on the right 
		(1) and the hole on the left (2). Both of them reflect from the 
		non-transparent INI double barrier at corresponding energies.
		(b)~After reflection each of the states split further into the electron 
		and hole components. Interference of the latter states results 
		in the electron reflection amplitude~${\tilde r}_{\rm ee}$
		and the electron-to-hole transmission amplitude~${\tilde t}_{\rm eh}$.
	}
	\label{fig:setup_nsn}
\end{figure}

\section{Preliminaries} \label{sec:preliminaries}

We quantify the entangler efficiency by transparency i.e. by the probability for the incident electron to reach a superconductor through one terminal having created a Cooper pair in the superconductor and a hole that left the entangler through the other terminal. Our idea is to control the resonance structure of the Bogoliubov states by setting double-barrier potentials in such a way that for specific resonance energies in a hybrid structure~\cite{Sadovskyy:2007,Sadovskyy:2012} the electron-to-hole transmission probability became equal to unity, in full analogy with Fabry-P\'erot interferometer. To gain a feeling how the perfect transmission can be reached, we consider a special resonance structure of the barriers associated with terminals. Let the left scatterer have a very narrow resonance $E_\rF + \varepsilon$ above Fermi energy $E_\rF$, and the right one have the same resonance below Fermi energy at $E_\rF - \varepsilon$, see Fig.~\ref{fig:setup_nsn}. Suppose further that the transparency assumes the value of unity at these energies and is zero otherwise. The electron with the energy $E_\rF + \varepsilon$ incidents from the left. The electron-to-hole reflection (with amplitude $r_{\rm eh}$) and electron-to-electron transmission ($t_{\rm ee}$) are blocked, so incident electron can ether reflect back as an electron ($r_{\rm ee}$) or transmit as a hole ($t_{\rm eh}$). The latter process occurs with the probability of unity for certain energy~$\varepsilon$ and the superconductor length~$L$. The described configuration is, in fact, realization of Mach-Zehnder interferometer, involving just two trajectories.

\section{Scattering matrix approach} \label{sec:scattering_matrix}

Let us consider an electron-like quasiparticle with the energy $E_\rF + \varepsilon$ incident at the one-dimensional X-S-X structure shown in Fig.~\ref{fig:setup_nsn}. Both X-parts stand for the energy-dependent barriers realized via the insulator-normal metal-insulator (INI) scatterers, which exhibit sharp resonances with the unity transparency for identical $\delta$-function-like insulators. To find the transmission probability ${\tilde T}_{\rm eh} = |{\tilde t}_{\rm eh}|^2$ of such an X-S-X system, we solve Bogoliubov-de Gennes (BdG)~\cite{Bogoliubov:1968,GennesBook:1968,Lesovik:2011} equations with energy~$\varepsilon$ below the superconducting gap~$\Delta$, $0 \leqslant \varepsilon \leqslant \Delta$. We take the piecewise potential in BdG equations so that $\hat\Delta(x) = \Delta$ in the superconductor and $\hat\Delta(x)=0$ outside. The electron- ($u$) and hole-like ($v$) components of the wave function of the superconductor are then
\begin{align}
	\left[ u, v \right]^{\rs T}
	& = [ e^{i\alpha}, 1 ]^{\rs T} 
		\bigl( .\, e^{ipx - qx} + .\, e^{-ipx + qx} \bigr) \nonumber \\
	& + [ e^{-i\alpha}, 1 ]^{\rs T}
		\bigl( .\, e^{ipx + qx} + .\, e^{-ipx - qx} \bigr),
	\nonumber
\end{align}
where dots stand for some constants. The normal state solution of BdG between superconductor and left (L) [right (R)] scatterer is given by the linear combination of plane waves:
\begin{equation}
	\left[\! \begin{array}{c} u \\ v \end{array} \!\right] =
	\left[\! \begin{array}{c} 
		c_\rLR^{\rm e\rightarrow} e^{ik_{\rm e}x} + c_\rLR^{\rm e\leftarrow} e^{-ik_{\rm e}x} \\
		c_\rLR^{\rm h\leftarrow} e^{ik_{\rm h}x} + c_\rLR^{\rm h\rightarrow} e^{-ik_{\rm h}x}
	\end{array} \!\right] \!.
	\nonumber
\end{equation}
The central superconducting part with the ideal NS boundaries couples the incident and reflected states, $[c_\rL^{\rm e\leftarrow}, c_\rR^{\rm e\rightarrow}, c_\rL^{\rm h\leftarrow}, c_\rR^{\rm h\rightarrow}]^{\rs T} = {\cal S}^\rS [c_\rL^{\rm e\rightarrow}, c_\rR^{\rm e\leftarrow}, c_\rL^{\rm h\rightarrow}, c_\rR^{\rm h\leftarrow}]^{\rs T}$ via the scattering matrix
\begin{equation}
	{\cal S}^\rS = \left[ \begin{array}{llll}
		0 & t_{\rm ee} & r_{\rm he} & 0 \\
		t_{\rm ee} & 0 & 0 & r_{\rm he} \\
		r_{\rm eh} & 0 & 0 & t_{\rm hh} \\
		0 & r_{\rm eh} & t_{\rm hh} & 0
	\end{array} \right]\!\!
	\label{eq:S_NSN}
\end{equation}
with nonzero amplitudes
\begin{equation}
	t_{\rm ee(hh)} 
	= \frac{e^{\pm ipL} \sin\alpha}{\sin(\alpha-iqL)}, \quad
	r_{\rm eh(he)} = \frac{\sinh(qL)}{i\sin(\alpha-iqL)}.
	\label{eq:t_r_NSN}
\end{equation}
Hereafter we will be using subscript e(h) to denote electron (hole) component of the wave function (e.g. $t_{\rm eh}$ is the electron-to-hole transmission amplitude). The transmission amplitudes $t_{\rm ee(hh)}$ and corresponding transparencies
$
	T_{\rm ee(hh)} 
	= {\sin^2\!\alpha} / [\sin^2\!\alpha + \sinh^2(qL)]
$
describe co-tunnetilng~\cite{Averin:1989,Averin:1990} in an ideal N-S-N contact. The inverse coherence length~$q$ and the wave vector~$p$ naturally appear from the solution of BdG equation with the fixed modulus of the superconducting gap $\Delta$ and are defined as $p^2 - q^2 = k_\rF^2$ and $2pq = (2m/\hbar^2) \Delta \sin\alpha$, where $\alpha = \arccos(\varepsilon/\Delta)$ is the auxiliary phase $\alpha \in [0 \ldots \pi/2]$, $k_\rF = \sqrt{2mE_\rF}/\hbar$ is the Fermi wave vector, and $m$ is the mass of the electron. For $\Delta \ll E_\rF$ one finds $q \approx (k_\rF\Delta/2E_\rF) \sin\alpha$ and $p \approx k_\rF$. We count the energy $\varepsilon$ from the Fermi energy $E_\rF$.

The left and right hand side energy-dependent X barriers form energy dependent barriers with scattering matrices ${\cal S}^\rLR = {\rm diag} \{ {\cal S}^\rLR_{\rm e}, {\cal S}^\rLR_{\rm h} \}$. The electron and hole subparts are given by
\begin{equation}
	{\cal S}^\rLR_{\rm e} = \left[ \begin{array}{ll}
		t_\rLR^{\rm e} & r_\rLR^{\rm e} \\
		r_\rLR^{\rm e} & t_\rLR^{\rm e}
	\end{array} \right]\!, \quad
	{\cal S}^\rLR_{\rm h} = \left[ \begin{array}{ll}
		t_\rLR^{\rm h} & r_\rLR^{\rm h} \\
		r_\rLR^{\rm h} & t_\rLR^{\rm h}
	\end{array} \right]\!,
\end{equation}
where $t^\reh = t(\pm\varepsilon)$ and $r^\reh = r(\pm\varepsilon)$. Experimentally, one can control the position of the resonances with respect to the Fermi energy by the external gate voltage.

\begin{figure}[tb]
	\begin{center}
		\includegraphics[width=8.5cm]{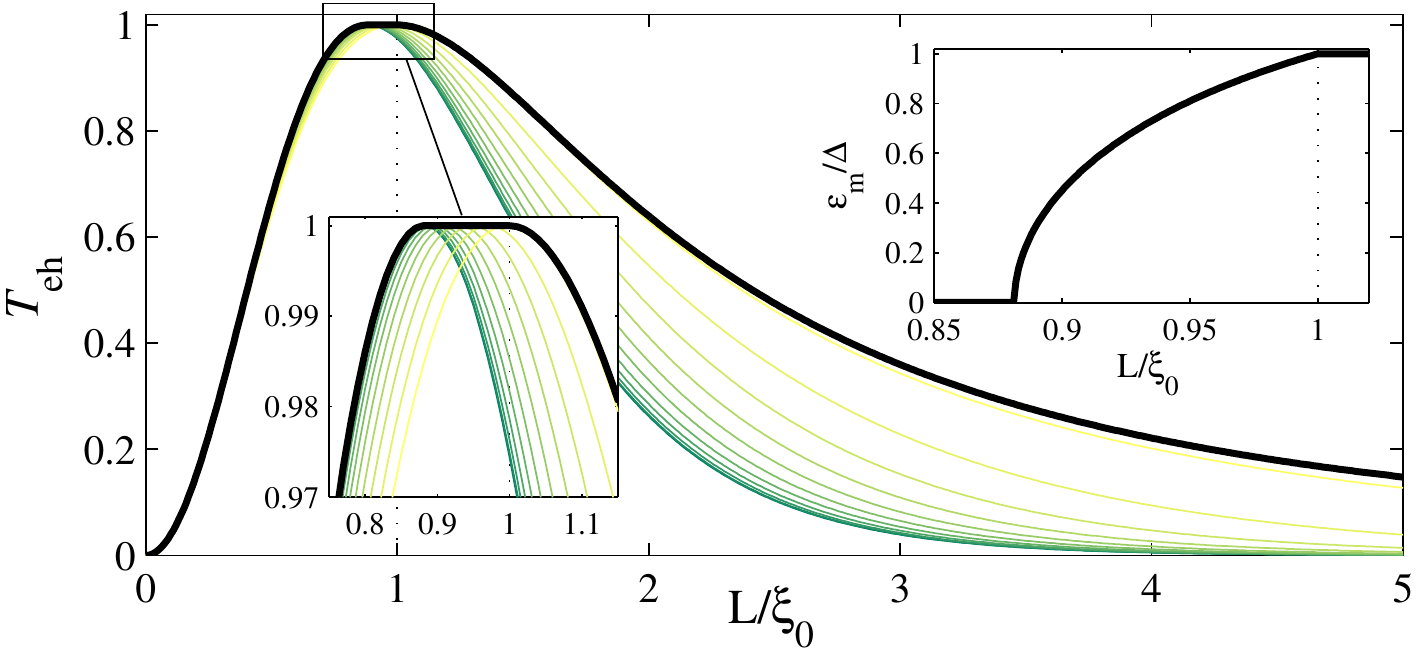}
	\end{center} \vspace{-5mm}
	\caption{
		Transparency ${\tilde T}_{\rm eh}$ as a function of $L/\xi_0$ 
		for $\theta = \pi n$ and different energies of the incident electron 
		from $\varepsilon = 0$ (green) to $\varepsilon = \Delta$ (yellow).
		Maximal transparency ${\tilde T}_{\rm eh}^\varepsilon = 
		\max_\varepsilon\{{\tilde T}_{\rm eh}\}$ is shown by the black envelope curve.
		Right inset: Energy~$\varepsilon_{\rm m}$, where this maximum was reached.
	}
	\label{fig:Tmax}
\end{figure}

\section{Ideal transparency case} \label{sec:ideal_transparency}

Now we are equipped for finding components of the scattering matrix ${\cal \tilde S}^\rS$ of the X-S-X system. We concentrate on the electron-to-hole transmission amplitude ${\tilde t}_{\rm eh}$ of this matrix and transparency ${\tilde T}_{\rm eh}$ as a measure of the entangler efficiency. We consider a narrow resonance with the width much smaller then $\varepsilon$ and spacings between resonances being much lager then $\Delta$. If 
$t_\rL^{\rm e} = 1$, 
$r_\rL^{\rm h} = e^{i\varphi_\rL^{\rm h}}$,
$r_\rR^{\rm e} = e^{i\varphi_\rR^{\rm e}}$, and $t_\rR^{\rm h} = 1$, 
the X-S-X transparency becomes ideal, ${\tilde T}_{\rm eh} = 1$. In this case, both, X-S-X transmission and reflection amplitudes, are defined by two pairs of paths shown in Fig.~\ref{fig:setup_nsn}, and the corresponding expression for~${\tilde t}_{\rm eh}$ assumes a simple form
\begin{equation}
	{\tilde t}_{\rm eh} = t_\rL^{\rm e}
		(t_{\rm ee} r_\rR^{\rm e} r_{\rm eh} + r_{\rm eh} r_\rL^{\rm h} t_{\rm hh}) t_\rR^{\rm h}. 
	\label{eq:teh_INISINI_res}
\end{equation}
Combining this relation with Eq.~\eqref{eq:t_r_NSN} one finds the X-S-X transparency as 
\begin{equation}
	\! {\tilde T}_{\rm eh} =
	{
		4 \sinh^2(qL) \, \sin^2\!\alpha \, \cos^2\!\theta
	}/{
		[\sin^2\!\alpha + \sinh^2(qL)]^2
	},
	\label{eq:Teh_INISINI_res}
\end{equation}
where $\theta = pL + (\varphi_\rR^{\rm e} - \varphi_\rL^{\rm h})/2$. We see that the transparency becomes ideal, ${\tilde T}_{\rm eh} = 1$, provided $\theta = \pi n$ and $\sinh(qL)=\sin\alpha$. Since both $q$ and $\alpha$ are energy dependent, the latter equality implicitly defines the energy at which the transparency becomes unity (requirement $\theta = \pi n$ can be attained by adjusting $\varphi_\rR^{\rm e}$ and $\varphi_\rL^{\rm h}$). The maximal transparency ${\tilde T}_{\rm eh}^\varepsilon = \max_\varepsilon\{{\tilde T}_{\rm eh}\}$ as a function of the dimensionless parameter $L/\xi_0$, where $\xi_0=\hbar v_{\rs F}/\Delta$ and $v_{\rs F}=\hbar k_{\rs F}/m$, is shown in Fig.~\ref{fig:Tmax}. Note that ${\tilde T}_{\rm eh}$ is small in both limits of (i) a short superconductor, $L \ll \xi_0$, where electron-to-hole reflection amplitudes are small $r_{\rm eh(he)}\propto L/\xi_0$ and electron passes the superconductor freely and of (ii) a long superconductor, $L\gg\xi_0$, because the transmission amplitude of the N-S-N part decays exponentially, $t_{\rm ee(hh)}\propto e^{-L/\xi_0}$. The unity value ${\tilde T}_{\rm eh}^\varepsilon = 1$ is achieved in the interval $L/\xi_0 \in [\arcsinh 1 \ldots 1]$. According to Eq.~\eqref{eq:t_r_NSN}, each point at the flat top corresponds to the different energy~$\varepsilon$ and is a result of the competition between $q(\varepsilon)$ and $\alpha(\varepsilon)$ dependencies.

\section{Asymmetric INI parts and arbitrary resonance positions} \label{sec:arbitrary_resonances}

To understand how robust the unitary limit of CPS is and derive to which degree one can deviate from the ideal resonances condition still maintaining the nearly unitary limit, let us model the energy-dependent X parts as non-ideal dots (asymmetric INI double barriers) with arbitrary resonance positions. We first choose INI double barriers such that both of them had identical pairs of inner and outer point scatterers as shown in Fig.~\ref{fig:setup_nsn}. The inner point scatterers are described by transmission~$t_{\rm i}$ and reflection~$r_{\rm i}$ amplitudes, and outer point scatterers have~$t_{\rm o}$ and $r_{\rm o}$ correspondingly. Then the transmission and reflection coefficients for each INI part assume the form 	
$
	t_\rLR^\reh 
	= t_{\rm i} t_{\rm o} e^{ik_\reh d_\rLR} / (1 - r_{\rm i} r_{\rm o} e^{ 2ik_\reh d_\rLR})
$
and
$
	r_{\rL{\rm i}{\scriptscriptstyle (}\rR{\rm i}{\scriptscriptstyle )}}^\reh 
	= r_{\rm i} + r_{\rm o} t_{\rm i}^2 e^{2ik_\reh d_\rLR} / (1 - r_{\rm i} r_{\rm o} e^{2ik_\reh d_\rLR}),
$
where $k_\reh = \sqrt{2m(E_\rF \pm \varepsilon)}/\hbar$ and $d_\rLR$ are the lengths of left and right INI double barriers, respectively. The energies of the resonance at the left ($\varepsilon_\rL$) and right ($\varepsilon_\rR$) dots with respect to the Fermi level are controlled by adjusting corresponding gate voltages. Taking $|\varepsilon_\rLR| \ll E_\rF$ we parametrize amplitudes of INI double barriers by phase differences $2(k_\reh - k_\rLR)d_\rLR = (\pm\varepsilon - \varepsilon_\rLR) / \delta_\rLR$, where `$\pm$' correspond to the electron (hole) excitations, $k_\rLR = \sqrt{2m(E_\rF + \varepsilon_\rLR)} / \hbar$, and $\delta_\rLR =\hbar v_\rF / 2 d_\rLR$ are spacings between consequent resonances. The resonance half-widths are $\Gamma_\sLR = \delta_\rLR (1 - |r_{\rm i} r_{\rm o}|) / \sqrt{|r_{\rm i} r_{\rm o}|}$. Let us suppose further that both point barriers are $\delta$-function-like barriers with the transmission $t_\rio = 1 / (1 + iZ_\rio)$ and reflection $r_\rio = -iZ_\rio / (1 + iZ_\rio)$ amplitudes, respectively, and inner, $Z_{\rm i}$, and outer, $Z_{\rm o}$, strengths. 

The electron-to-hole transmission amplitude of the hybrid INI-S-INI system is then found to be
\begin{subequations}
\begin{align}
	{\tilde t}_{\rm eh} & = 
		t_\rL^{\rm e} [
			t_{\rm ee} r_{\rR{\rm i}}^{\rm e} r_{\rm eh}
			+ r_{\rm eh} r_{\rL{\rm i}}^{\rm h} t_{\rm hh}
		] t_\rR^{\rm h}
	\, / {\cal D},
	\label{eq:teh_INISINI}
\end{align}
where the denominator
\begin{multline}
	{\cal D}
	= 1
	- t_{\rm ee}^2 r_{\rL{\rm i}}^{\rm e} r_{\rR{\rm i}}^{\rm e} 
	- t_{\rm hh}^2 r_{\rL{\rm i}}^{\rm h} r_{\rR{\rm i}}^{\rm h}
	- r_{\rm eh} r_{\rm he} (r_{\rL{\rm i}}^{\rm e} r_{\rL{\rm i}}^{\rm h} + r_{\rR{\rm i}}^{\rm e} r_{\rR{\rm i}}^{\rm h}) \\
	+ (t_{\rm ee} t_{\rm hh} - r_{\rm eh} r_{\rm he})^2 \, r_{\rL{\rm i}}^{\rm e} r_{\rR{\rm i}}^{\rm e} r_{\rL{\rm i}}^{\rm h} r_{\rR{\rm i}}^{\rm h}
	\nonumber
\end{multline}
describes multiple reflections inside INI-S-INI structure. For ideal conditions of two isolated resonances, as in Fig.~\ref{fig:setup_nsn}, Eq.~\eqref{eq:teh_INISINI} reduces to Eq.~\eqref{eq:teh_INISINI_res}. If resonances are symmetric and energy-independent, $t_\rL^\reh = t_\rR^\reh = t$ and $r_{\rL{\rm i}}^\reh = r_{\rR{\rm i}}^\reh = r$, Eq.~\eqref{eq:teh_INISINI} gives the maximal possible transparency of the I-S-I junction as ${\tilde T}_{\rm eh}=0.5$.\cite{Chen:2015} In order to calculate the differential conductance and noise (see Sec.~\ref{sec:noise}), we also find the electron-to-electron transmission amplitude ${\tilde t}_{\rm ee}$ responsible for elastic co-tunneling, 
\begin{align}
	{\tilde t}_{\rm ee} & = 
		t_\rL^{\rm e} [
			t_{\rm ee} (1 - t_{\rm hh}^2 r_{\rL{\rm i}}^{\rm h} r_{\rR{\rm i}}^{\rm h})
			+ r_{\rm eh} r_{\rL{\rm i}}^{\rm h} t_{\rm hh} r_{\rR{\rm i}}^{\rm h} r_{\rm he}
		] t_\rR^{\rm e}
	\, / {\cal D}.
	\label{eq:tee_INISINI}
\end{align} 
The corresponding electron-to-hole and electron-to-electron reflection amplitudes are given by
\begin{align}
	{\tilde r}_{\rm eh} 
	& = t_\rL^{\rm e} r_{\rm eh} \bigl[
		1 + (t_{\rm ee} t_{\rm hh} - r_{\rm eh} r_{\rm he}) r_{\rR{\rm i}}^{\rm e} r_{\rR{\rm i}}^{\rm h}
	\bigr] t_\rL^{\rm h}
	\, / {\cal D},
	\label{eq:reh_INISINI} \\
	{\tilde r}_{\rm ee} 
	& = r_{\rL{\rm o}}^{\rm e} +
	t_\rL^{\rm e} \bigl[
		r_{\rm eh} r_{\rL{\rm i}}^{\rm h} r_{\rm he}
		+ t_{\rm ee} r_{\rR{\rm i}}^{\rm e} t_{\rm ee} \nonumber\\
		& \qquad\qquad\quad - (t_{\rm ee} t_{\rm hh} - r_{\rm eh} r_{\rm he})^2 \, r_{\rL{\rm i}}^{\rm h} r_{\rR{\rm i}}^{\rm h} r_{\rR{\rm i}}^{\rm e}
	\bigr] t_\rL^{\rm e}
	\, / {\cal D}.
	\label{eq:ree_INISINI}
\end{align}
\end{subequations}
Note, that the symmetry of the BdG equations $(\varepsilon, u, v) \to (-\varepsilon, -v^*, u^*)$ leads to the following relations among the transmission coefficients, ${\tilde t}_{\rm hh} = {\tilde t}_{\rm ee}^{*\diamond}$, ${\tilde t}_{\rm he} = - {\tilde t}_{\rm eh}^{*\diamond}$, ${\tilde r}_{\rm he} = - {\tilde r}_{\rm eh}^{*\diamond}$, and ${\tilde r}_{\rm hh} = {\tilde r}_{\rm ee}^{*\diamond}$, where `$\diamond$' stands for $(\varepsilon, \varepsilon_\rL, \varepsilon_\rR) \to - (\varepsilon, \varepsilon_\rL, \varepsilon_\rR)$. The obtained relations, Eqs.~\eqref{eq:teh_INISINI}--\eqref{eq:ree_INISINI} and BdG symmetry, enable us to write down a general expression for a scattering matrix of the asymmetric INI-S-INI junction (cf. Eq.~\eqref{eq:S_NSN} for ideal N-S-N junction):
\begin{equation}
	{\cal \tilde S}^\rS = \left[ \begin{array}{llll}
		{\tilde r}_{\rm ee} & {\tilde t}_{\rm ee}^{\scriptscriptstyle\square} & {\tilde r}_{\rm he} & {\tilde t}_{\rm he}^{\scriptscriptstyle\square} \vspace{1mm} \\
		{\tilde t}_{\rm ee} & {\tilde r}_{\rm ee}^{\scriptscriptstyle\square} & {\tilde t}_{\rm he} & {\tilde r}_{\rm he}^{\scriptscriptstyle\square} \vspace{1mm} \\
		{\tilde r}_{\rm eh} & {\tilde t}_{\rm eh}^{\scriptscriptstyle\square} & {\tilde r}_{\rm hh} & {\tilde t}_{\rm hh}^{\scriptscriptstyle\square} \vspace{1mm} \\
		{\tilde t}_{\rm eh} & {\tilde r}_{\rm eh}^{\scriptscriptstyle\square} & {\tilde t}_{\rm hh} & {\tilde r}_{\rm hh}^{\scriptscriptstyle\square}
	\end{array} \right]\!\!,
	\label{eq:S_INISINI}
\end{equation}
where `$\scriptstyle\square$' denotes left-to-right reflection.

\begin{figure}[tb]
	\vspace{-2mm} \begin{center}
		\subfloat{
			\hspace{-3mm} \includegraphics[width=4.3cm]{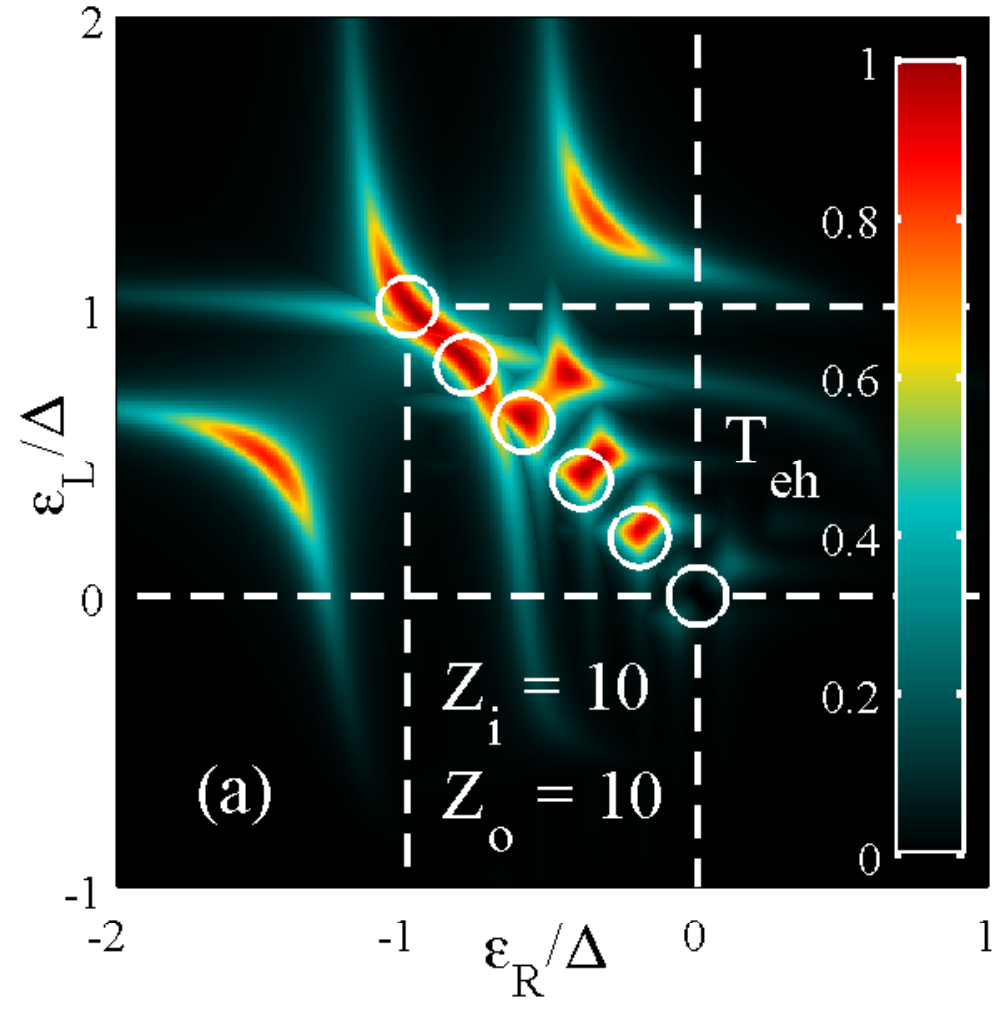}
			\label{fig:T_eLeR_fixE_10}
		}
		\subfloat{
			\hspace{-2mm} \includegraphics[width=4.3cm]{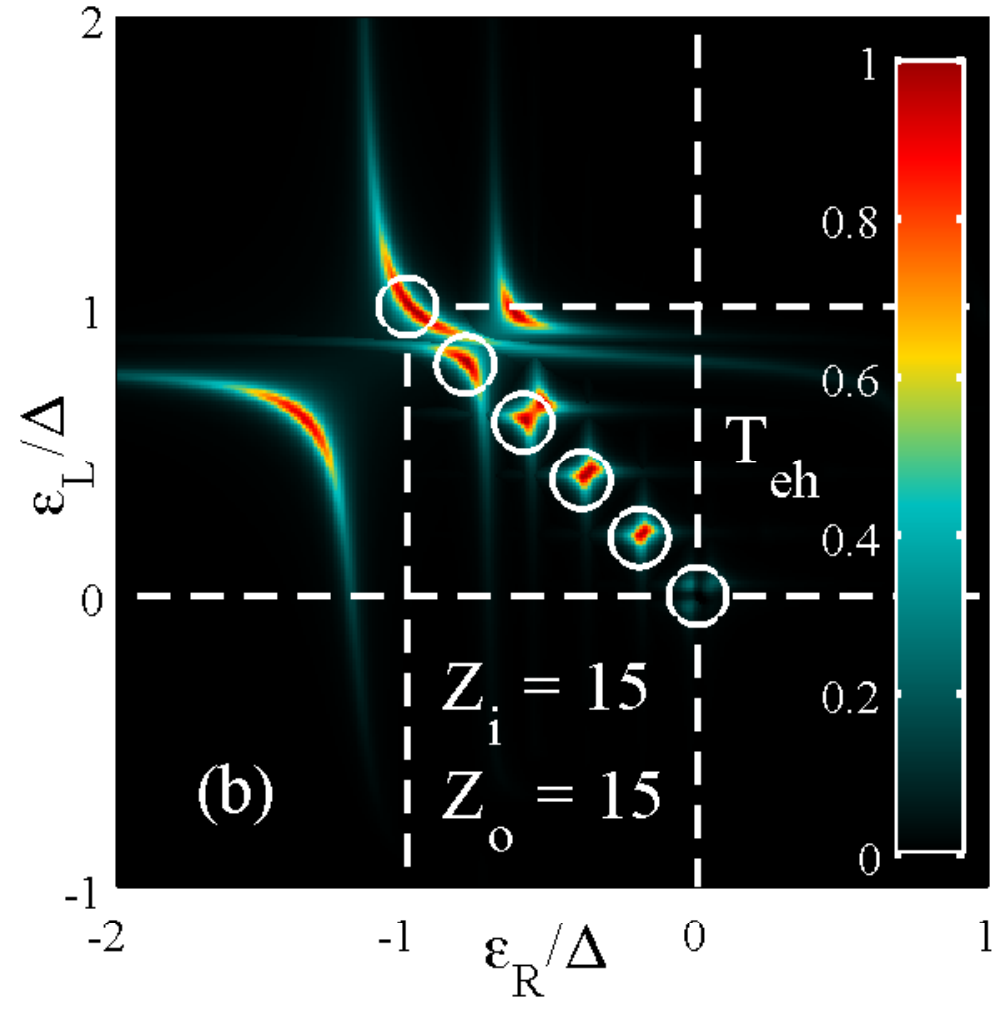}
			\label{fig:T_eLeR_fixE_15}
		}
	\end{center} \vspace{-5mm}
	\caption{
		Color plots of transparency ${\tilde T}_{\rm eh}$ as functions 
		of~$\varepsilon_\rR$ and $\varepsilon_\rL$ for $L/\xi_0 = 1$, 
		$\delta_\rLR/\Delta = 1$, and $\theta = \pi n$. Plots for different 
		energies~$\varepsilon$ of incident electron are placed on top 
		of each other. Resonances corresponding to 
		$\varepsilon/\Delta = 0.01$, $0.2$, $0.4$, $0.6$, $0.8$, and $0.99$ 
		are marked with circles with centers at $(\varepsilon_\rL,\varepsilon_\rR) 
		= (\varepsilon, -\varepsilon)$.
		(a)~$Z_\rio = 10$ ($\Gamma_\sLR / \Delta = 0.1$).
		(b)~$Z_\rio = 15$ ($\Gamma_\sLR / \Delta = 0.045$).
	}
	\label{fig:T_eLeR_fixE}
\end{figure}

The analysis of the transparency ${\tilde T}_{\rm eh}$ as a function of~$\varepsilon_\rL$ and $\varepsilon_\rR$ for different energies of an incident electron~$\varepsilon$ is presented in a form of the color plots in Fig.~\ref{fig:T_eLeR_fixE}. We choose resonance half-widths $\Gamma_{\!\rL} = \Gamma_{\!\rR} = 0.1\Delta$ in Fig.~\subref{fig:T_eLeR_fixE_10} and $\Gamma_{\!\rL} = \Gamma_{\!\rR} = 0.045\Delta$ in Fig.~\subref{fig:T_eLeR_fixE_15} to be smaller than the superconducting gap. Typically, the transparency for each energy $\varepsilon$ has a pronounced peak at $(\varepsilon_\rL,\varepsilon_\rR) \sim (\varepsilon, -\varepsilon)$. The peaks at energies $\varepsilon/\Delta = 0.01$, $0.2$, $0.4$, $0.6$, $0.8$, and $0.99$ are well separated, so one can absorb all the dependencies of ${\tilde T}_{\rm eh}$ of $\varepsilon_\rL$ and $\varepsilon_\rR$ in the same figure. These peaks are marked with white circles. For energies near the Fermi level $\varepsilon \sim 0$ the resonance is expected to be at $(\varepsilon_\rL,\varepsilon_\rR) \sim (0, 0)$, but, as follows from Eq.~\eqref{eq:teh_INISINI}, at this point $\varepsilon_\rL = \varepsilon_\rR = \varepsilon = 0$ and as for symmetric barriers $r_{\rm i} = r_{\rm o}$, the transparency is suppressed, ${\tilde T}_{\rm eh} = 0$. For larger energies $0 \lesssim \varepsilon \leqslant \Delta$ the locus of resonances is about the `diagonal' $(\varepsilon_\rL,\varepsilon_\rR) \sim (\varepsilon, -\varepsilon)$. At $(\varepsilon_\rL,\varepsilon_\rR) = (\varepsilon, -\varepsilon)$ and symmetric INI barriers $r_{\rm i} = r_{\rm o}$ the transmission amplitude reduces to the `ideal' case given by Eq.~\eqref{eq:teh_INISINI_res}. In addition, the transparency has one extra peak at `off-diagonal' line $\varepsilon_\rL - \varepsilon_\rR = 2\varepsilon$ for any given energy $\varepsilon$, especially for $\varepsilon$ close to $\Delta$. At this line the electron reflection amplitude to the left coincides with the hole reflection amplitude to the right, $r_{\rL{\rm i}}^{\rm e} = r_{\rR{\rm i}}^{\rm h}$. This resembles the symmetric Fabry-P\'erot interferometer with the unity transparency at the resonances and gives rise to the `off-diagonal' resonances in the INI-S-INI system. 

\begin{figure}[tb]
	\vspace{-2mm} \begin{center}
		\subfloat{
			\hspace{-3mm} \includegraphics[width=4.3cm]{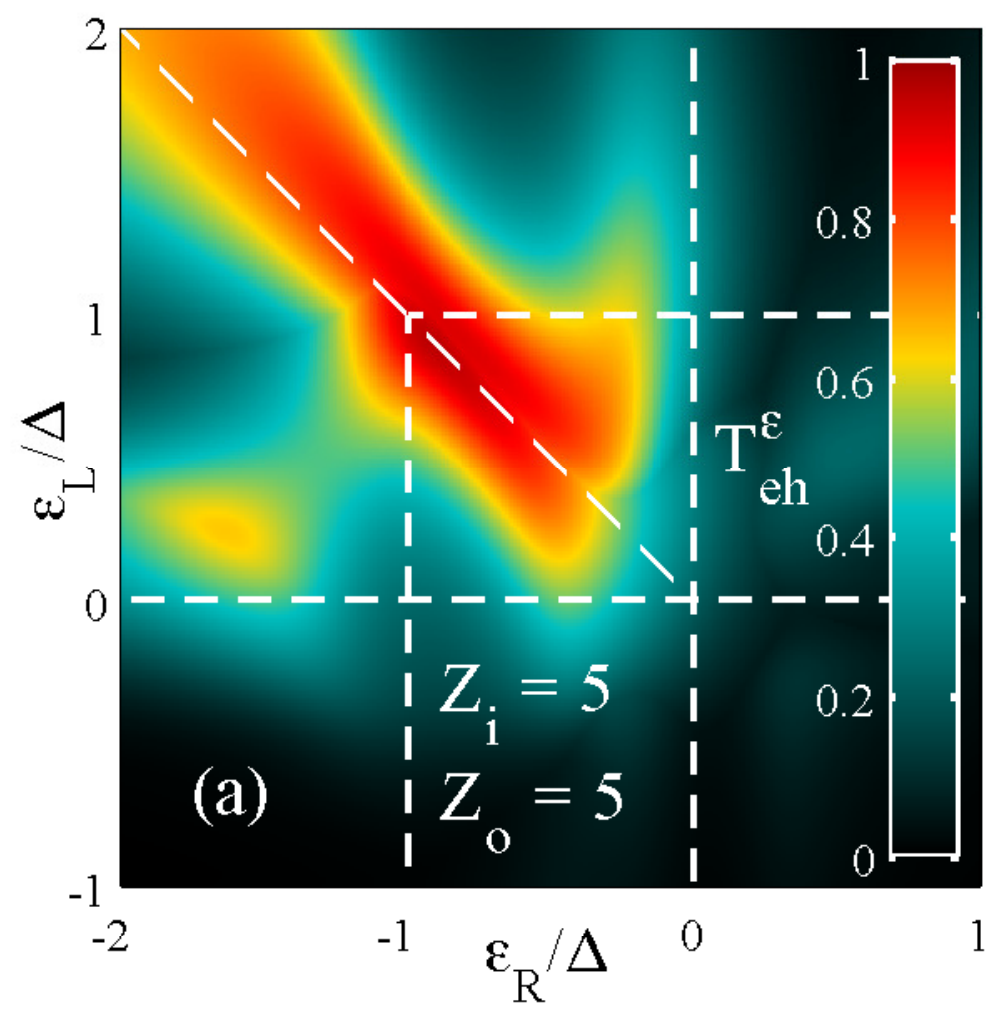}
			\label{fig:T_eLeR_maxE_55}
		}
		\subfloat{
			\hspace{-2mm} \includegraphics[width=4.3cm]{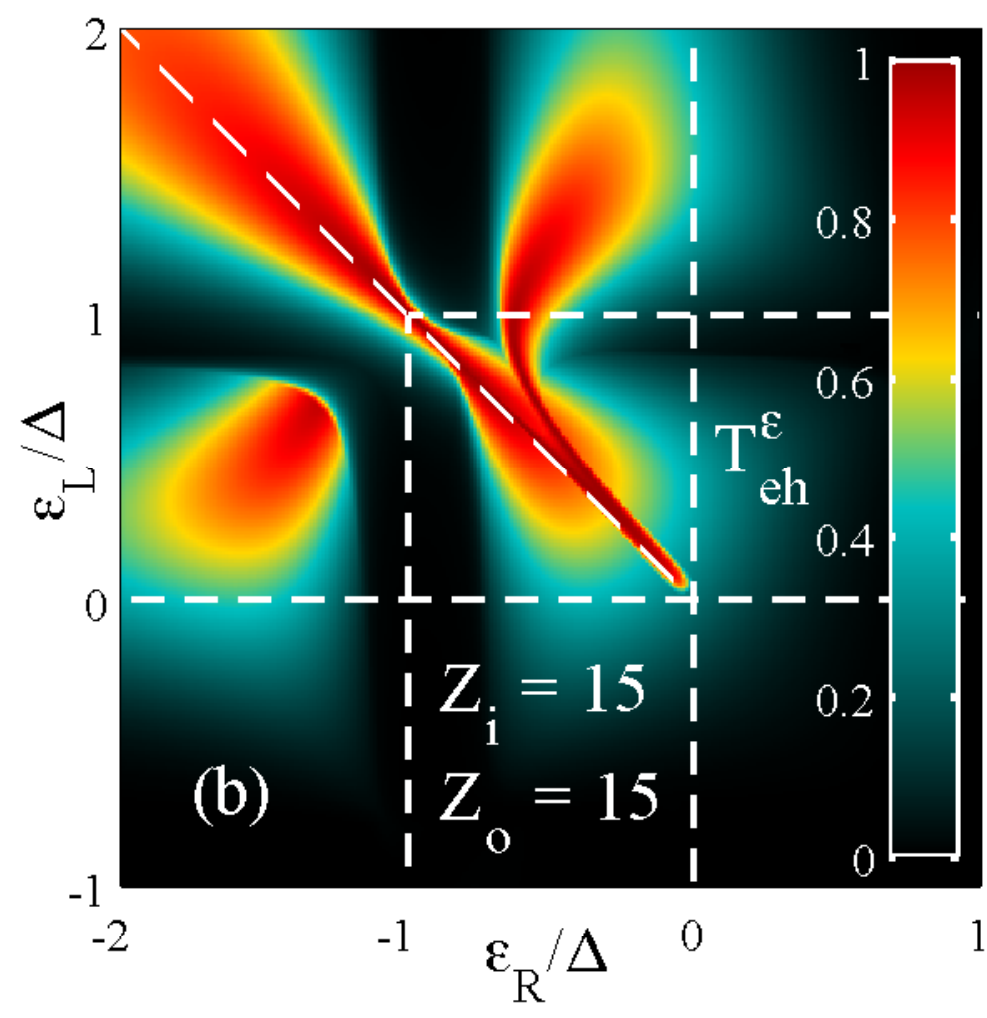}
			\label{fig:T_eLeR_maxE_1515}
		} \\ \vspace{-3mm}
		\subfloat{
			\hspace{-3mm} \includegraphics[width=4.3cm]{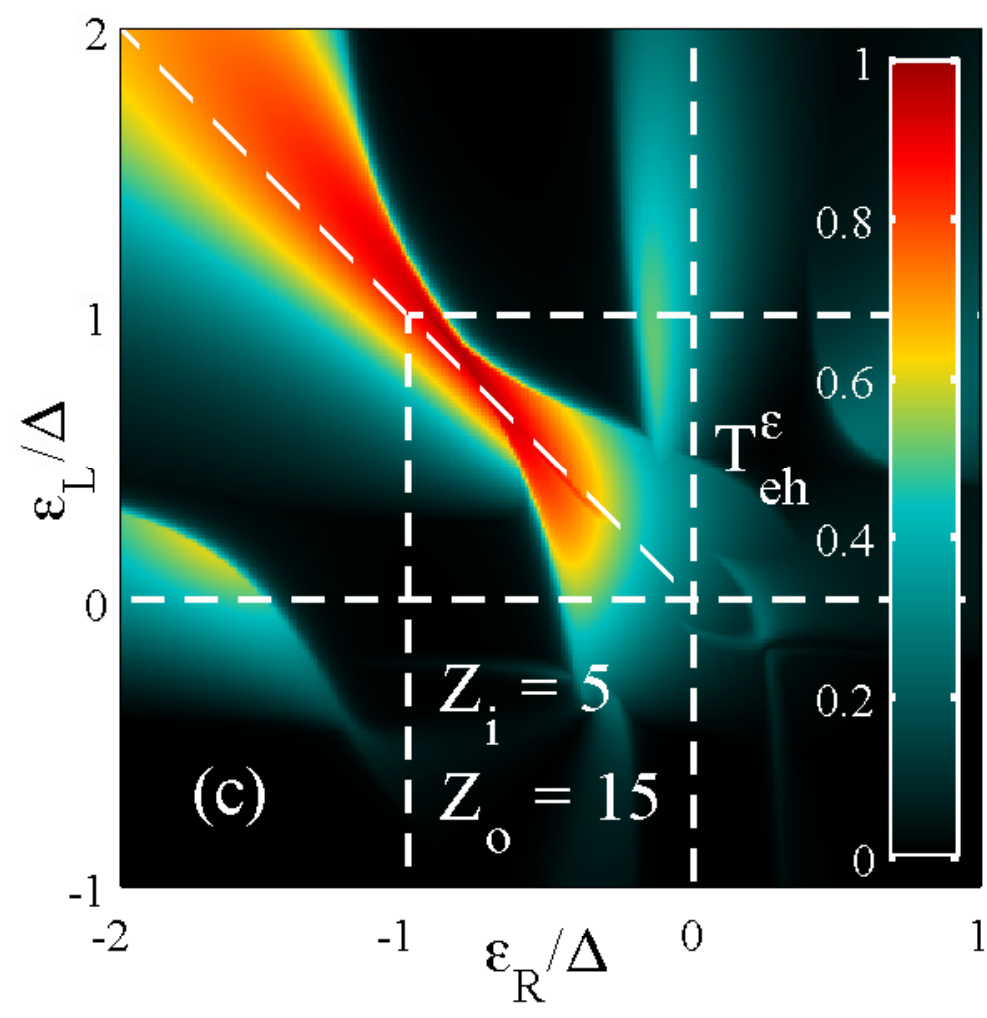}
			\label{fig:T_eLeR_maxE_515}
		}
		\subfloat{
			\hspace{-2mm} \includegraphics[width=4.3cm]{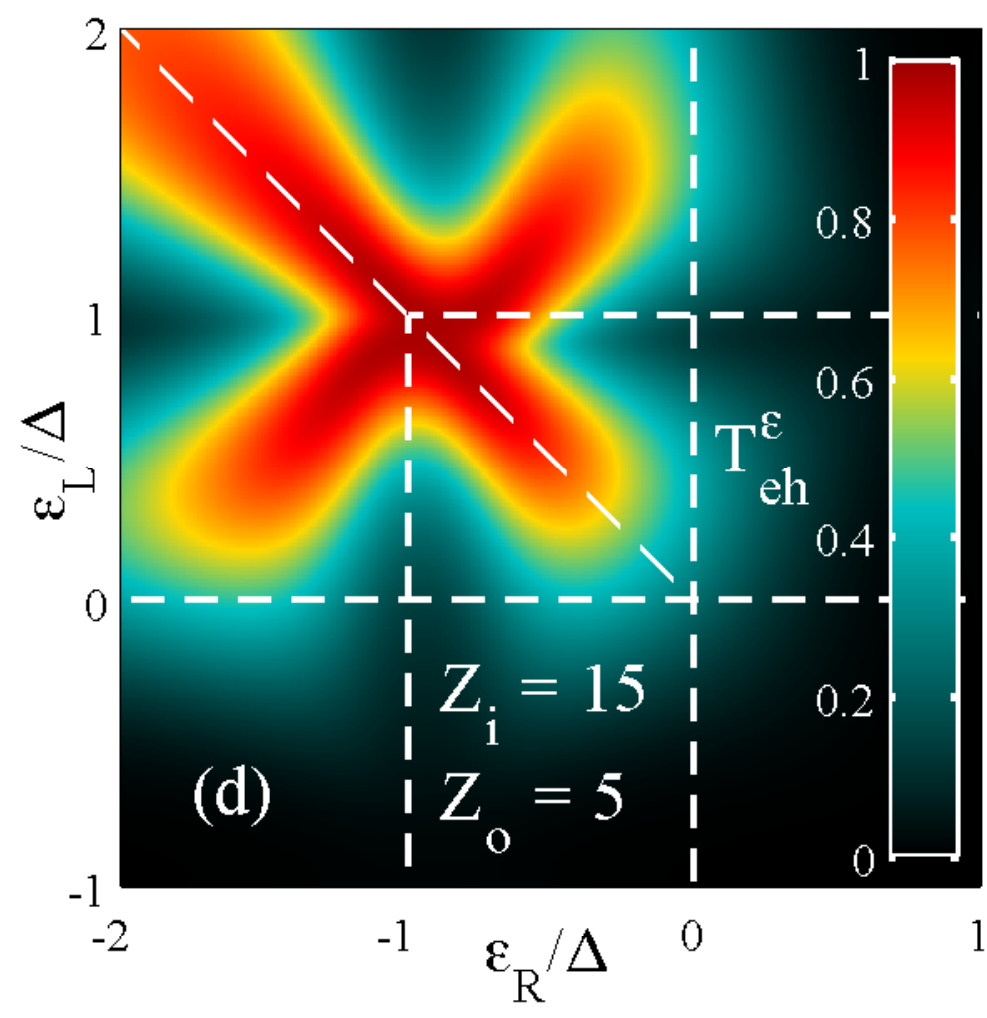}
			\label{fig:T_eLeR_maxE_155}
		} 
		\\ \vspace{-3mm}
		\subfloat{
			\hspace{-3mm} \includegraphics[width=4.3cm]{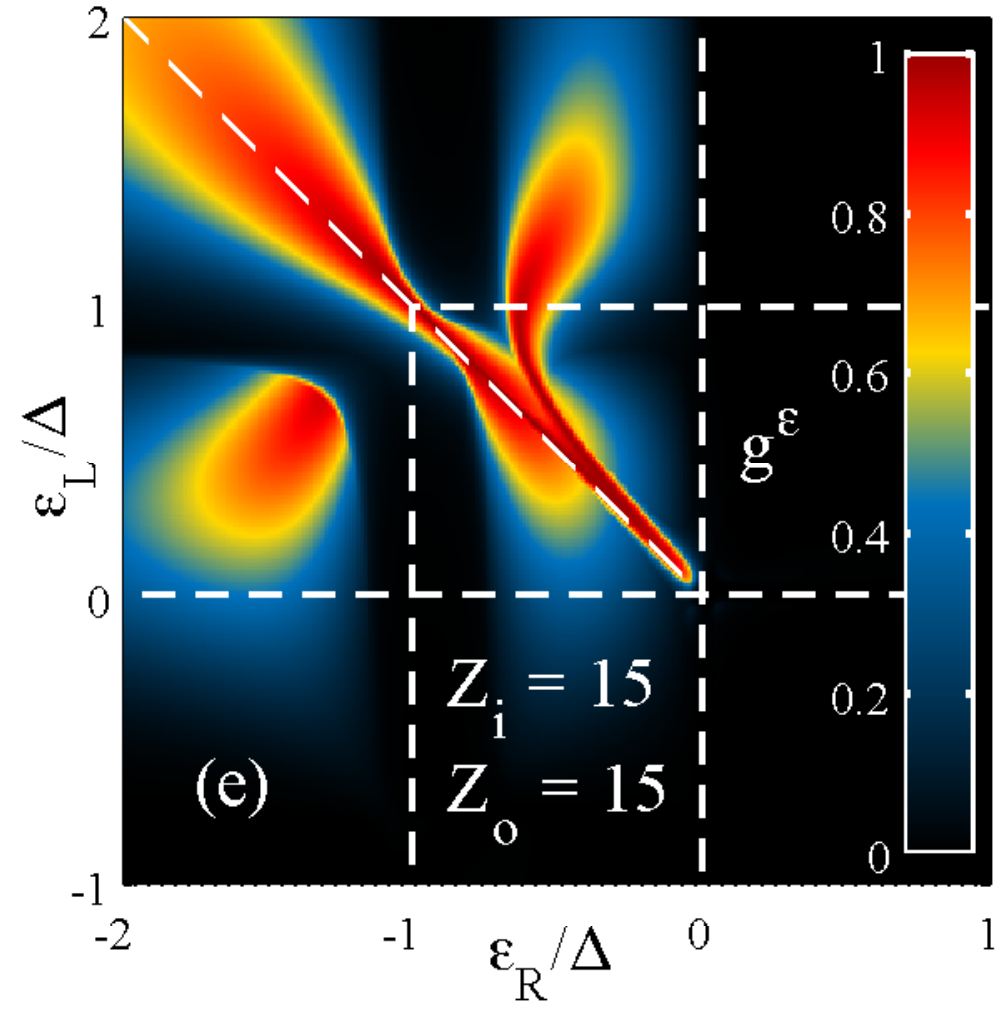}
			\label{fig:G_eLeR_maxE_1515}
		}
		\subfloat{
			\hspace{-2mm} \includegraphics[width=4.3cm]{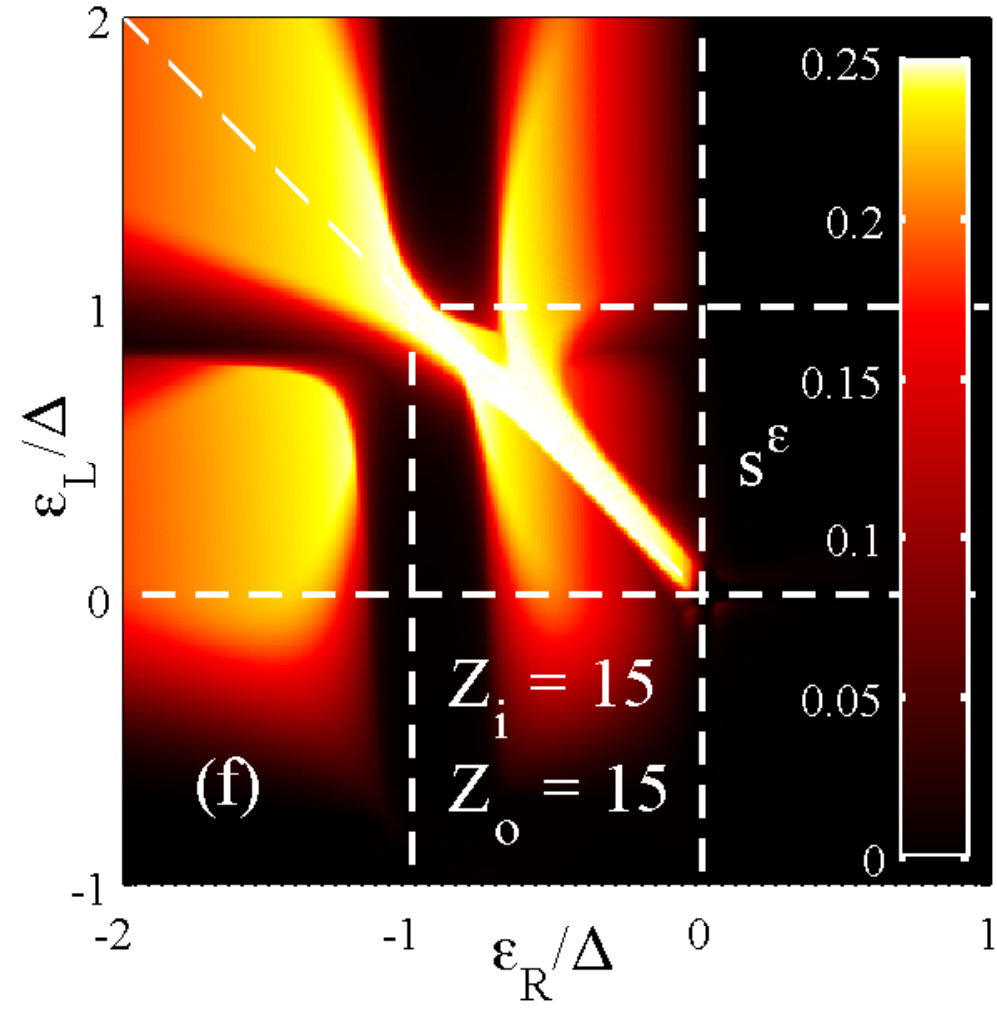}
			\label{fig:S_eLeR_maxE_1515}
		}
	\end{center} \vspace{-5mm}
	\caption{
		(a) Color plot of the maximal transparency ${\tilde T}_{\rm eh}^\varepsilon 
		= \max_\varepsilon\{{\tilde T}_{\rm eh}\}$
		as function of~$\varepsilon_\rR$ and $\varepsilon_\rL$ 
		for $Z_\rio = 5$ ($\Gamma_\sLR / \Delta = 0.39$).
		(b)~${\tilde T}_{\rm eh}^\varepsilon$ for $Z_\rio = 15$ ($\Gamma_\sLR / \Delta = 0.045$).
		(c)~${\tilde T}_{\rm eh}^\varepsilon$ for $Z_{\rm i} = 5$ and $Z_{\rm o} = 15$ 
		($\Gamma_\sLR / \Delta = 0.22$).
		(d)~${\tilde T}_{\rm eh}^\varepsilon$ for $Z_{\rm i} = 15$ and $Z_{\rm o} = 5$.
		(e)~Maximal differential conductance $g^\varepsilon 
		= \max_\varepsilon\{{\tilde T}_{\rm eh} - {\tilde T}_{\rm ee}\}$ for $Z_\rio = 15$.
		(f)~Maximal differential noise $s^\varepsilon 
		= \max_\varepsilon\{{\tilde T}_{\rm eh}(1-{\tilde T}_{\rm eh}) - 
		{\tilde T}_{\rm ee}(1-{\tilde T}_{\rm ee})\}$ for $Z_\rio = 15$.
		We choose $L/\xi_0 = 1$, $\delta_\rLR/\Delta = 10$, and $\theta = \pi n$ for all plots.
	}
	\label{fig:T_eLeR_maxE}
\end{figure}

Further information about the maximum CPS outcome can be obtained by considering the maximal transparency ${\tilde T}_{\rm eh}^\varepsilon = \max_\varepsilon\{{\tilde T}_{\rm eh}\}$ behavior. Figures~\subref{fig:T_eLeR_maxE_55} and \subref{fig:T_eLeR_maxE_1515} demonstrate color plots of ${\tilde T}_{\rm eh}^\varepsilon = \max_\varepsilon\{{\tilde T}_{\rm eh}\}$ as functions of $\varepsilon_\rL$ and $\varepsilon_\rR$ for symmetric INI barriers for $Z_\rio = 5$ ($\Gamma_\sLR = 0.39\Delta$) and $Z_\rio = 15$ ($\Gamma_\sLR = 0.045\Delta$), correspondingly (increasing resonance sharpness). One sees the permanent resonance along the tilted white line corresponding to the ideal case defined by Eq.~\eqref{eq:Teh_INISINI_res}. In Fig.~\subref{fig:T_eLeR_maxE_515} the asymmetric case is shown for the inner barrier strength $Z_{\rm i} = 5$ smaller then outer barrier strength $Z_{\rm o} = 15$ ($\Gamma_\sLR = 0.22\Delta$): the maximum is determined not by the internal INI resonance, but by the resonance between the outer walls. The reversed situation, $Z_{\rm i} = 15$ and $Z_{\rm o} = 5$ is presented in Fig.~\subref{fig:T_eLeR_maxE_155}. One observes additional resonances, e.g. loci of ${\tilde T}_{\rm eh}^\varepsilon$ form a cross-like configuration comprising the diagonal and the segment of the line $\varepsilon_\rL - \varepsilon_\rR = 2\Delta$ in Fig.~\subref{fig:T_eLeR_maxE_155}. While at the `diagonal' the resonances (along line $\varepsilon_\rL = - \varepsilon_\rR$) originate from the ideal case~\eqref{eq:teh_INISINI_res}, the additional `crossbar' loci of maximal transparencies stem from the `off-diagonal' resonances in Fig.~\ref{fig:T_eLeR_fixE}. The regimes with strong and nearly symmetric resonances are stable against added randomness so that~${\tilde T}_{\rm eh}$ as a function of $\varepsilon$, $\varepsilon_\rL$, and $\varepsilon_\rR$ has a maximum value of unity. For example, for $Z_\rio = 15$ [Fig.~\subref{fig:T_eLeR_maxE_1515}] unitary limit retains up to $\pm 50\%$ of independent random change in $Z_{\rm i}$ and $Z_{\rm o}$.

\section{Differential conductance and noise} \label{sec:noise}

The measurements of the tunneling probability, in particular, those that correspond to the conversion of an electron into a hole, ${\tilde T}_{\rm eh}$, are indirect and are carried out via the measurements of various transport characteristics of CPS such as differential conductance, ${\rm d}I/{\rm d}V$, and noise, ${\rm d}S/{\rm d}V$. Following the approach outlined in the review Ref.~\onlinecite{Lesovik:2011}, we obtain expressions for the differential conductance and noise through the channels providing electron-to-electron, ${\tilde T}_{\rm ee}$, and electron-to-hole, ${\tilde T}_{\rm eh}$, scatterings and discuss their dependences of the gate potentials, $\varepsilon_\rR$ and $\varepsilon_\rL$.

We start with the differential conductance 
\begin{equation}
	{\rm d}I/{\rm d}V = (2e^2/h)g, \quad
	g = {\tilde T}_{\rm eh} - {\tilde T}_{\rm ee},
\end{equation}
where $h$ is the Planck constant, and transmission probabilities ${\tilde T}_{\rm eh} = |{\tilde t}_{\rm eh}|^2$ and ${\tilde T}_{\rm ee} = |{\tilde t}_{\rm ee}|^2$ are defined by expressions~\eqref{eq:teh_INISINI} and \eqref{eq:tee_INISINI}. The direction of the current $I$ from the right terminal to the superconductor was chosen as a positive direction. The color plot Fig.~\subref{fig:G_eLeR_maxE_1515} of the maximal conductance for the exemplary case $Z_\rio = 15$ reproduces in the main those of ${\tilde T}_{\rm eh}$ [Fig.~\subref{fig:T_eLeR_maxE_1515}], besides that the differential conductance has slightly sharper peaks and is almost suppressed at $\varepsilon_\rR > 0$. 

Next, we calculate the differential noise 
\begin{equation}
	{\rm d}S/{\rm d}V = (2e^3/h) s, \quad
	s = {\tilde T}_{\rm eh}(1-{\tilde T}_{\rm eh}) - {\tilde T}_{\rm ee}(1-{\tilde T}_{\rm ee})
\end{equation}
at the zero frequency corresponding to the cross-correlator of the current in the left and right terminals (positive directions are chosen towards the superconductor), where $V$ is the voltage at the left terminal, while at the superconductor and at the right terminal the voltage is zero. Figure~\subref{fig:S_eLeR_maxE_1515} shows that the locus of the maximal value $s = 1/4$ is similar to that of ${\tilde T}_{\rm eh} = 1$. (The locus of $s = 1/4$ almost corresponds to ${\tilde T}_{\rm eh} = 1/2$, which, in turn, is obtained from ${\tilde T}_{\rm eh} = 1$ by a slight shift in energy.) The cross-correlators can be used for characterization of the effectiveness of the entangler.\cite{Chtchelkatchev:2002,Hofstetter:2009,Wei:2010} Appendix~\ref{sec:additional_plots} presents more color plots of ${\tilde T}_{\rm eh}$, $g$, and $s$ for variety of parameters.

\section{Effect of the Y-geometry} \label{sec:Y_geometry}

\begin{figure}[b]
	\begin{center}
		\subfloat{\includegraphics[width=7.6cm]{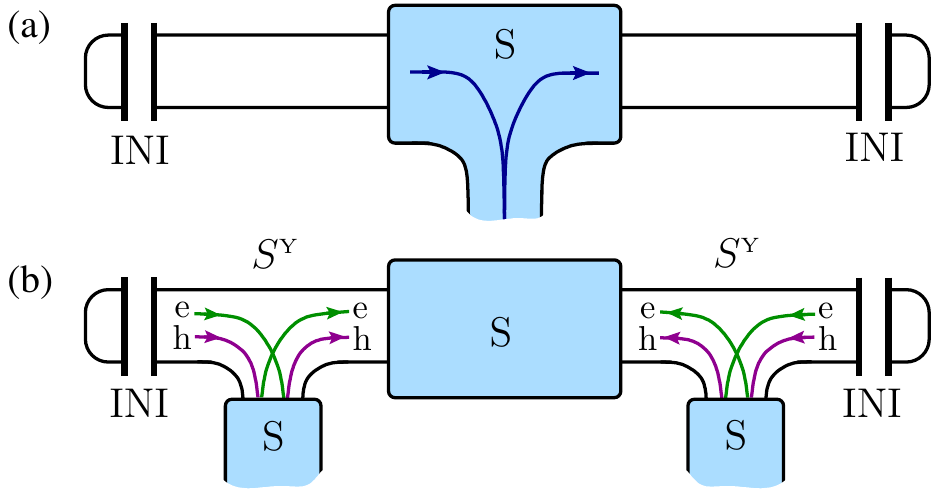} \label{fig:setup_nsn_ground}} 
		\subfloat{\label{fig:setup_nsn_ground_model}} \\ \vspace{-2mm}
		\subfloat{\includegraphics[width=8.5cm]{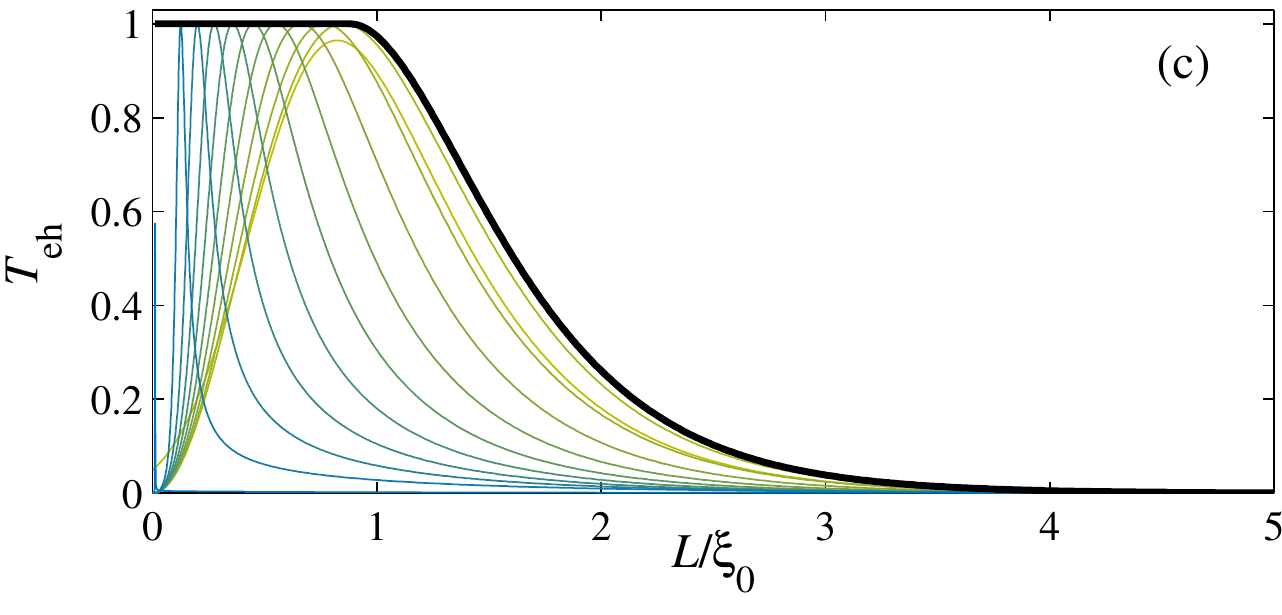} \label{fig:Tmax_grounded}}
	\end{center} \vspace{-6mm}
	\caption{
		(a)~Sketch of the Cooper pair splitter based on INI-S-INI 
		structure with grounded superconducting part. 
		(b)~Model of the grounded entangler. Green and magenta 
		lines show additional trajectories of electrons and holes.
		(c)~Transparency ${\tilde T}_{\rm eh}$ as a function of $L/\xi_0$ 
		for $\theta = \pi n$, maximally connected infinite superconductors 
		($\gamma = \pi/2$ in ${\cal S}^\rY$), 
		and different energies of the incident electron $\varepsilon$.
	}
	\label{fig:nsn_ground}
\end{figure}

In the typical experimental setups the central superconducting island is grounded,~\cite{Hofstetter:2009,Hofstetter:2011,Herrmann:2010,Schindele:2012,Tan:2015} see the corresponding sketch in Fig.~\subref{fig:setup_nsn_ground}. We thus discuss now how the tunneling probabilities, ${\tilde T}_{\rm eh}$, change upon grounding the central island. A superconducting island of the finite length $L$ is characterized by the Andreev reflecting amplitudes $r_{\rm ee(hh)}$ with the non-unity moduli, enabling thus splitting a Cooper pair into separate normal leads. At the perfect interface between the metal and an infinite one-dimensional superconductor a full Andreev reflection, where electron converts into a hole and vice versa, occurs. We, hereafter, will be referring to this kind of a superconductor as to the grounded one. To incorporate the grounding into our one-dimensional problem, let us wedge two infinite superconductors between the quantum dots and the central superconductor, see Fig.~\subref{fig:setup_nsn_ground_model}. This scheme accounts for all possible scattering processes in the system with the grounded central part. The electron-to-hole transmission probability can be found by the reduction of the grand scattering matrix of the system.\cite{Sadovskyy:2015} We assume that all superconductors in our circuit have the same phase. Then both electron-to-hole and hole-to-electron reflections at the interfaces between the terminals and the infinite superconductors result in gaining of the same phase, i.e., $r_{\rm eh(he)} = e^{-i\alpha}$ [see green and red arrows in Fig.~\subref{fig:setup_nsn_ground_model}]. We connect grounded superconductors through the normal Y-junctions. Scattering matrices ${\cal S}^\rY$ of the normal metal Y-junctions, 
\begin{equation}
	\left[ \begin{array}{l}
		b_\rL^{\rm e\leftarrow} \\ 
		b_\rR^{\rm e\rightarrow} \\ 
		b_\rB^{\rm e\downarrow} 
	\end{array} \right]
	= {\cal S}^\rY
	\left[ \begin{array}{l}
		b_\rL^{\rm e\rightarrow} \\ 
		b_\rR^{\rm e\leftarrow} \\
		b_\rB^{\rm e\uparrow}
	\end{array} \right],
	\quad
	\left[ \begin{array}{l}
		b_\rL^{\rm h\leftarrow} \\ 
		b_\rR^{\rm h\rightarrow} \\ 
		b_\rB^{\rm h\downarrow} 
	\end{array} \right]
	= {\cal S}^\rY
	\left[ \begin{array}{l}
		b_\rL^{\rm h\rightarrow} \\ 
		b_\rR^{\rm h\leftarrow} \\
		b_\rB^{\rm h\uparrow}\,,
	\end{array} \right]
	\nonumber
\end{equation}
are assumed to be the same and energy independent. They can be parametrized as follows\cite{Jarlskog:2005}
\begin{equation}
	{\cal S}^\rY
	= \left[ \begin{array}{ccc}
		- \frac{(1-\cos\gamma)}{2}	e^{i(\delta_1-\delta_2)} &	\frac{1+\cos\gamma}{2}	&	\frac{\sin\gamma}{\sqrt{2}} e^{i\delta_1} \\
		\frac{1+\cos\gamma}{2}	&	- \frac{1-\cos\gamma}{2} e^{i(\delta_2-\delta_1)}	&	\frac{\sin\gamma}{\sqrt{2}} e^{i\delta_2} \\
		- \frac{\sin\gamma}{\sqrt{2}} e^{-i\delta_2}	&	- \frac{\sin\gamma}{\sqrt{2}} e^{-i\delta_1}	&	\cos\gamma
	\end{array} \right]\!\!,
	\nonumber
\end{equation}
where we assumed the left-right symmetry. To avoid extra resonances we set additional phases to zero, $\delta_1 = \delta_2 = 0$, so that the Y-junction scattering matrix becomes a function of single parameter $\gamma$ only, which is the measure of coupling to the infinite (grounded) superconductors,
\begin{equation}
	\! {\cal S}^\rY
	\! = \! \left[\!\! \begin{array}{ccc}
		- (1-\cos\gamma)/2	&		\;\; (1+\cos\gamma)/2	&		\; \sin\gamma/\sqrt{2} \\
		(1+\cos\gamma)/2	&		- (1-\cos\gamma)/2		&	\; \sin\gamma/\sqrt{2} \\
		\!\!\! - \sin\gamma /\sqrt{2}	&	-\sin\gamma/\sqrt{2}		&	\; \cos\gamma
	\end{array} \!\right]\!\!.
\end{equation}
The combination of Y-splitter and infinite superconductor results in the additional mixing between electron- and hole-like states. However, number of quasiparticles conserves during this mixing, so, this process does not break down the possibility of the 100\% efficiency. This is shown in Fig.~\subref{fig:Tmax_grounded} which displays the dependence of the electro-to-hole transmission probability as a function of central superconductor length $L$. Comparison with Fig.~\ref{fig:Tmax} shows that the positions of maxima of ${\tilde T}_{\rm eh}$ merely shift.

\section{Discussion and conclusion} \label{sec:conclusion}

We expect the efficient CPS to hold in much broader class of systems than what we discussed. First, we have demonstrated that the unitary limit is strikingly robust against random deviation of scatterer's parameters from the perfect symmetry. Second, it is important that the interference process providing the unitary CPS limit can be achieved by contribution from two paths like in Mach-Zehnder interferometer, see Fig.~\ref{fig:setup_nsn}. This loosens the conditions for the system dimension tuning that ensures the maximal efficiency and suggests that the efficient CPS may hold in a higher dimensionality. Indeed, relevant semi-classical trajectories connecting two dots at the opposite terminals (like in the experiment of Ref.~\onlinecite{Tan:2015} with about 10\% efficiency) mimic a 1D situation and at the same time provide fair probability for dots to be joined, even though the traveling wave packets may laterally spread. However, straightforward mapping of the experimental situation onto the 2D dirty-superconductor model gives about two orders of magnitude less than in the experiment. Our findings suggest resolution of this controversy. Furthermore, our prediction that the maximal CPS efficiency is expected at $L \sim \xi_0$ and the plot of Fig.~\ref{fig:Tmax} compare nicely with the experimental result, provided we substitute $\xi_0$ by the coherence length $\xi$ corresponding to the experimental diffusive case.

To conclude, we have demonstrated that the outcome of Cooper pair splitting via the crossed Andreev transport in a one-dimensional hybrid system, comprising a superconductor sandwiched between the two normal metal terminals endowed with the double point scatterers, can achieve a robust unitary limit stable against asymmetry of the scatterers. We found that the electron-to-hole transmission probability (per a conducting channel) across such a system achieves its maximum if the width of the superconductor is of the order of the coherence length. Our study opens a route to reliable high-outcome procedure for creating entangled electrons.

\section*{Acknowledgments}

We thank P.\,Hakonen and D.\,S.\,Golubev for illuminating discussions which to large extent motivated this research. The work was supported by the U.S. Department of Energy, Office of Science, Materials Sciences and Engineering Division (V.\,V. and G.\,L.; G.\,L. was supported through Materials Theory Institute); by the Scientific Discovery through Advanced Computing (SciDAC) program funded by U.S. Department of Energy, Office of Science, Advanced Scientific Computing Research and Basic Energy Science (I.\,S.); and by the RFBR Grant No. 14-02-01287 (G.\,L.).

%\newpage

\clearpage 

\appendix
\begin{widetext} 
\section{Additional plots} \label{sec:additional_plots}
\pagenumbering{roman}

\subsection{Transparency ${\tilde T}_{\rm eh}$ as a function of~$\varepsilon_\rR$ and $\varepsilon_\rL$}

\begin{figure}[H]
	\vspace{-6mm} \begin{center}
		\subfloat{
			\hspace{-2mm} \includegraphics[width=4.4cm]{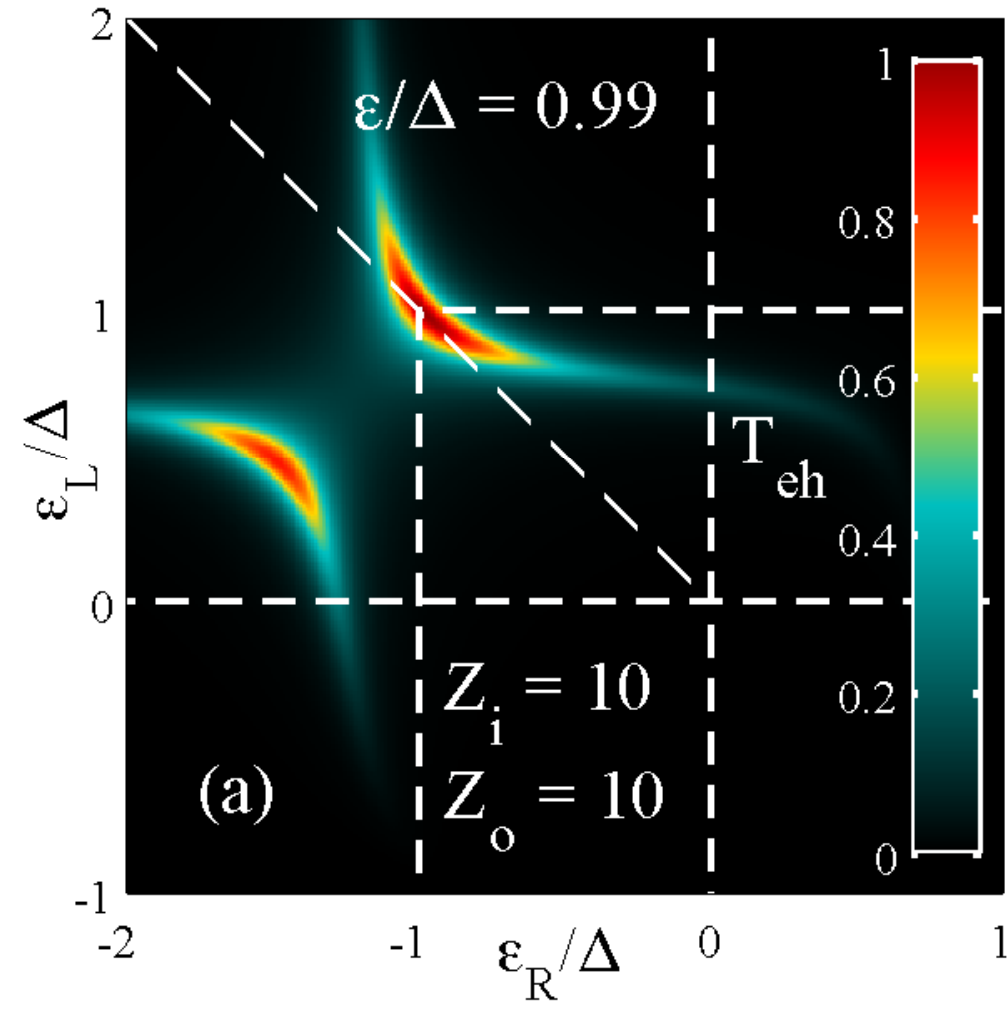}
		}
		\subfloat{
			\hspace{-2mm} \includegraphics[width=4.4cm]{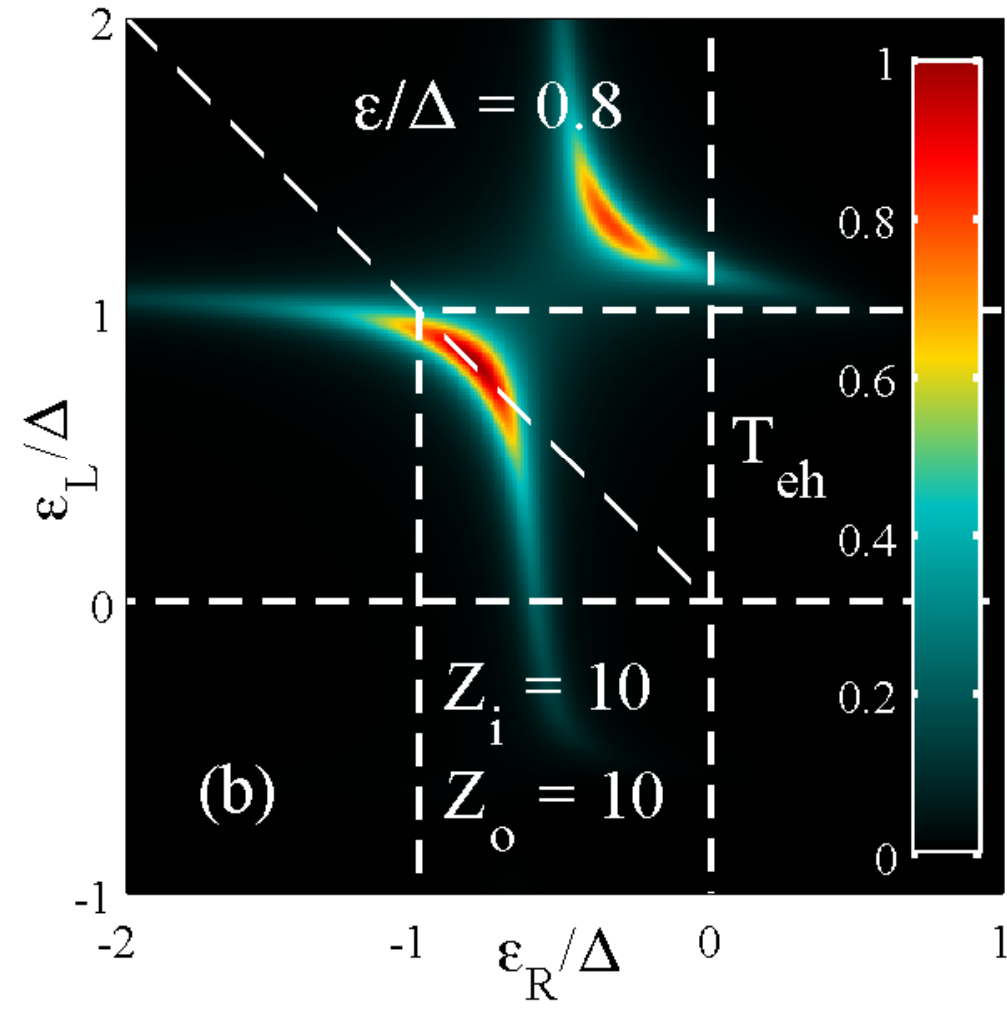}
		}
		\subfloat{
			\hspace{-2mm} \includegraphics[width=4.4cm]{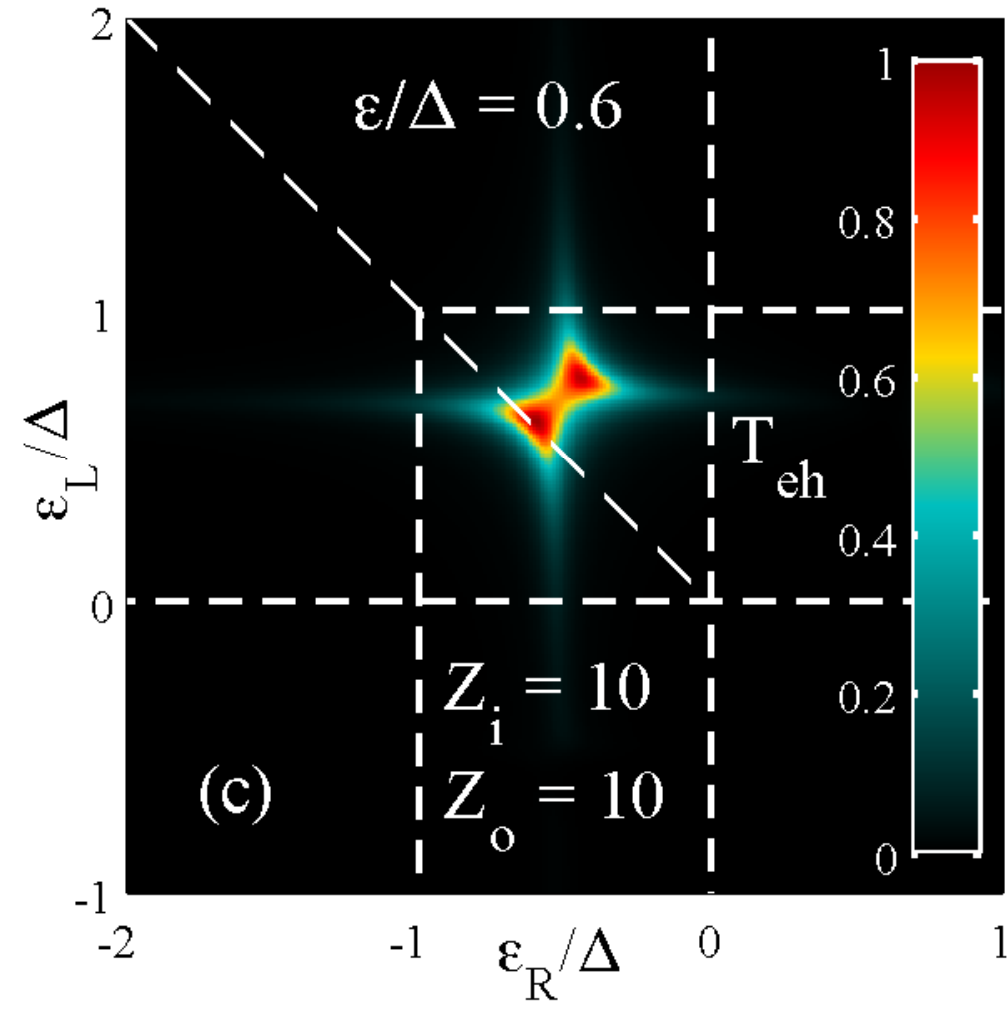}
		} \\ \vspace{-3mm}
		\subfloat{
			\hspace{-3mm} \includegraphics[width=4.4cm]{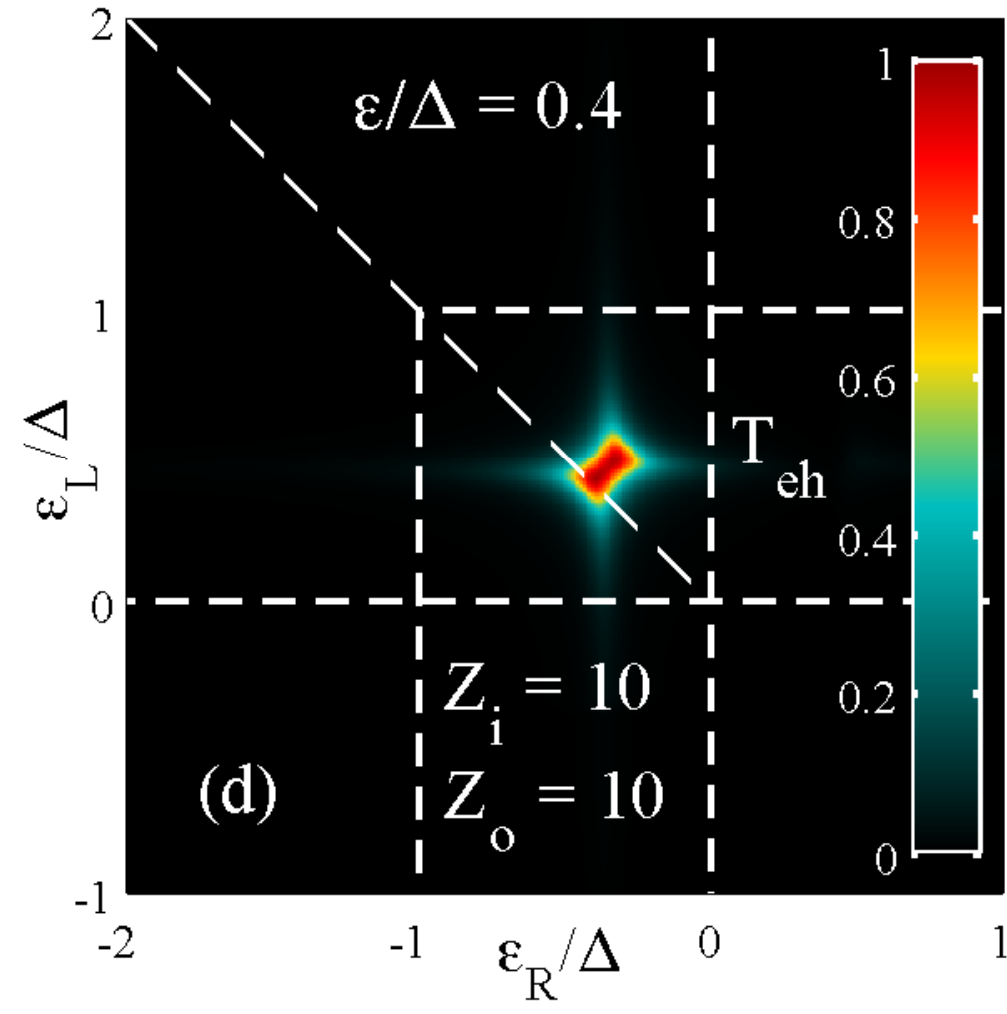}
		}
		\subfloat{
			\hspace{-2mm} \includegraphics[width=4.4cm]{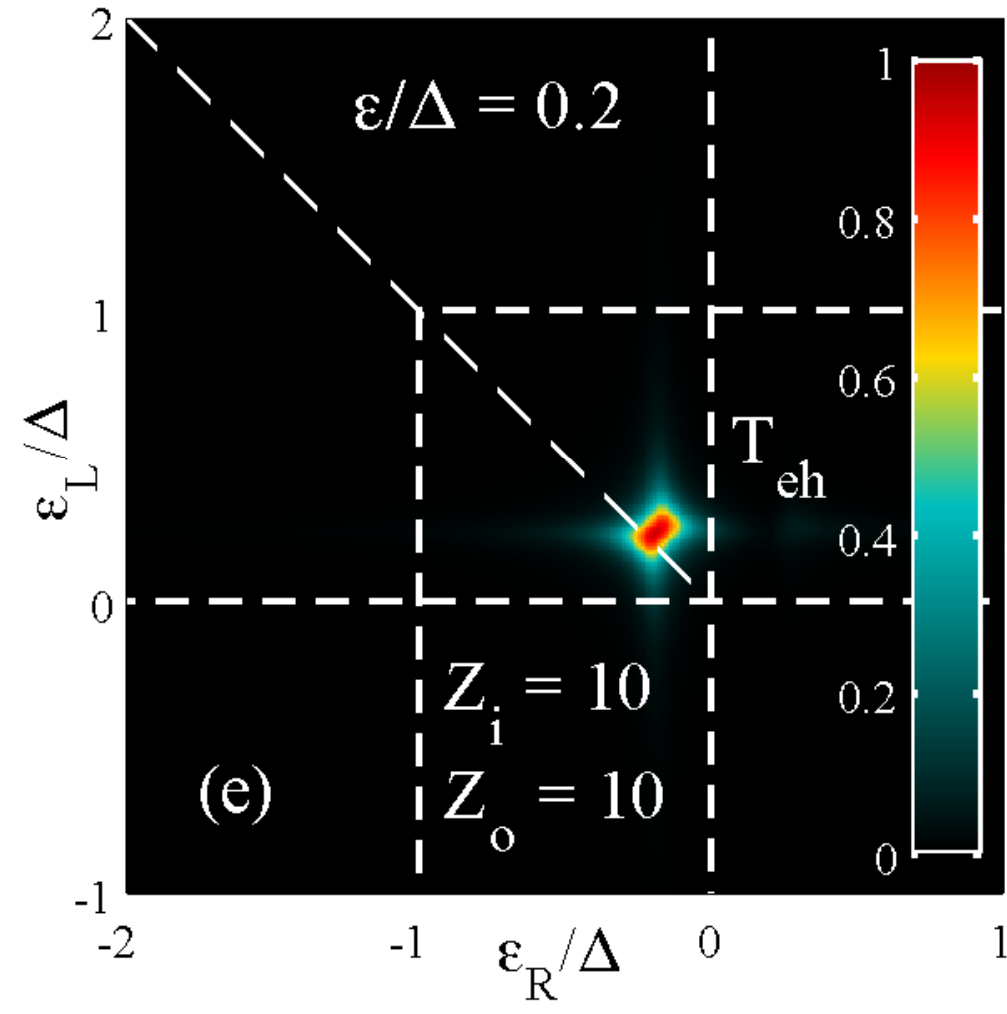}
		}
		\subfloat{
			\hspace{-2mm} \includegraphics[width=4.4cm]{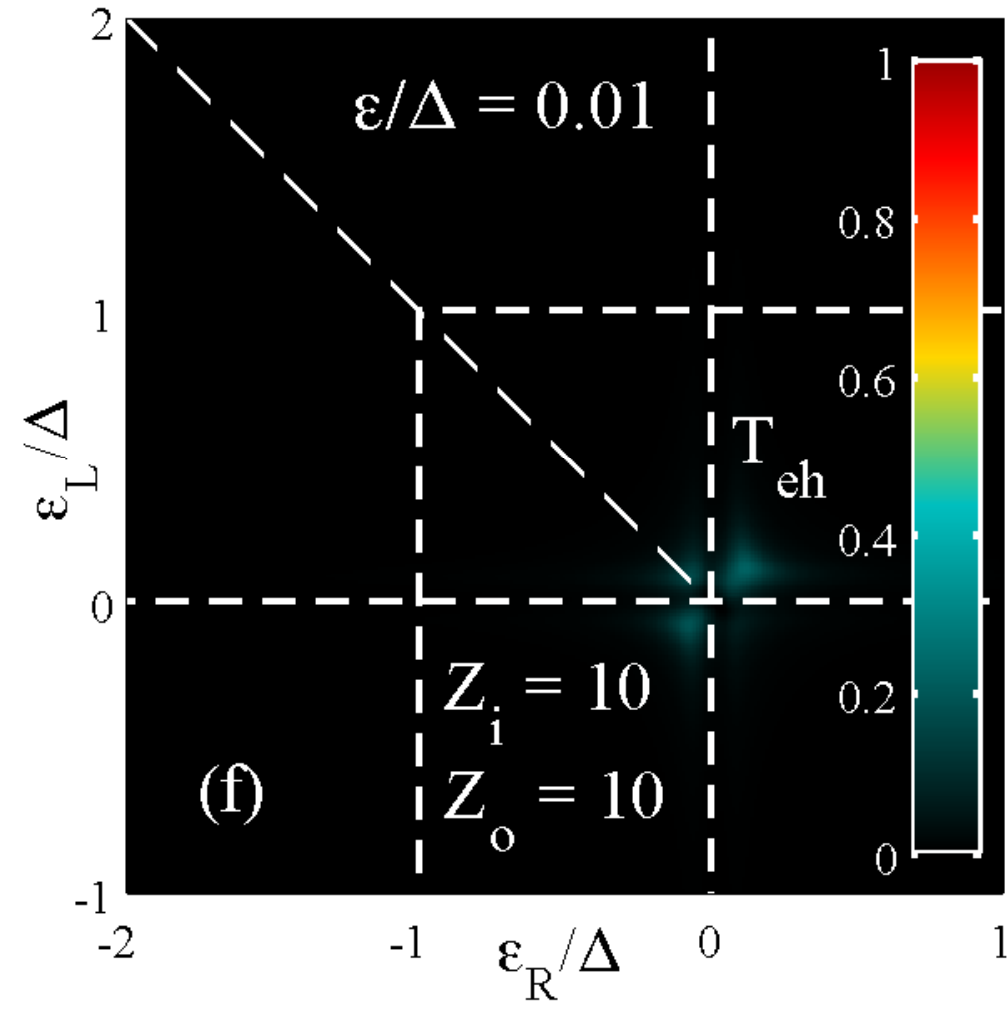}
		}
	\end{center} \vspace{-6mm}
	\caption{
		Transparency ${\tilde T}_{\rm eh}$ as a function of~$\varepsilon_\rR$ and $\varepsilon_\rL$ 
		for $L/\xi_0 = 1$, $\delta_\rLR/\Delta = 10$, and $Z_{\rm i} = Z_{\rm o} = 10$
		($\Gamma_{\!\rL,\rR} / \Delta = 0.1$).
		(a)~$\varepsilon = 0.99$.
		(b)~$\varepsilon = 0.8$.
		(c)~$\varepsilon = 0.6$.
		(d)~$\varepsilon = 0.4$.
		(e)~$\varepsilon = 0.2$.
		(f)~$\varepsilon = 0.01$.
	}
\end{figure}
\begin{figure}[H]
	\vspace{-6mm} \begin{center}
		\subfloat{
			\hspace{-2mm} \includegraphics[width=4.4cm]{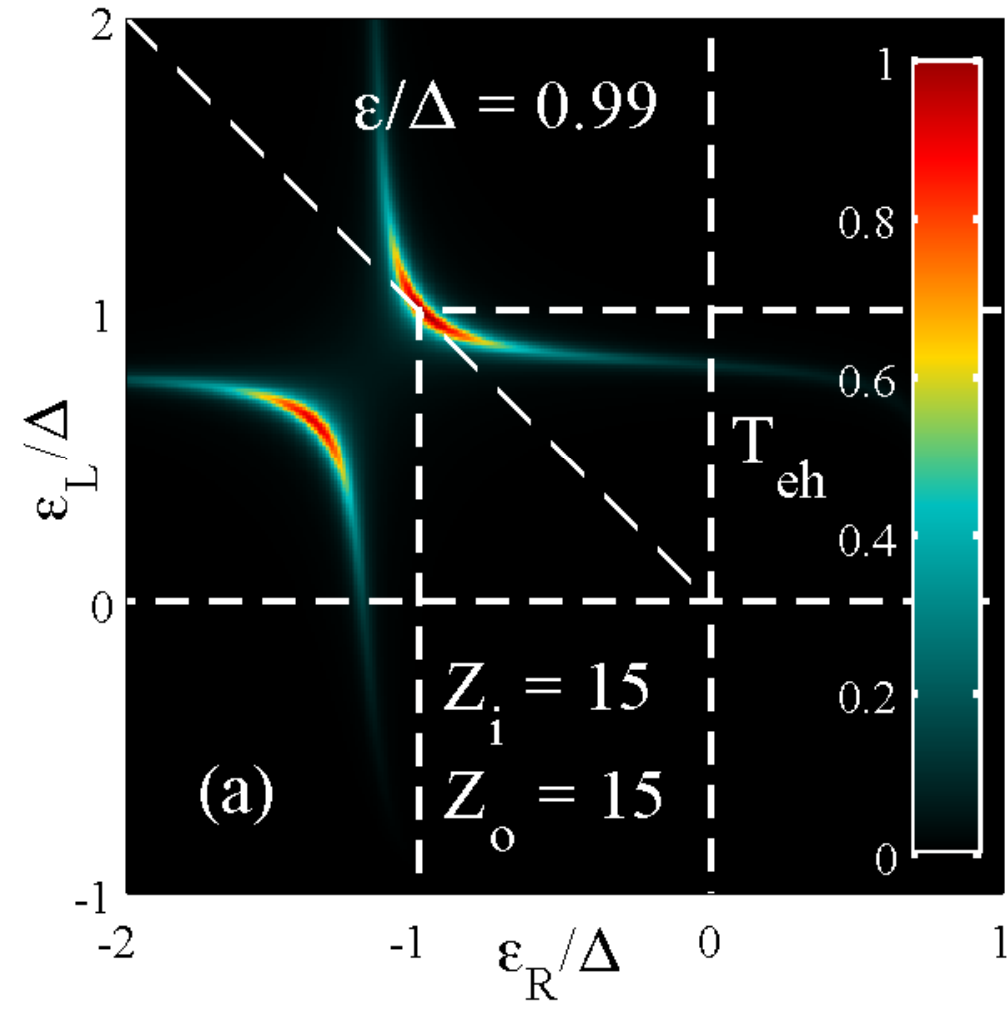}
		}
		\subfloat{
			\hspace{-2mm} \includegraphics[width=4.4cm]{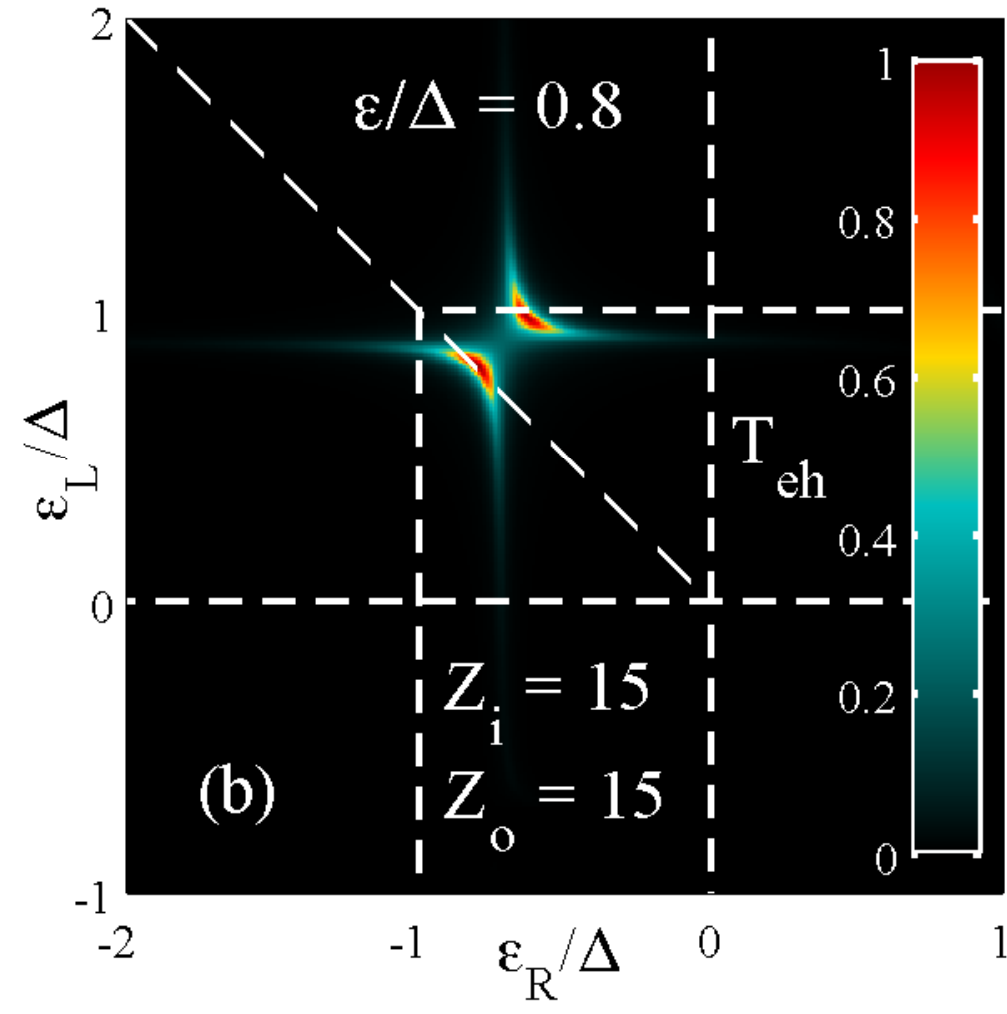}
		}
		\subfloat{
			\hspace{-2mm} \includegraphics[width=4.4cm]{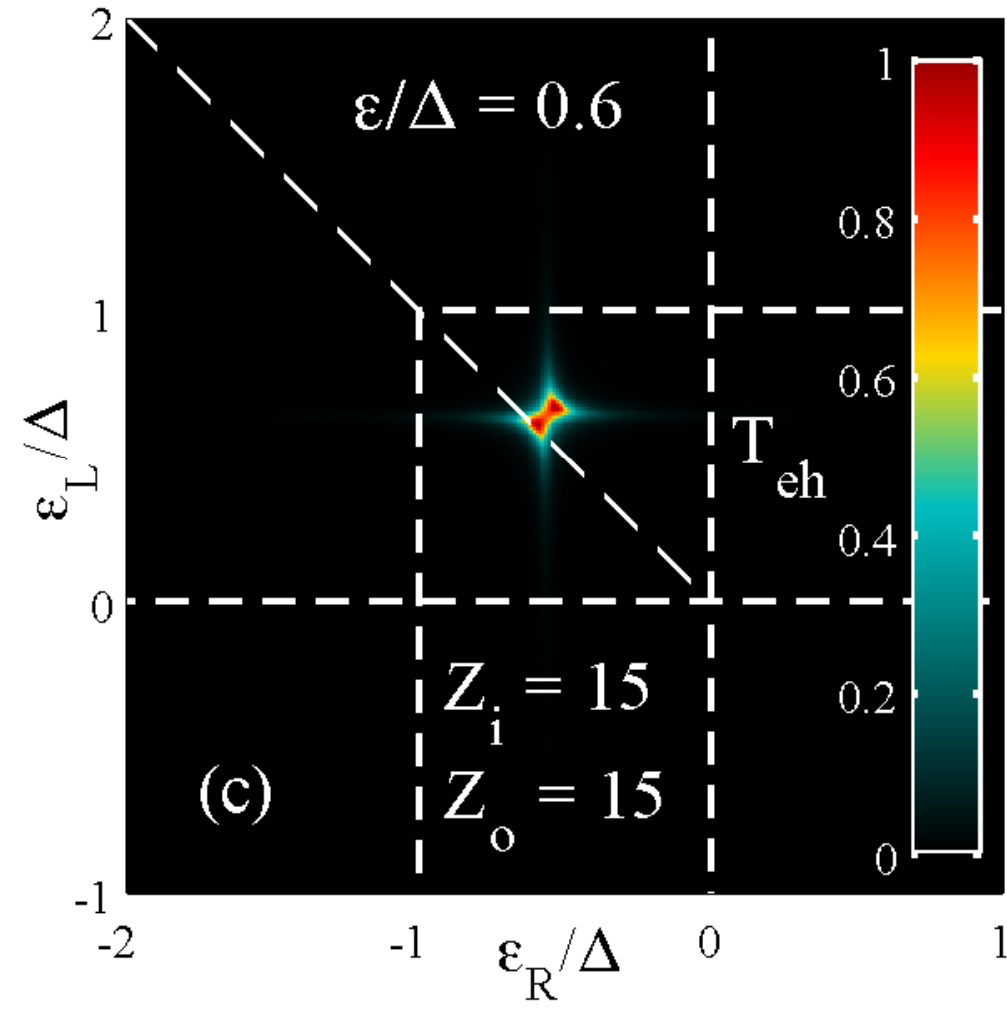}
		} \\ \vspace{-3mm}
		\subfloat{
			\hspace{-3mm} \includegraphics[width=4.4cm]{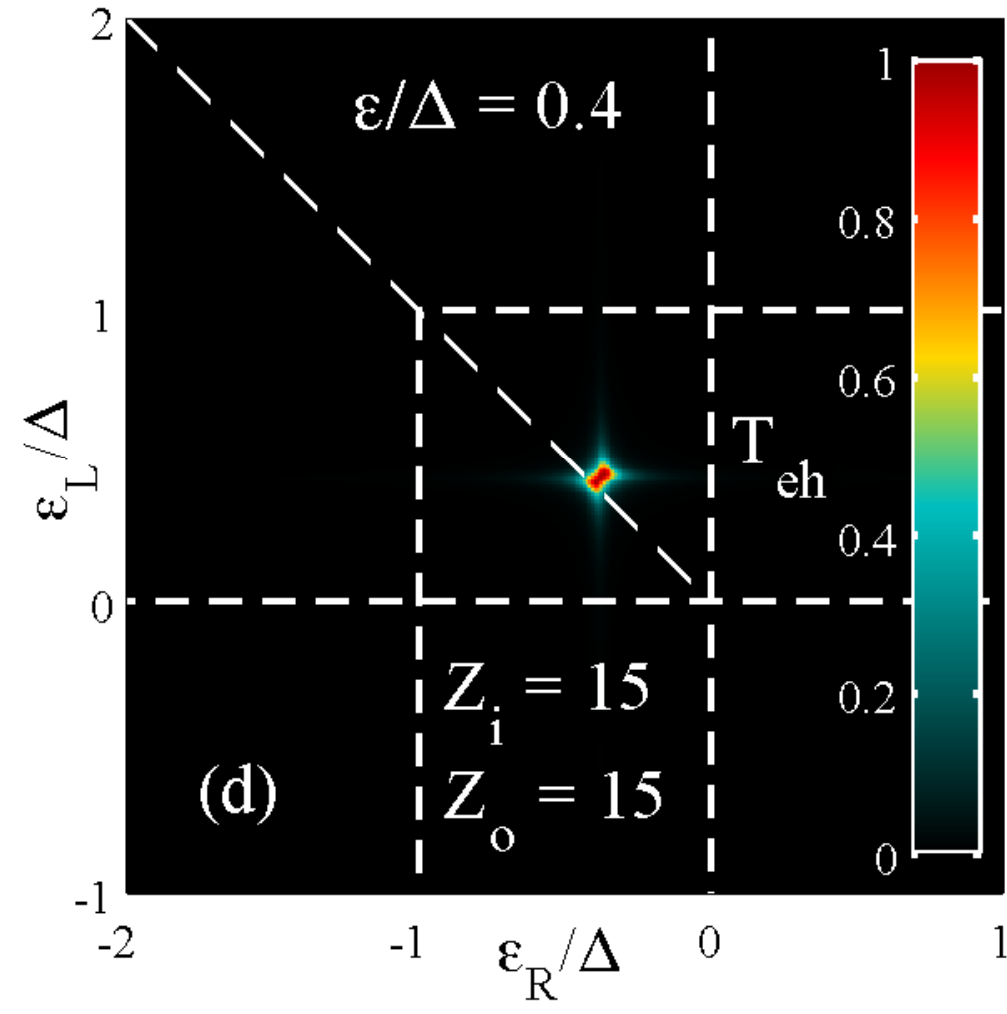}
		}
		\subfloat{
			\hspace{-2mm} \includegraphics[width=4.4cm]{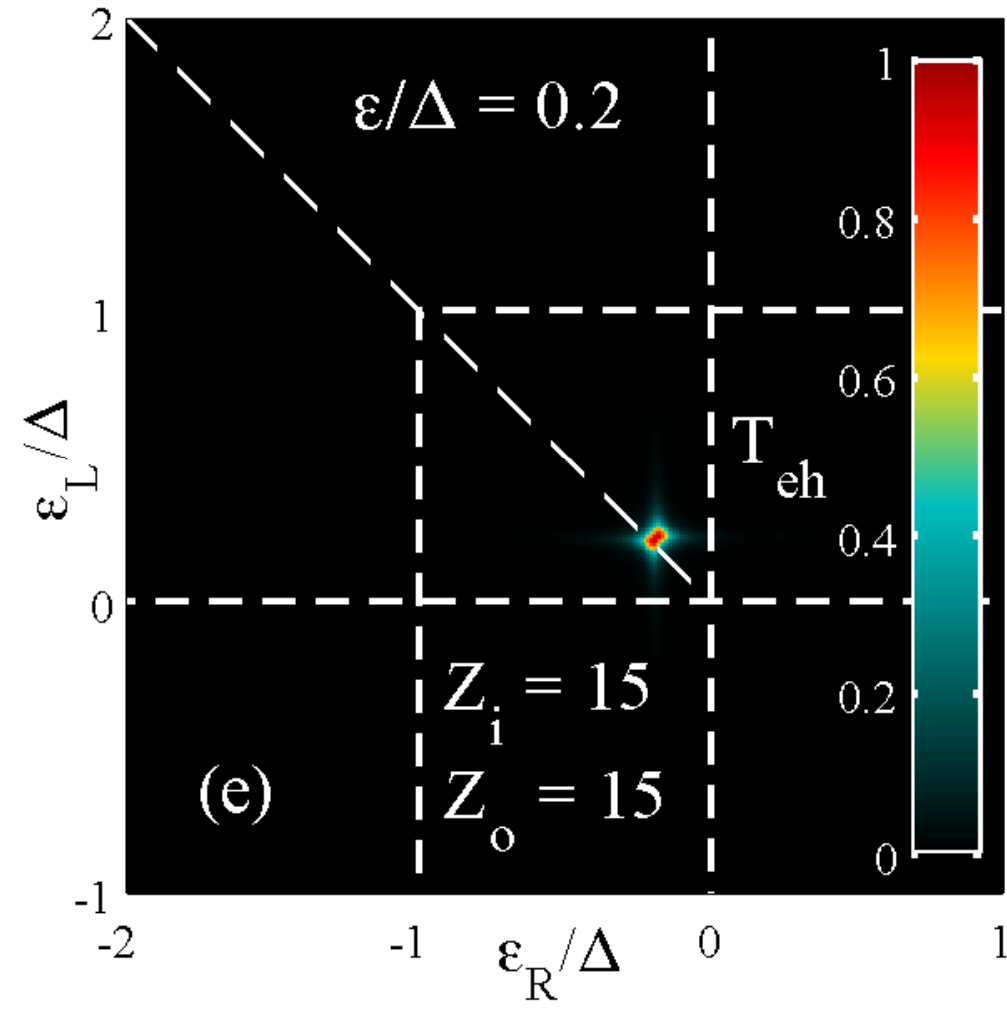}
		}
		\subfloat{
			\hspace{-2mm} \includegraphics[width=4.4cm]{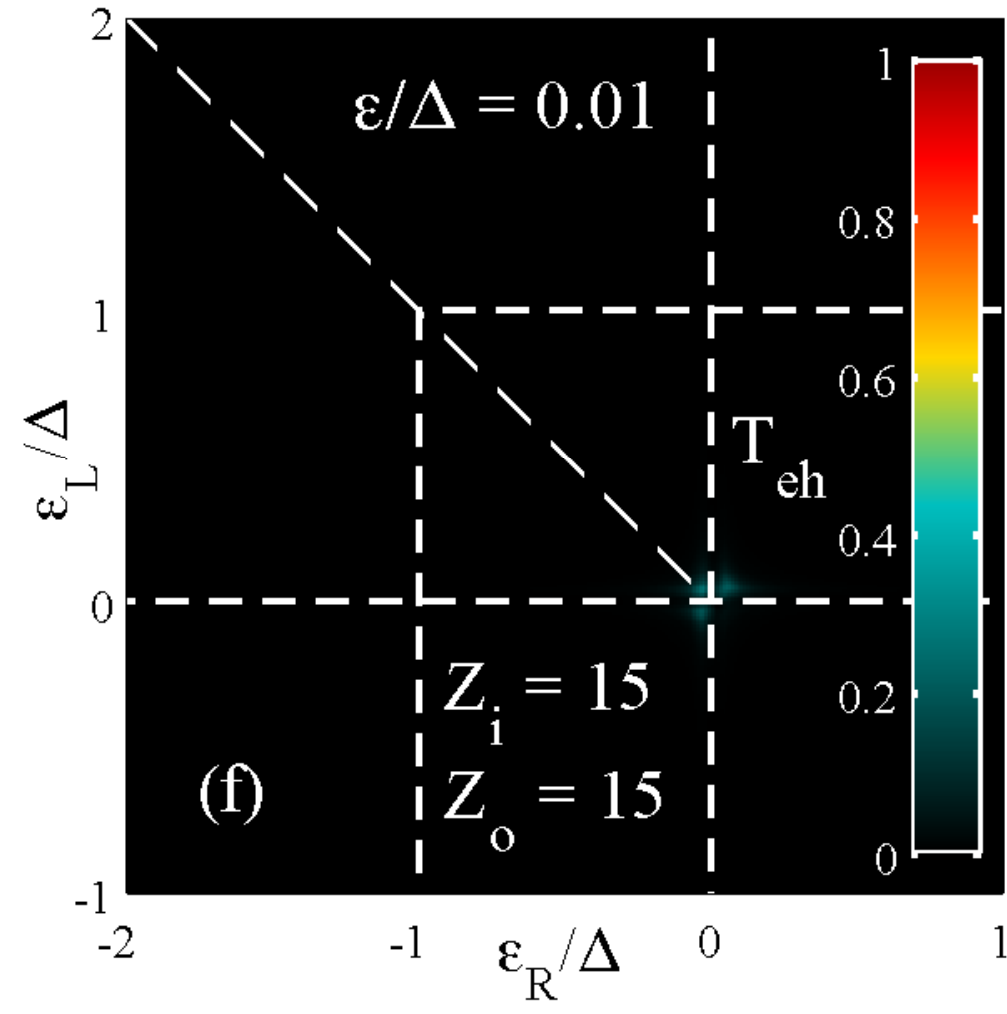}
		}
	\end{center} \vspace{-6mm}
	\caption{
		Transparency ${\tilde T}_{\rm eh}$ as a function of~$\varepsilon_\rR$ and $\varepsilon_\rL$ 
		for $L/\xi_0 = 1$, $\delta_\rLR/\Delta = 10$, and $Z_{\rm i} = Z_{\rm o} = 15$
		($\Gamma_{\!\rL,\rR} / \Delta = 0.045$).
		(a)~$\varepsilon = 0.99$.
		(b)~$\varepsilon = 0.8$.
		(c)~$\varepsilon = 0.6$.
		(d)~$\varepsilon = 0.4$.
		(e)~$\varepsilon = 0.2$.
		(f)~$\varepsilon = 0.01$.
	}
\end{figure}

\subsection{Maximal transparency $\max_\varepsilon\{{\tilde T}_{\rm eh}(\varepsilon)\}$ as a function of $\varepsilon_\rR$ and $\varepsilon_\rL$}

\begin{figure}[H]
	\vspace{-7mm} \begin{center}
		\subfloat{
			\hspace{-3mm} \includegraphics[width=4.3cm]{transparency_eLeR_maxE_kL=0pi_beta=1_sL=10_sR=10_Zi=5_Zo=5.pdf}
		}
		\subfloat{
			\hspace{-2mm} \includegraphics[width=4.3cm]{transparency_eLeR_maxE_kL=0pi_beta=1_sL=10_sR=10_Zi=15_Zo=15.pdf}
		} 
		\subfloat{
			\hspace{-2mm} \includegraphics[width=4.3cm]{transparency_eLeR_maxE_kL=0pi_beta=1_sL=10_sR=10_Zi=5_Zo=15.pdf}
		}
		\subfloat{
			\hspace{-2mm} \includegraphics[width=4.3cm]{transparency_eLeR_maxE_kL=0pi_beta=1_sL=10_sR=10_Zi=15_Zo=5.pdf}
		} \\ \vspace{-3mm}
		\subfloat{
			\hspace{-3mm} \includegraphics[width=4.3cm]{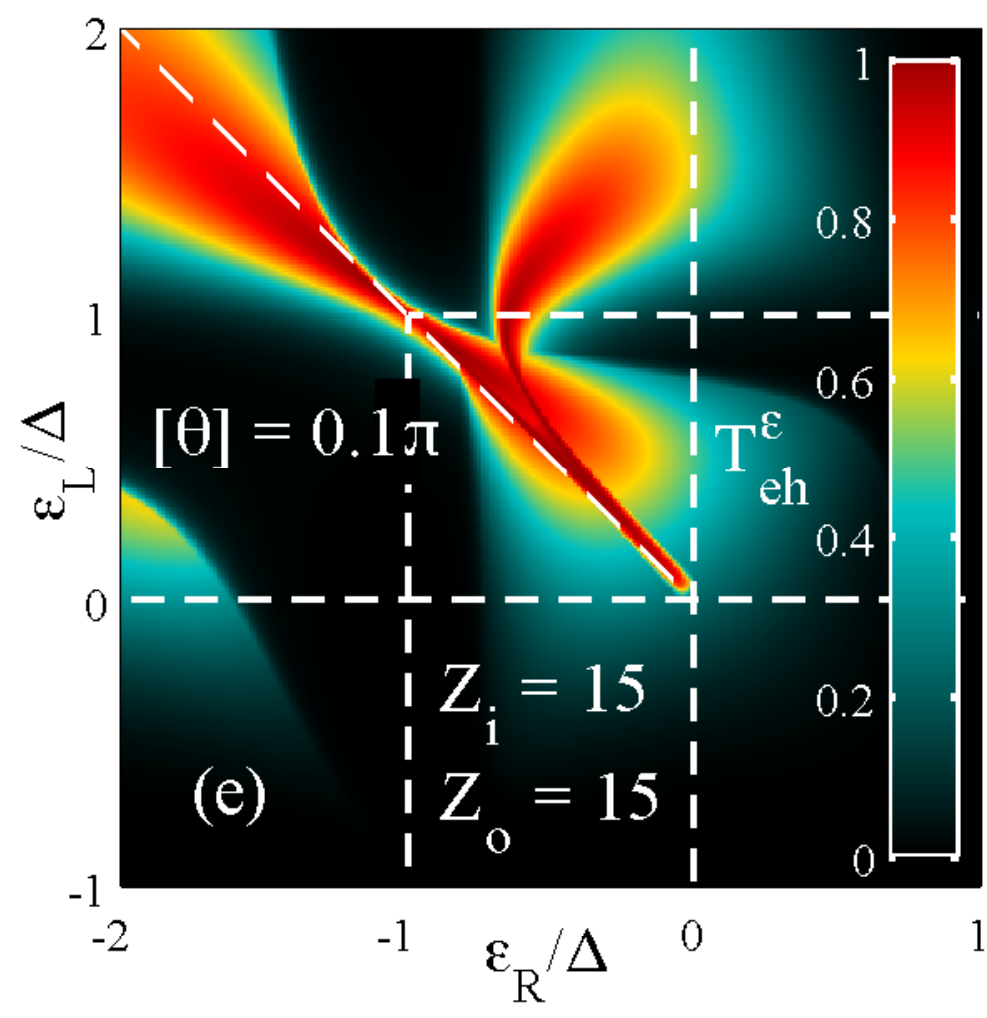}
		}
		\subfloat{
			\hspace{-2mm} \includegraphics[width=4.3cm]{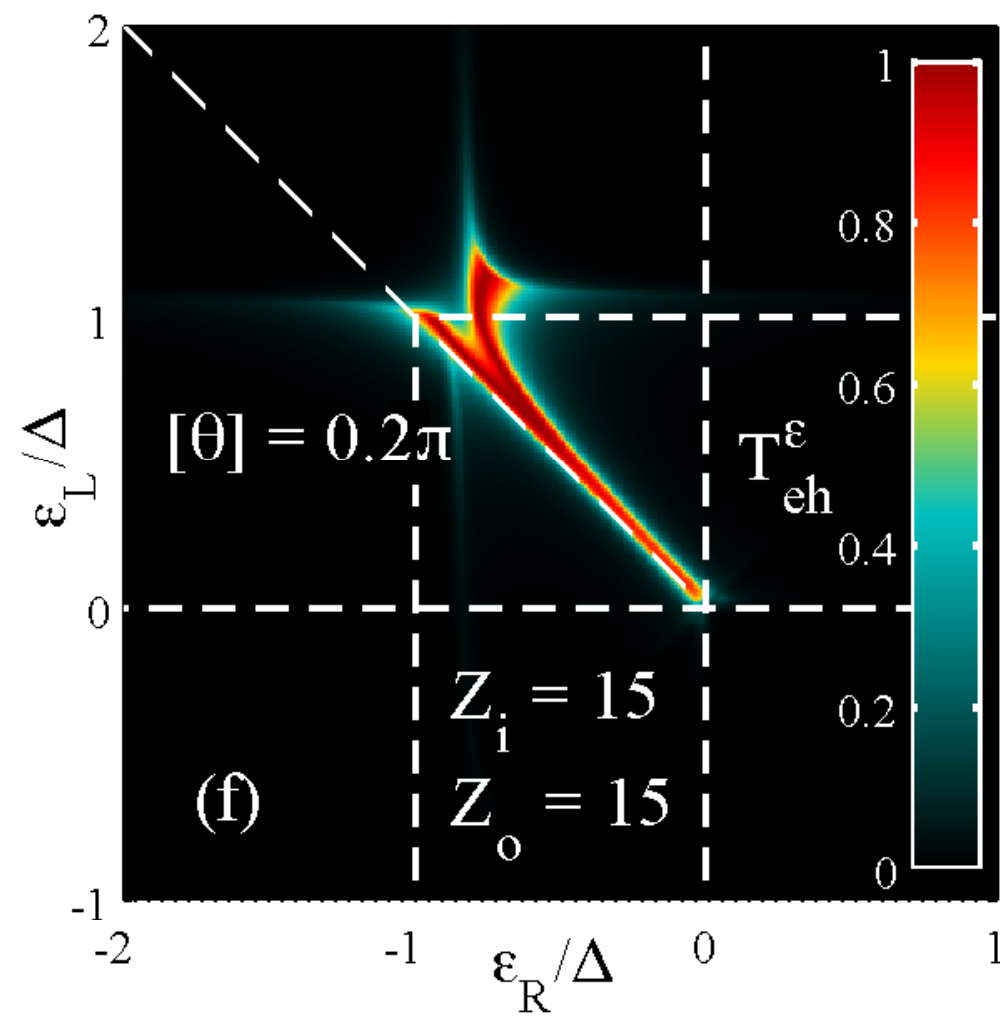}
		}
		\subfloat{
			\hspace{-2mm} \includegraphics[width=4.3cm]{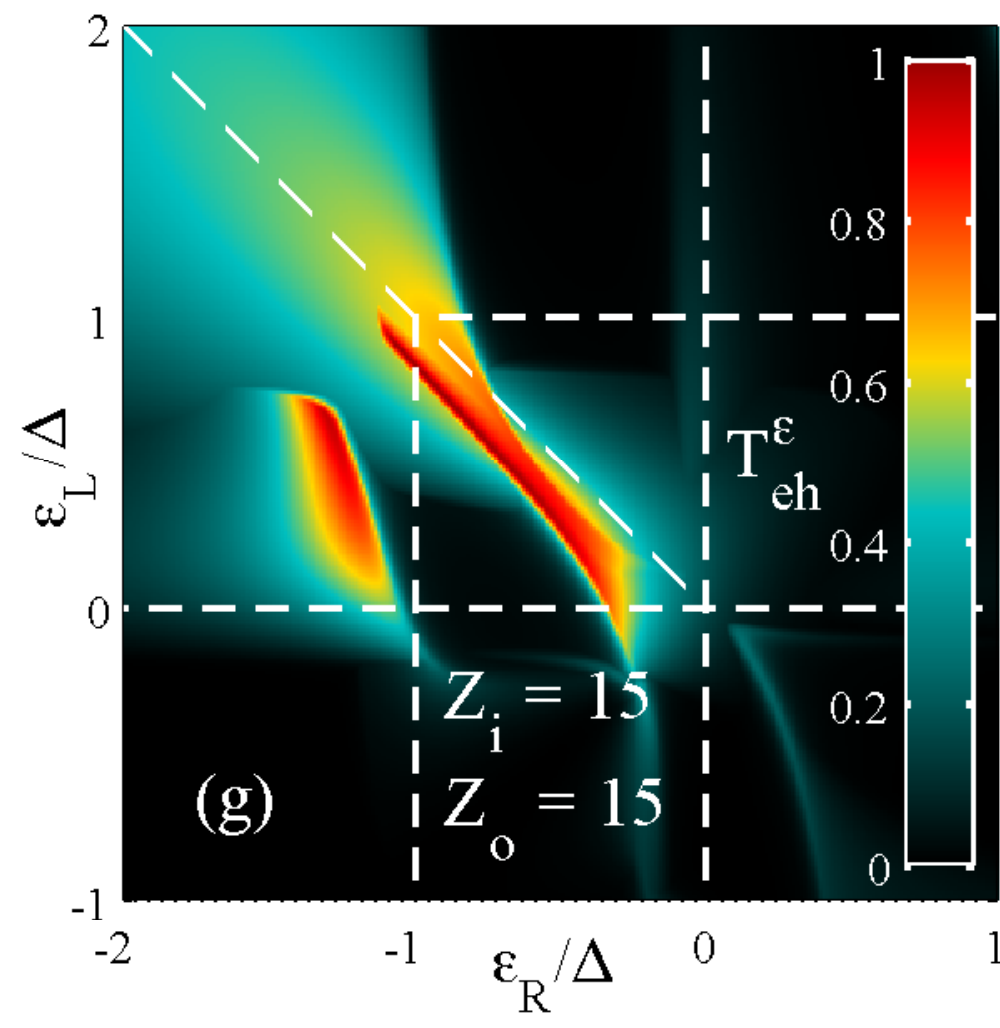}
		}
		\subfloat{
			\hspace{-2mm} \includegraphics[width=4.3cm]{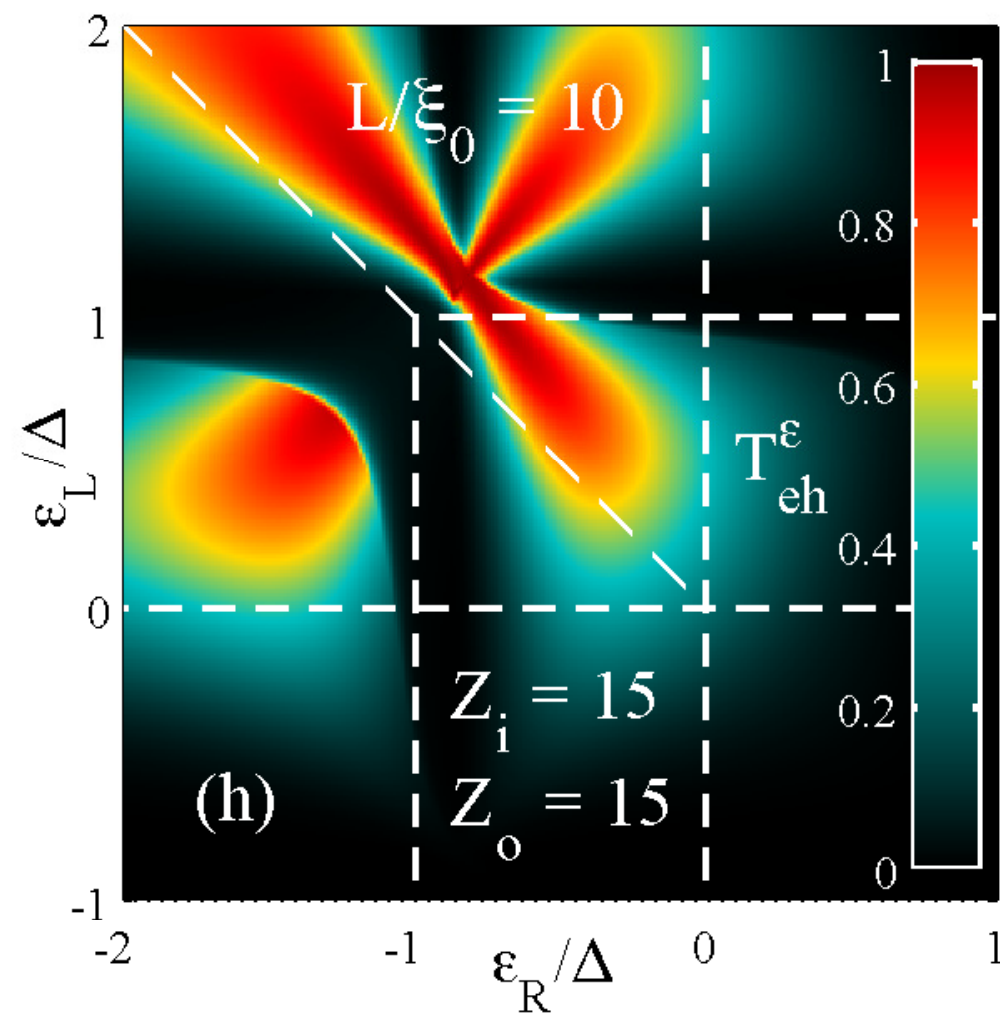}
		}
	\end{center} \vspace{-6mm}
	\caption{
		Maximal transparency ${\tilde T}_{\rm eh}^\varepsilon = \max_\varepsilon\{{\tilde T}_{\rm eh}\}$ 
		as a function of~$\varepsilon_\rR$ and $\varepsilon_\rL$ 
		for $L/\xi_0 = 1$, $\delta_\rLR/\Delta = 10$, and $\theta = \pi n$.
		(a)~$Z_{\rm i} = Z_{\rm o} = 5$ ($\Gamma_{\!\rL,\rR} / \Delta = 0.39$).
		(b)~$Z_{\rm i} = Z_{\rm o} = 15$ ($\Gamma_{\!\rL,\rR} / \Delta = 0.045$).
		(c)~$Z_{\rm i} = 5$ and $Z_{\rm o} = 15$ ($\Gamma_{\!\rL,\rR} / \Delta = 0.22$).
		(d)~$Z_{\rm i} = 15$ and $Z_{\rm o} = 5$. 
		(e)~$Z_{\rm i} = Z_{\rm o} = 15$ and $\theta = 0.1\pi + 2\pi n$.
		(f)~$Z_{\rm i} = Z_{\rm o} = 15$ and $\theta = 0.2\pi + 2\pi n$.
		(g)~$Z_{\rm i} = Z_{\rm o} = 15$ and $L/\xi_0 = 0.1$.
		(h)~$Z_{\rm i} = Z_{\rm o} = 15$ and $L/\xi_0 = 10$.
	}
\end{figure}

\subsection{Maximal transparency $\max_\varepsilon\{{\tilde T}_{\rm ee}(\varepsilon)\}$ as a function of $\varepsilon_\rR$ and $\varepsilon_\rL$}

\begin{figure}[H]
	\vspace{-7mm} \begin{center}
		\subfloat{
			\hspace{-3mm}
			\includegraphics[width=4.3cm]{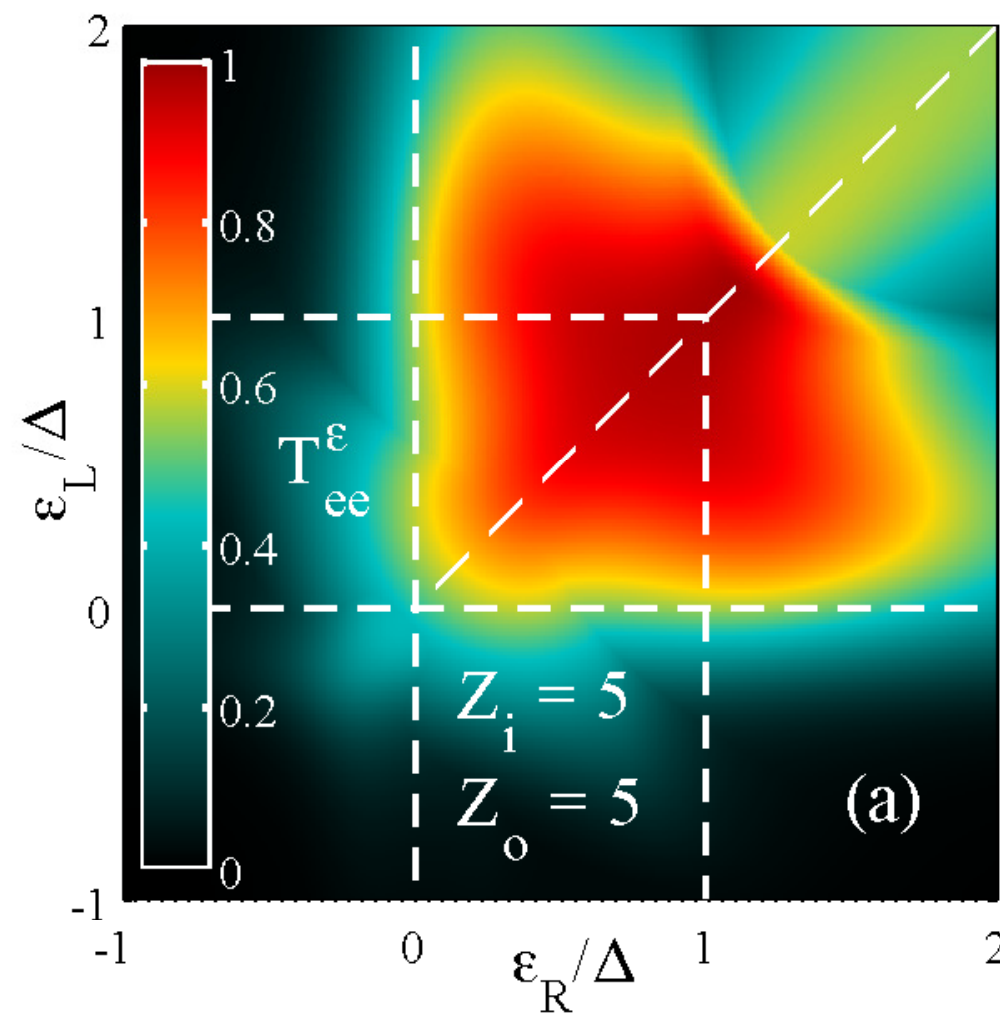}
		}
		\subfloat{
			\hspace{-2mm} \includegraphics[width=4.3cm]{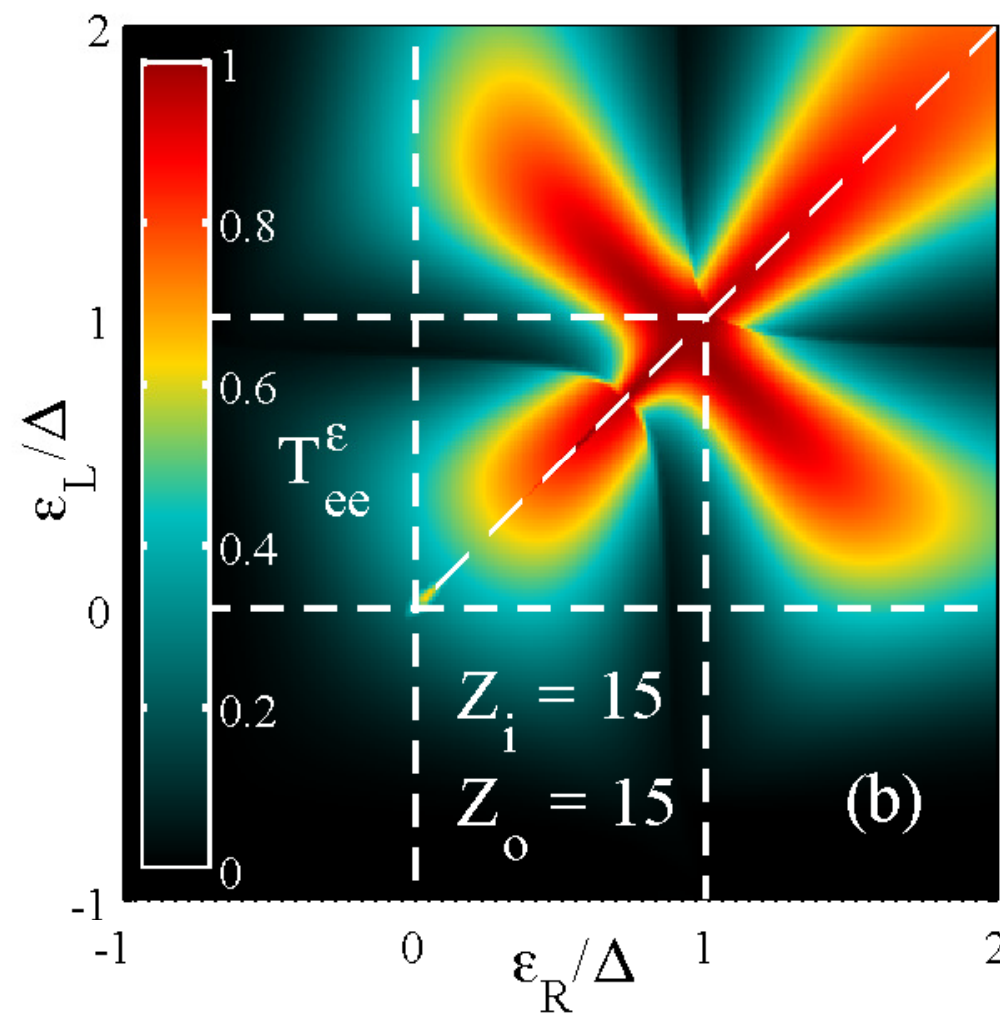}
		} 
		\subfloat{
			\hspace{-2mm} \includegraphics[width=4.3cm]{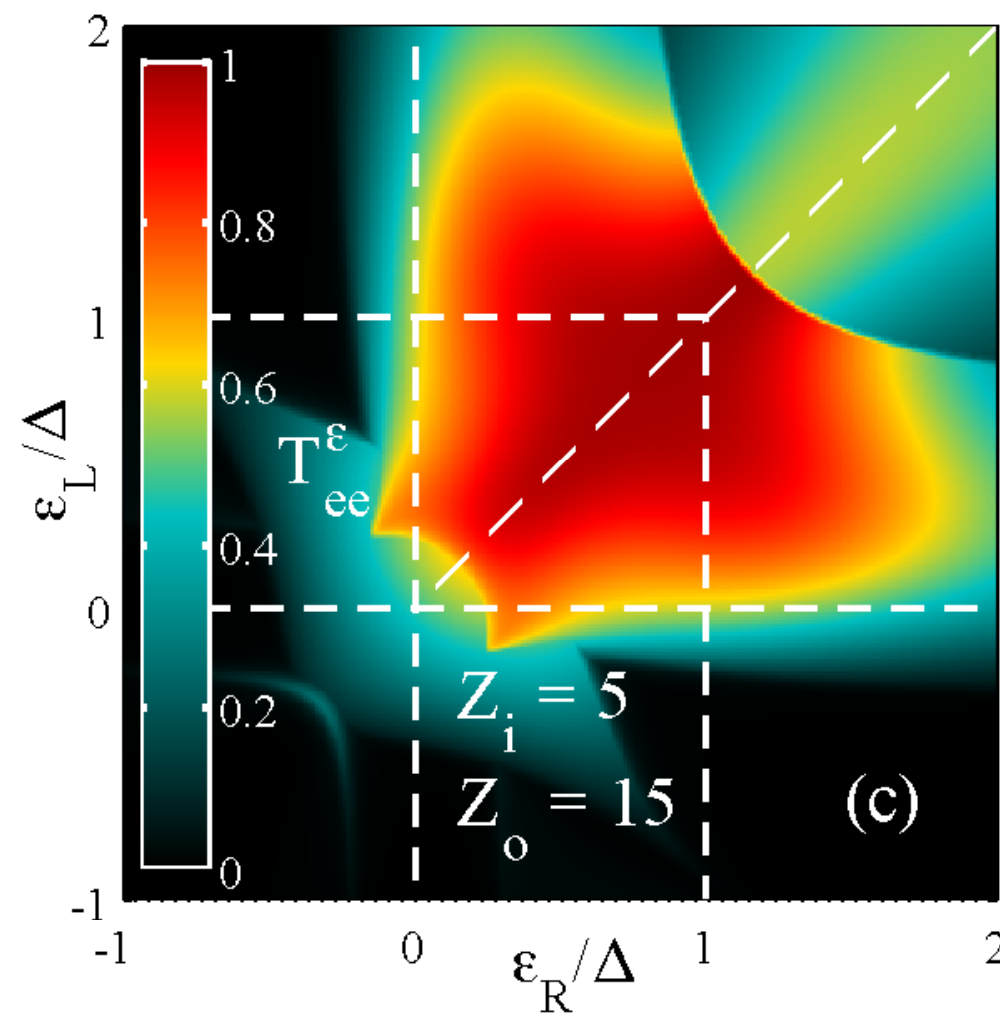}
		}
		\subfloat{
			\hspace{-2mm} \includegraphics[width=4.3cm]{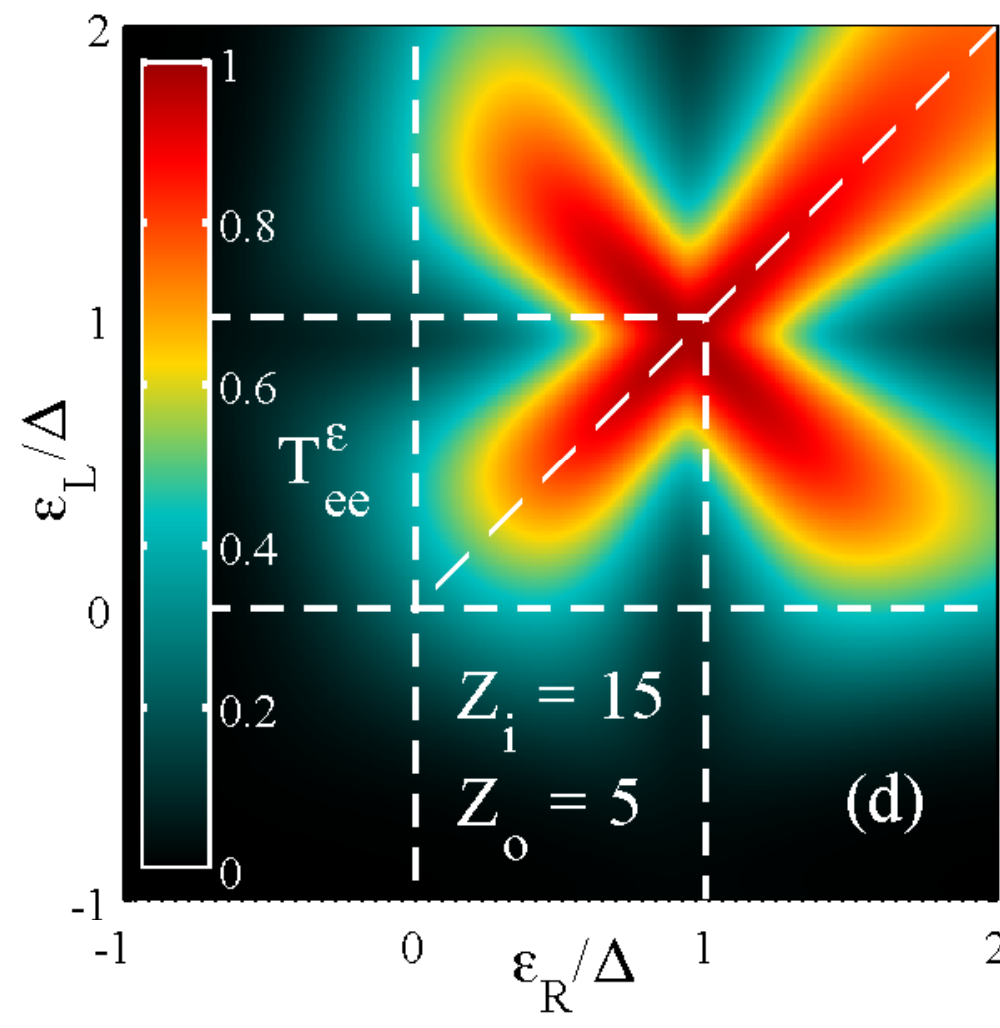}
		} \\ \vspace{-4mm}
		\subfloat{
			\hspace{-3mm} \includegraphics[width=4.3cm]{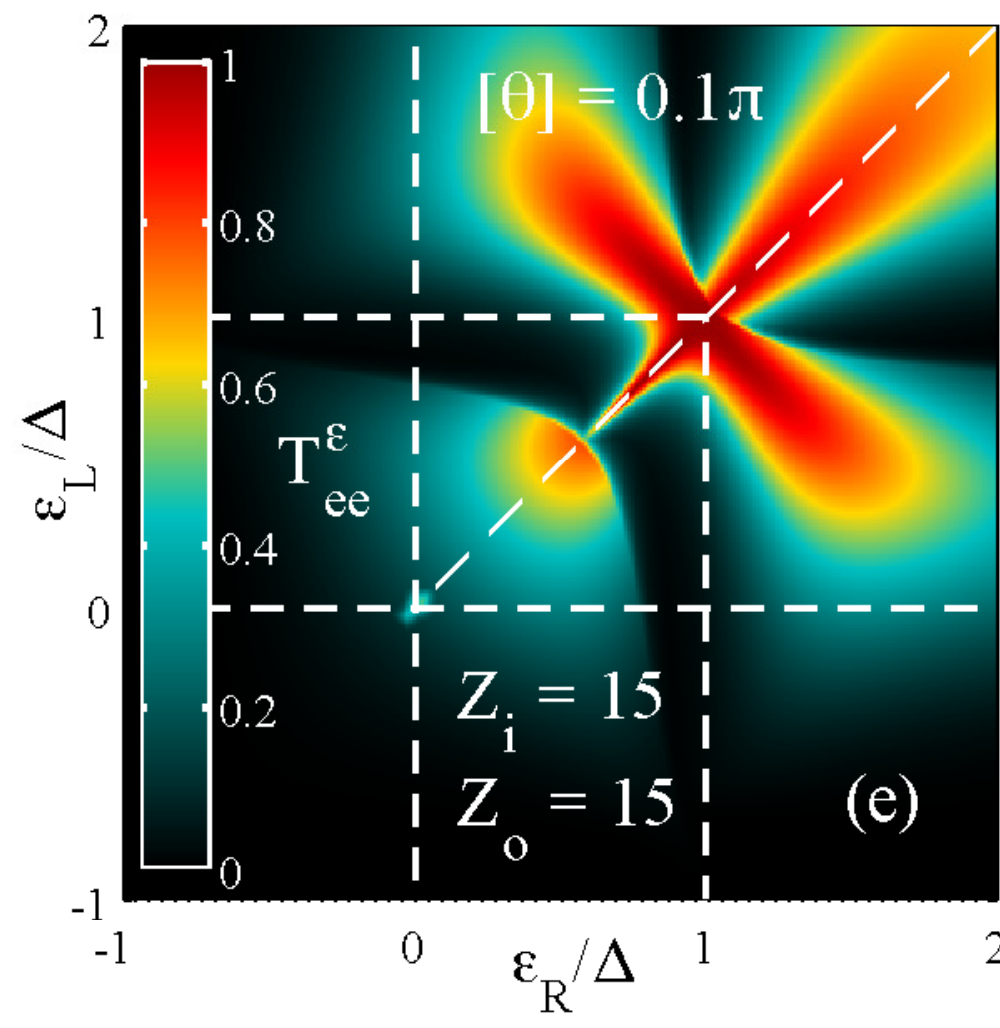}
		}
		\subfloat{
			\hspace{-2mm} \includegraphics[width=4.3cm]{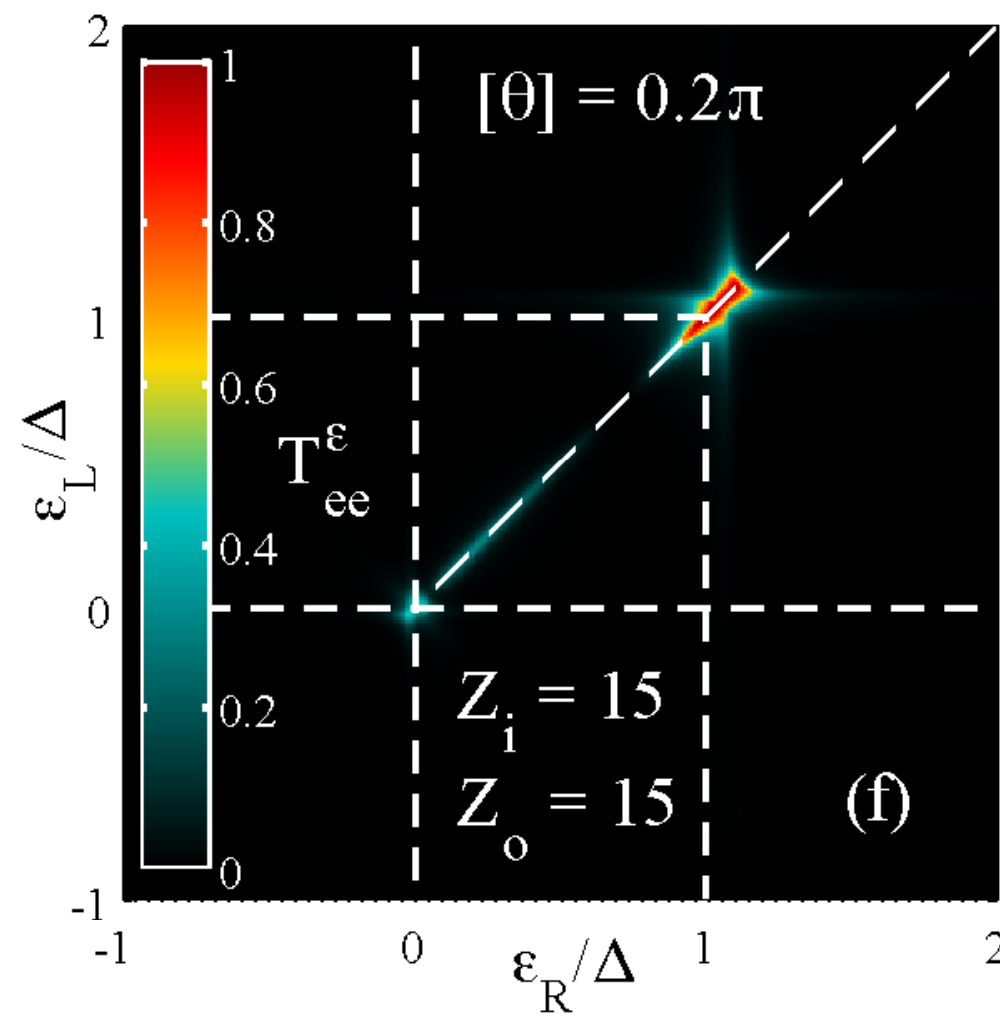}
		}
		\subfloat{
			\hspace{-2mm} \includegraphics[width=4.3cm]{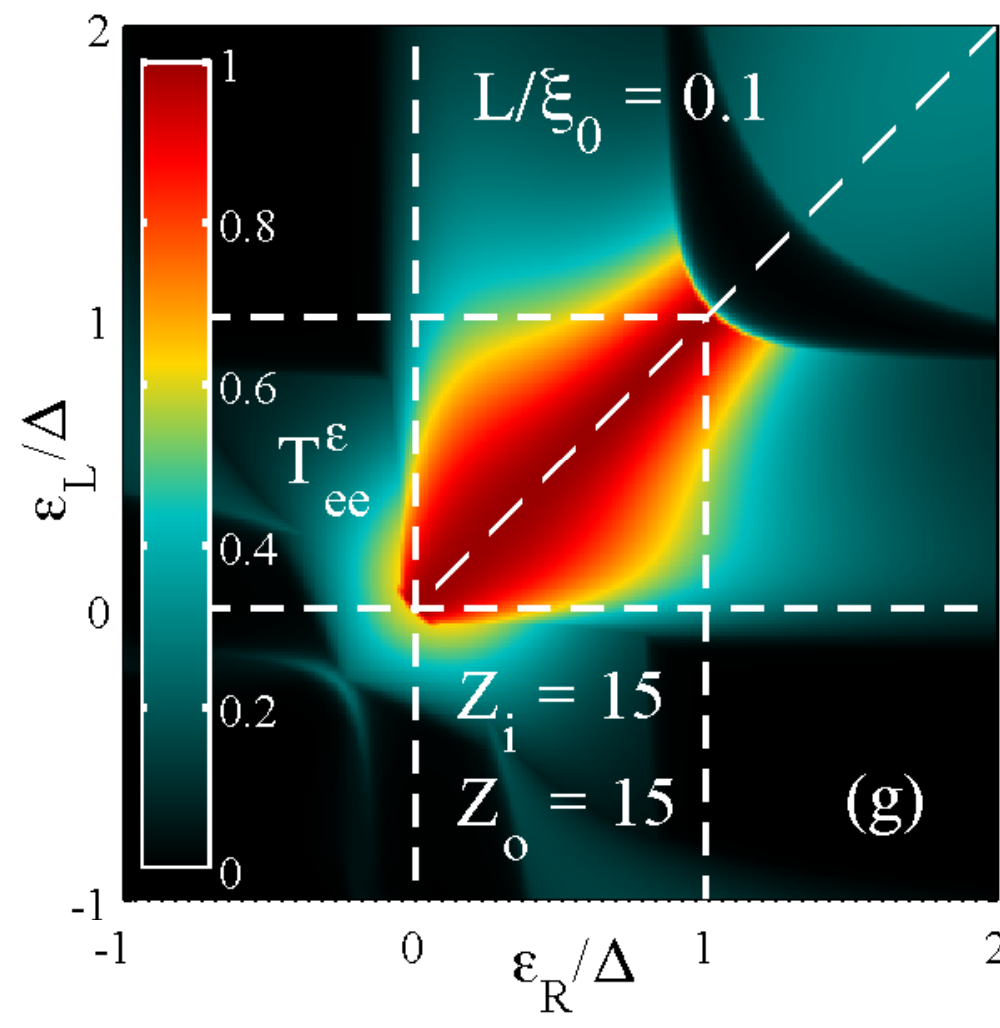}
		}
		\subfloat{
			\hspace{-2mm} \includegraphics[width=4.3cm]{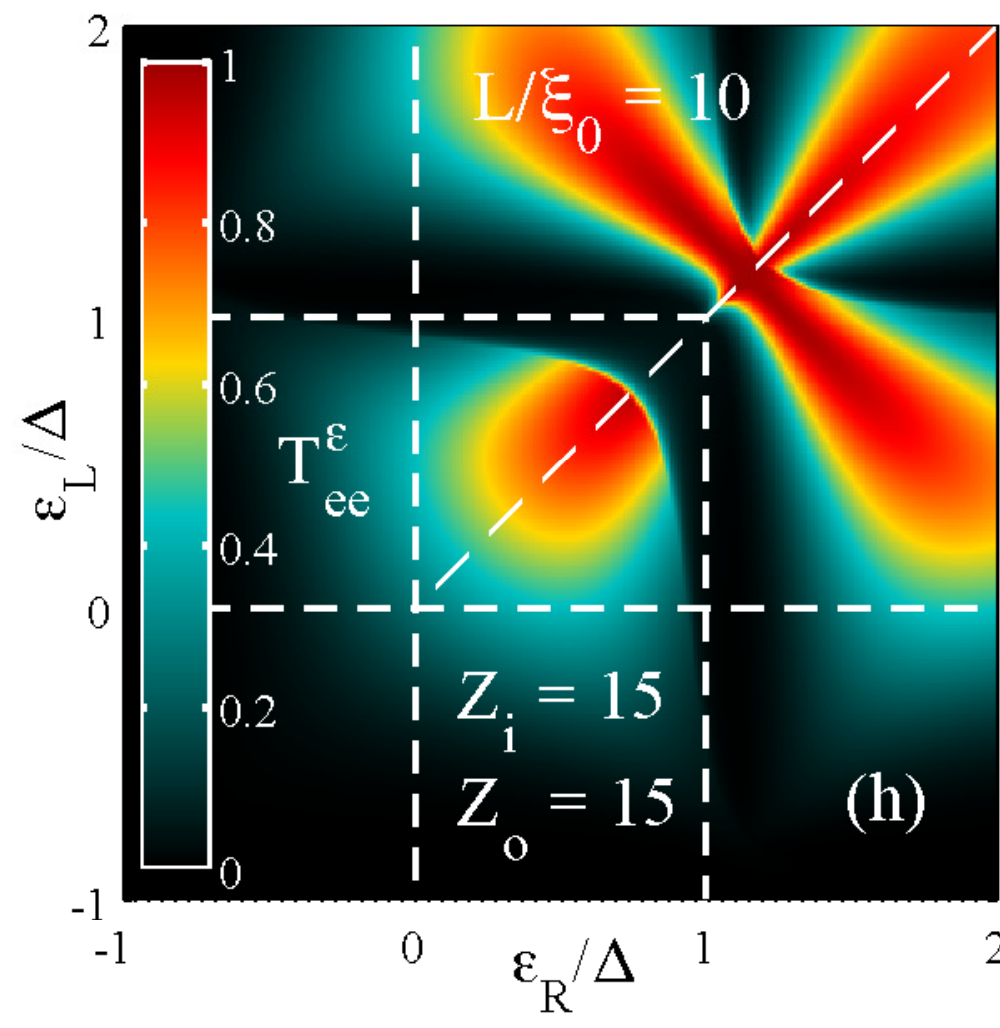}
		}
	\end{center} \vspace{-6mm}
	\caption{
		Color plots of maximal transparency ${\tilde T}_{\rm ee}^\varepsilon = \max_\varepsilon\{{\tilde T}_{\rm ee}\}$ 
		as functions of~$\varepsilon_\rR$ and $\varepsilon_\rL$ 
		for $L/\xi_0 = 1$, $\delta_\rLR/\Delta = 10$, and $\theta = \pi n$.
		(a)~$Z_{\rm i} = Z_{\rm o} = 5$ ($\Gamma_\sLR / \Delta = 0.39$).
		(b)~$Z_{\rm i} = Z_{\rm o} = 15$ ($\Gamma_\sLR / \Delta = 0.045$).
		(c)~$Z_{\rm i} = 5$ and $Z_{\rm o} = 15$ ($\Gamma_\sLR / \Delta = 0.22$).
		(d)~$Z_{\rm i} = 15$ and $Z_{\rm o} = 5$. 
		(e)~$Z_{\rm i} = Z_{\rm o} = 15$ and $\theta = 0.1\pi + 2\pi n$.
		(f)~$Z_{\rm i} = Z_{\rm o} = 15$ and $\theta = 0.2\pi + 2\pi n$.
		(g)~$Z_{\rm i} = Z_{\rm o} = 15$ and $L/\xi_0 = 0.1$.
		(h)~$Z_{\rm i} = Z_{\rm o} = 15$ and $L/\xi_0 = 10$.
	}
\end{figure}

\subsection{Maximal differential conductance $g^\varepsilon$ as a function of $\varepsilon_\rR$ and $\varepsilon_\rL$}

\begin{figure}[H]
	\vspace{-6mm} \begin{center}
		\subfloat{
			\hspace{-3mm}
			\includegraphics[width=4.3cm]{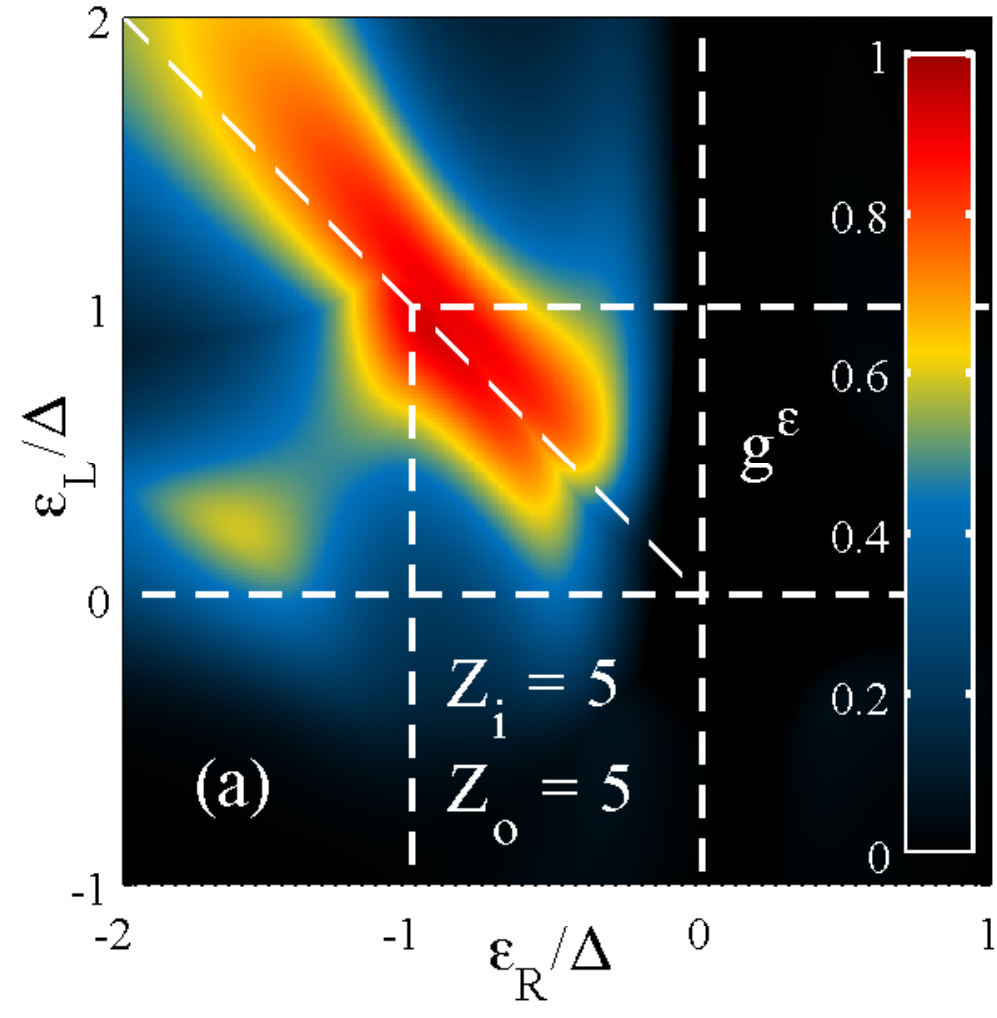}
		}
		\subfloat{
			\hspace{-2mm} \includegraphics[width=4.3cm]{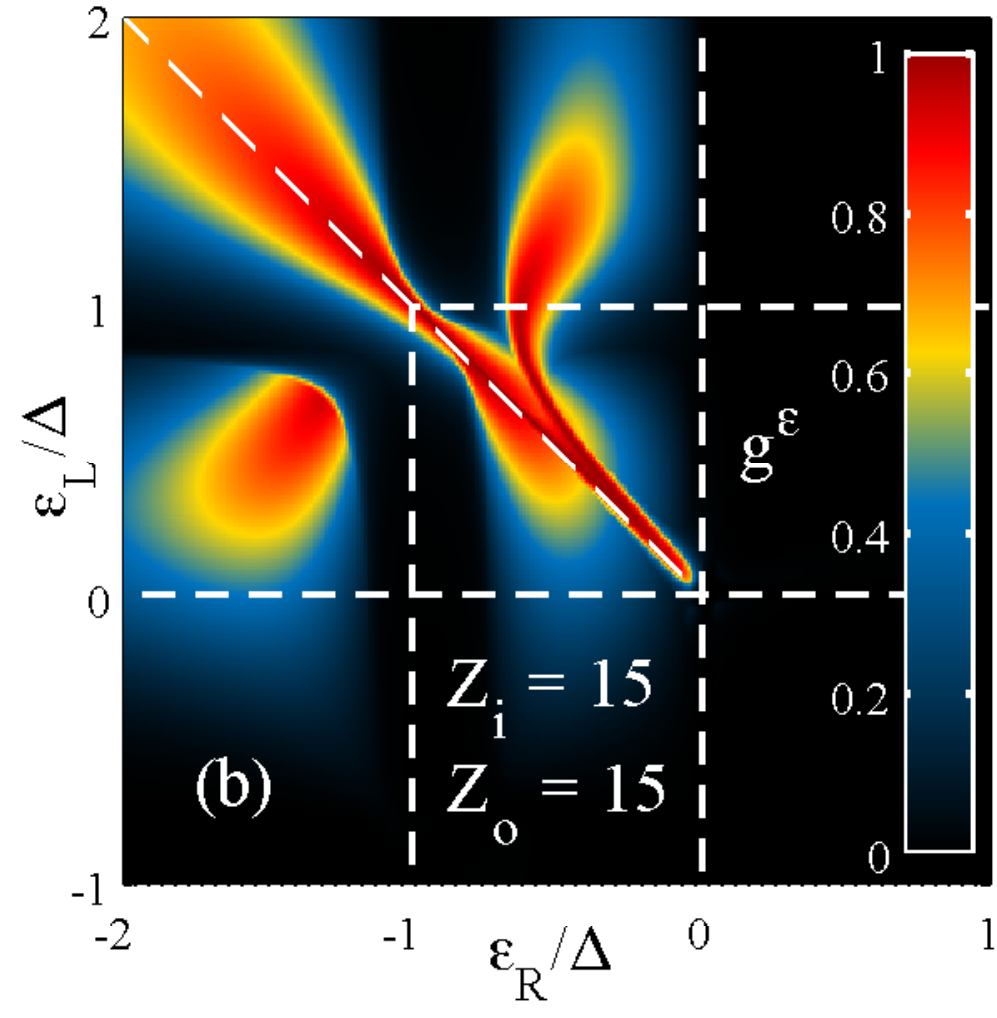}
		}
		\subfloat{
			\hspace{-2mm} \includegraphics[width=4.3cm]{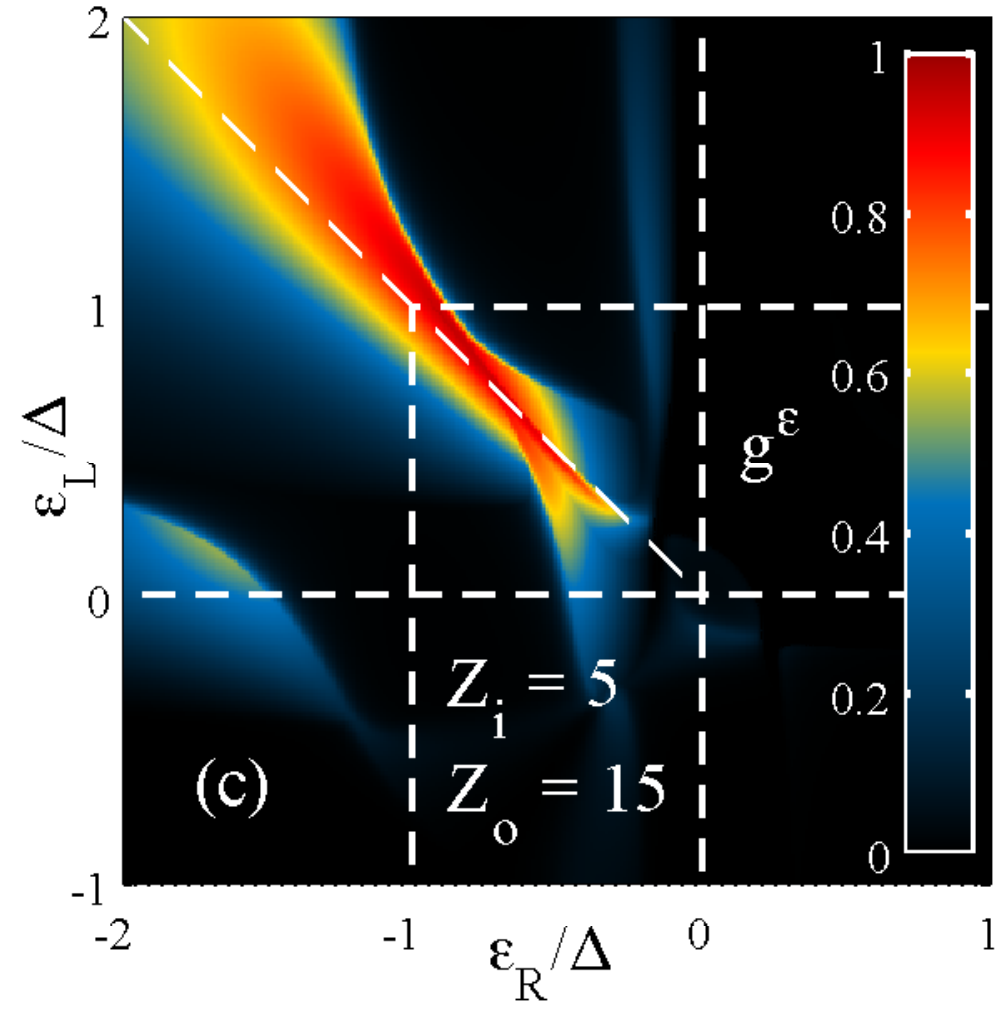}
		}
		\subfloat{
			\hspace{-2mm} \includegraphics[width=4.3cm]{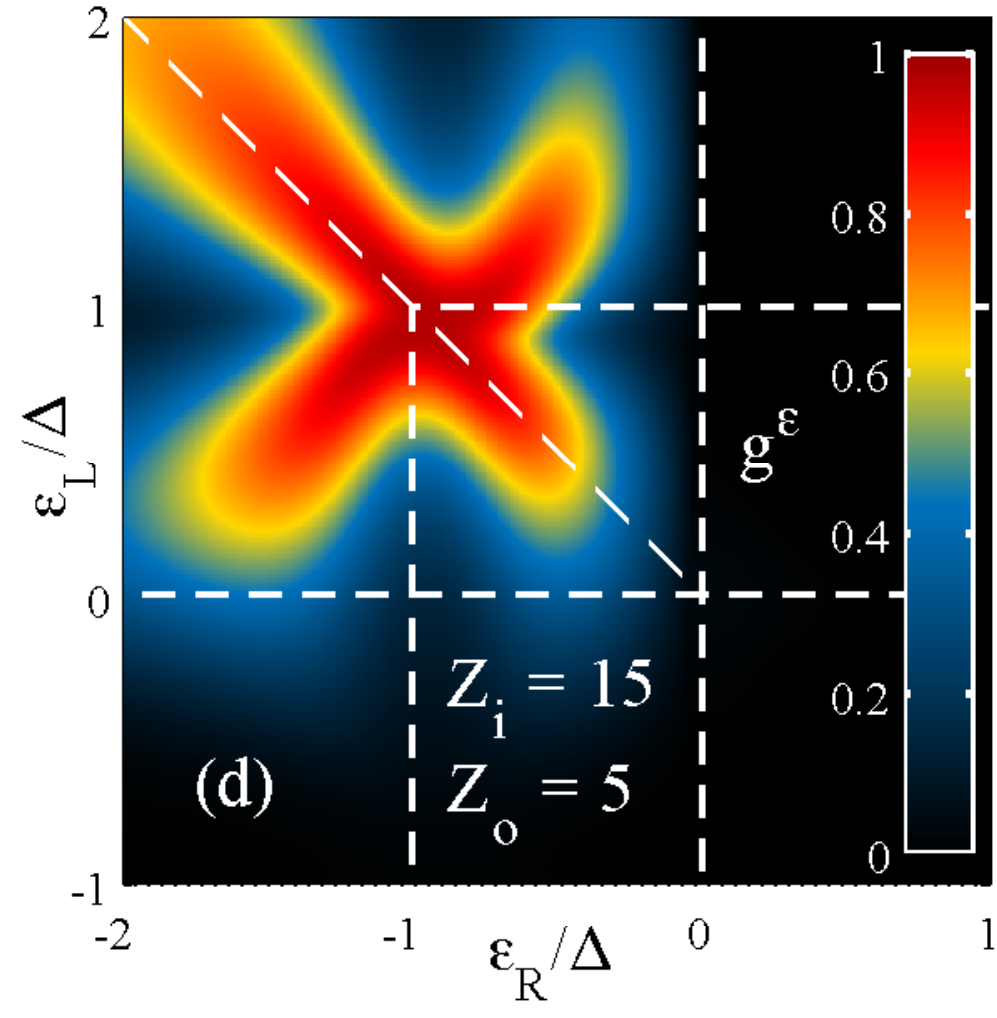}
		} \\ \vspace{-4mm}
		\subfloat{
			\hspace{-3mm} \includegraphics[width=4.3cm]{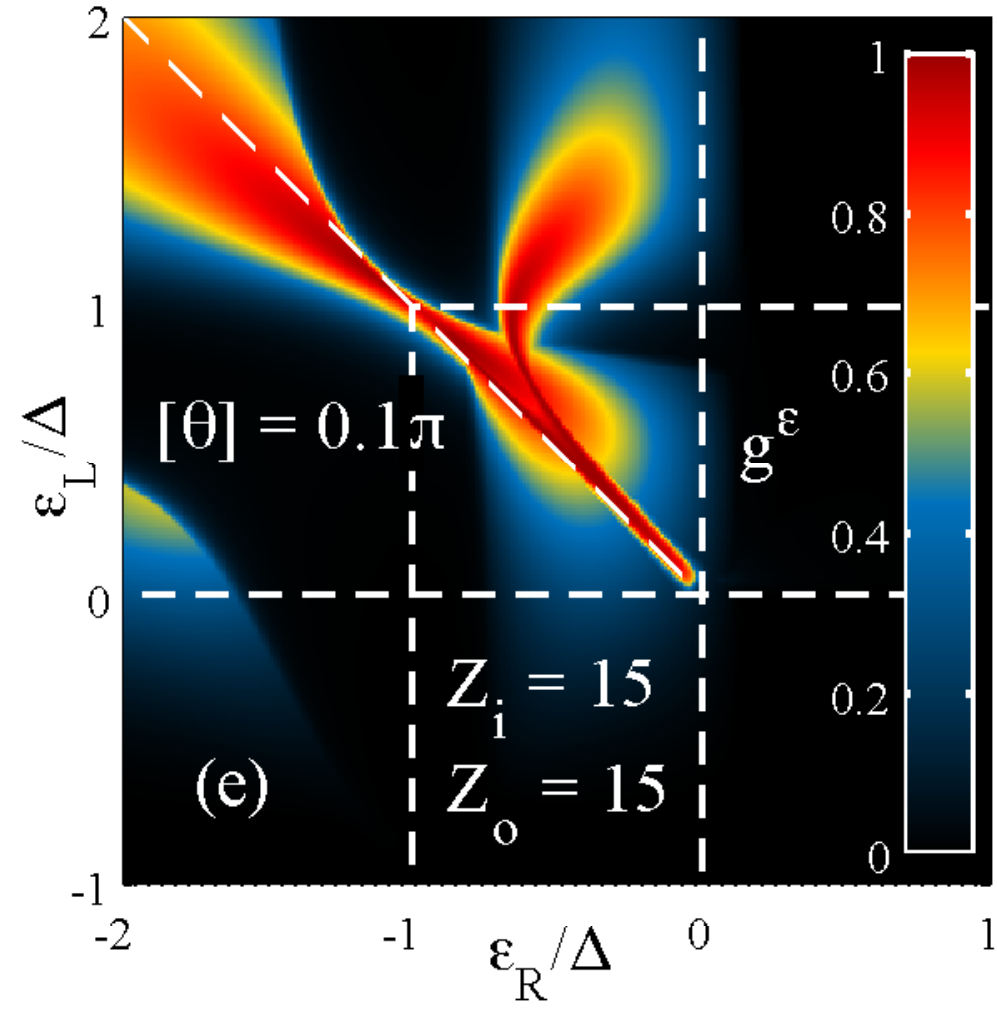}
		}
		\subfloat{
			\hspace{-2mm} \includegraphics[width=4.3cm]{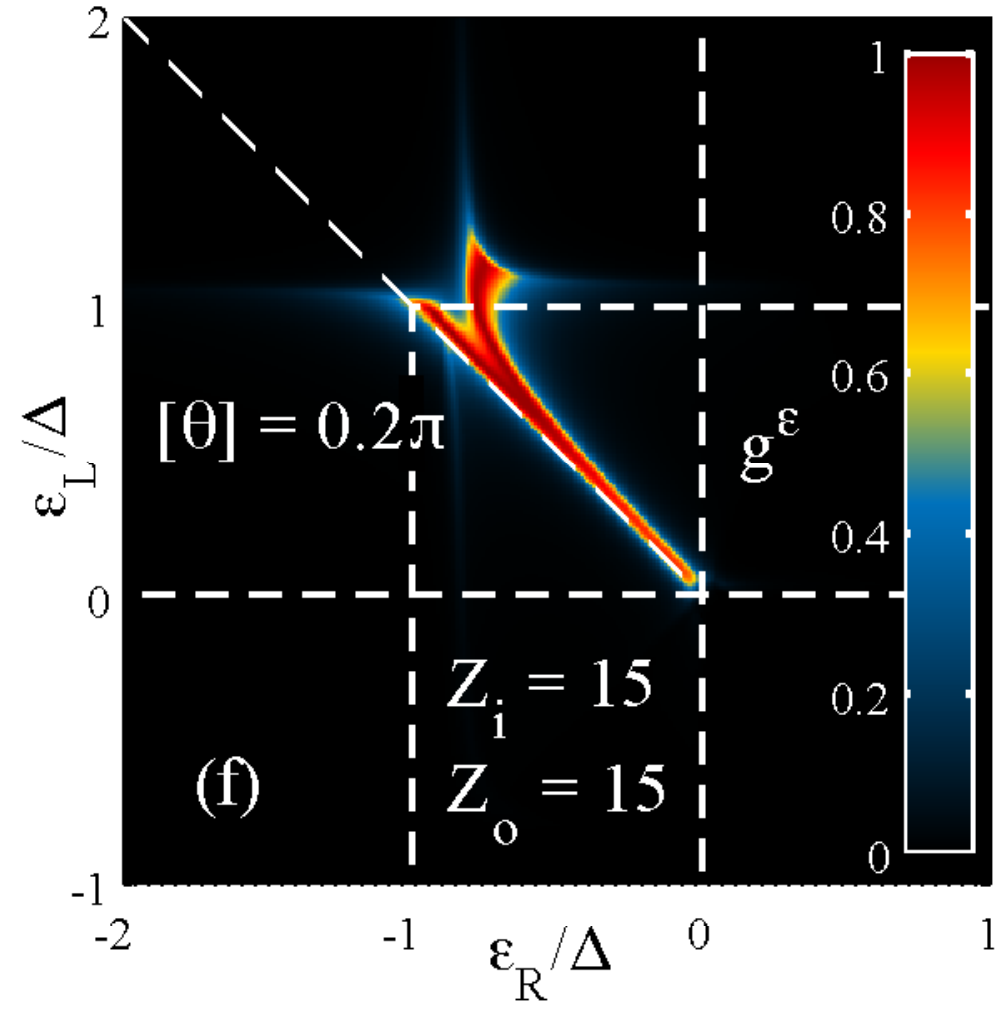}
		}
		\subfloat{
			\hspace{-2mm} \includegraphics[width=4.3cm]{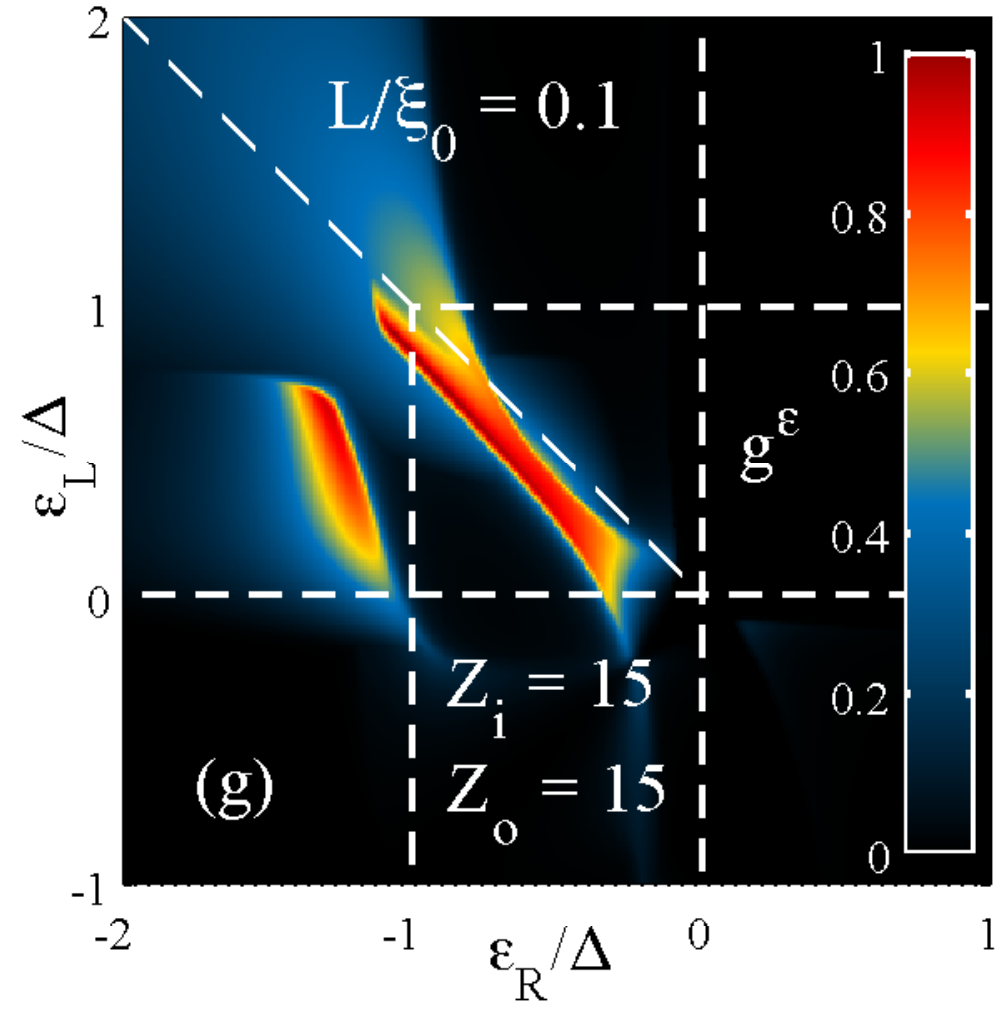}
		}
		\subfloat{
			\hspace{-2mm} \includegraphics[width=4.3cm]{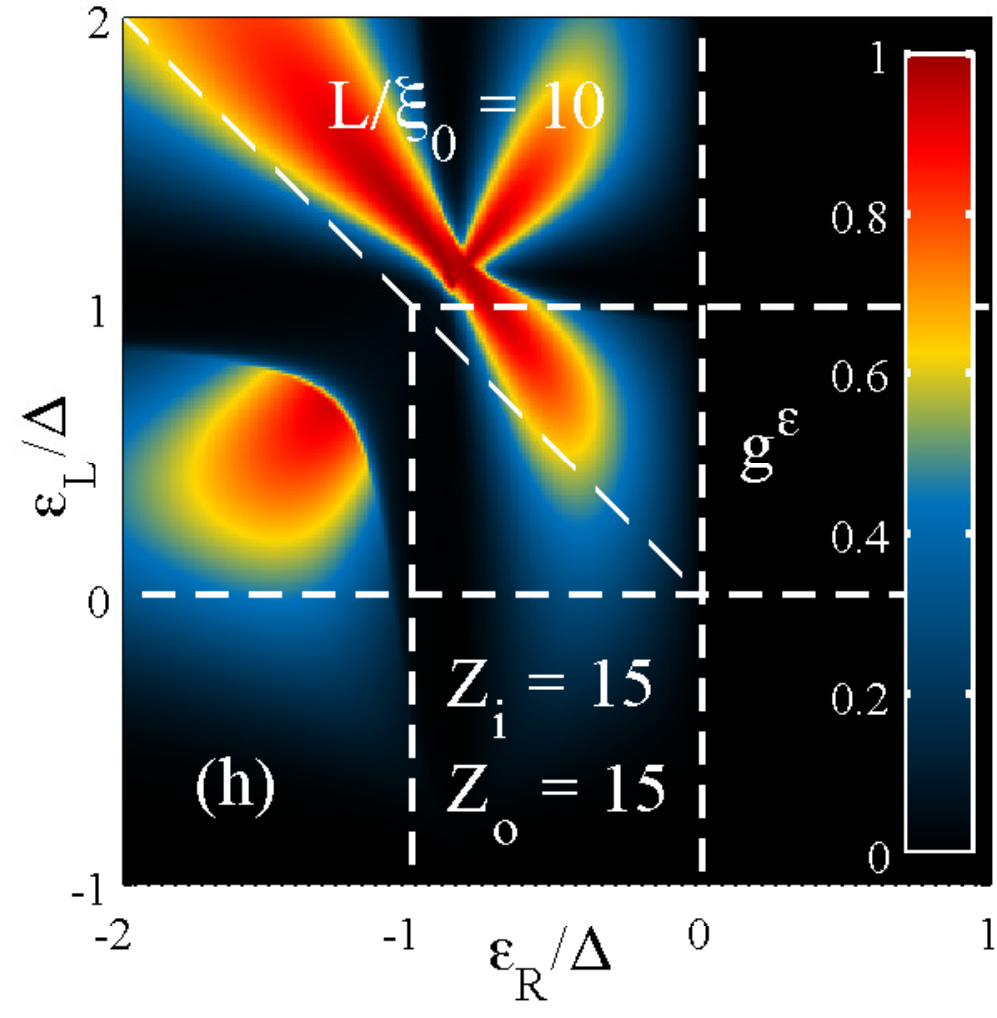}
		}
	\end{center} \vspace{-6mm}
	\caption{
		Color plots of maximal differential conductance $g^\varepsilon = \max_\varepsilon\{{\tilde T}_{\rm eh} - {\tilde T}_{\rm ee}\}$ 
		as functions of~$\varepsilon_\rR$ and $\varepsilon_\rL$ 
		for $L/\xi_0 = 1$, $\delta_\rLR/\Delta = 10$, and $\theta = \pi n$.
		(a)~$Z_{\rm i} = Z_{\rm o} = 5$ ($\Gamma_\sLR / \Delta = 0.39$).
		(b)~$Z_{\rm i} = Z_{\rm o} = 15$ ($\Gamma_\sLR / \Delta = 0.045$).
		(c)~$Z_{\rm i} = 5$ and $Z_{\rm o} = 15$ ($\Gamma_\sLR / \Delta = 0.22$).
		(d)~$Z_{\rm i} = 15$ and $Z_{\rm o} = 5$. 
		(e)~$Z_{\rm i} = Z_{\rm o} = 15$ and $\theta = 0.1\pi + 2\pi n$.
		(f)~$Z_{\rm i} = Z_{\rm o} = 15$ and $\theta = 0.2\pi + 2\pi n$.
		(g)~$Z_{\rm i} = Z_{\rm o} = 15$ and $L/\xi_0 = 0.1$.
		(h)~$Z_{\rm i} = Z_{\rm o} = 15$ and $L/\xi_0 = 10$.
	}
\end{figure}

\subsection{Maximal differential noise $s^\varepsilon$ as a function of $\varepsilon_\rR$ and $\varepsilon_\rL$}

\begin{figure}[H]
	\vspace{-6mm} \begin{center}
		\subfloat{
			\hspace{-3mm} \includegraphics[width=4.3cm]{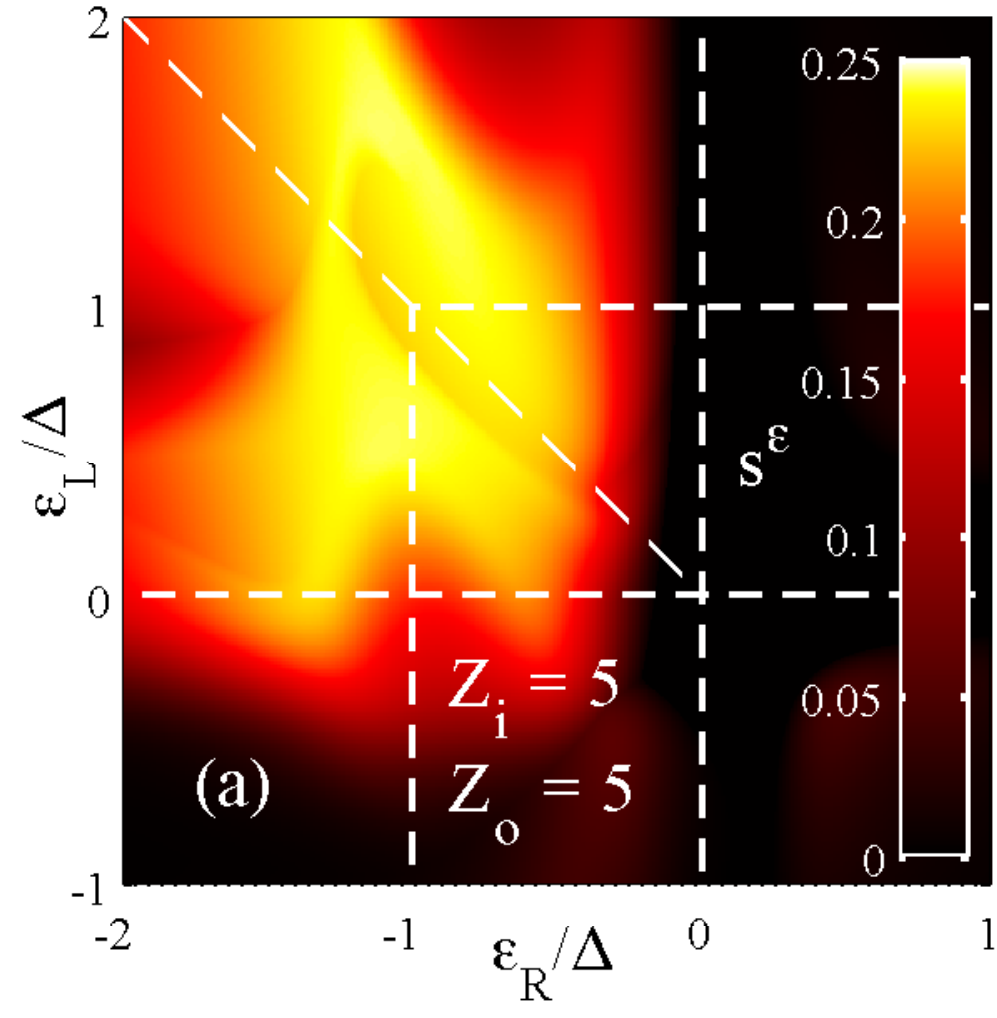}
		}
		\subfloat{
			\hspace{-2mm} \includegraphics[width=4.3cm]{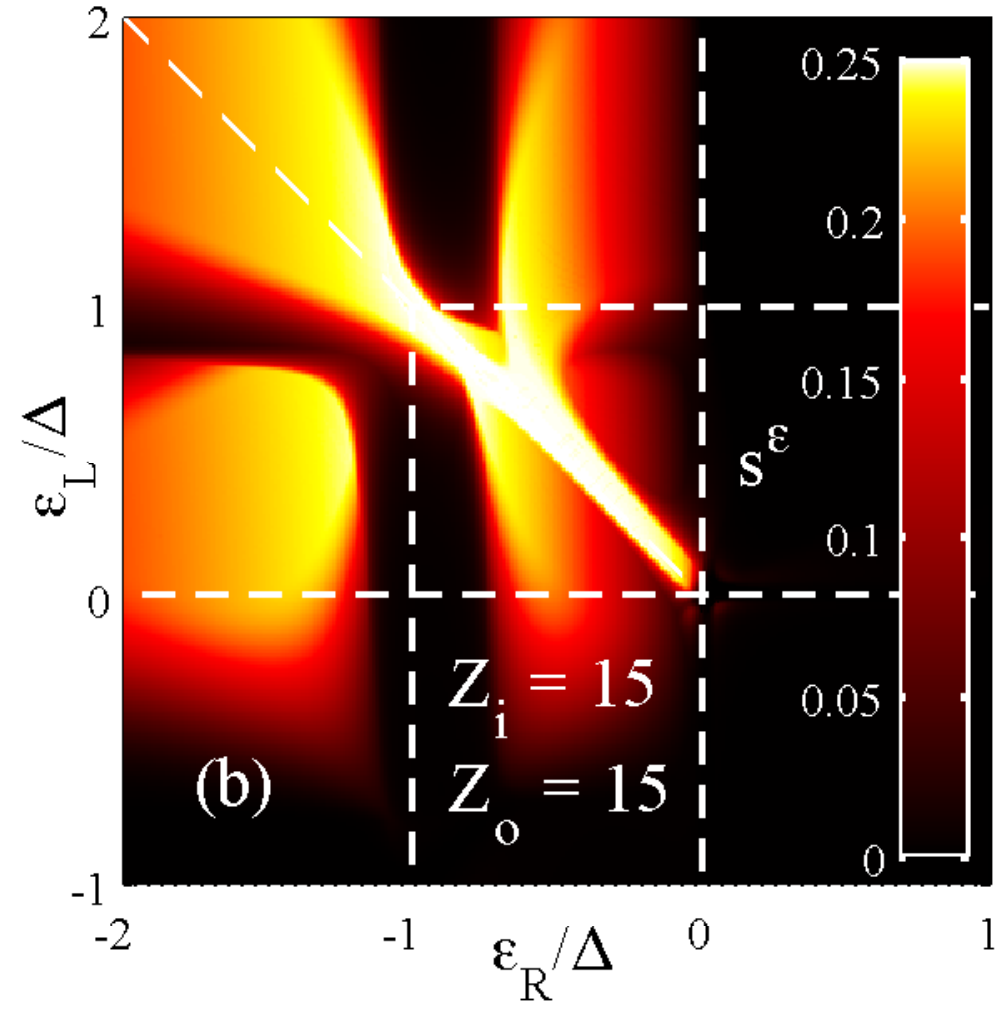}
		}
		\subfloat{
			\hspace{-2mm} \includegraphics[width=4.3cm]{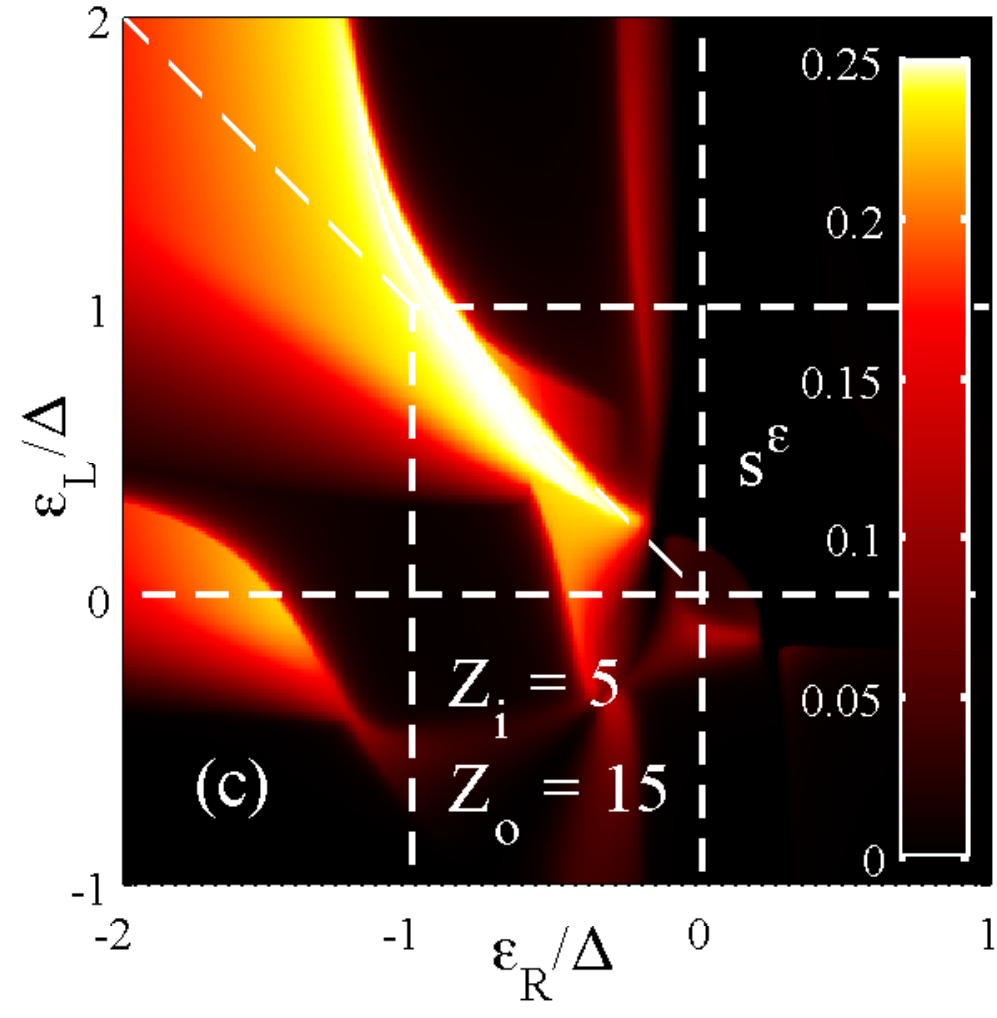}
		}
		\subfloat{
			\hspace{-2mm} \includegraphics[width=4.3cm]{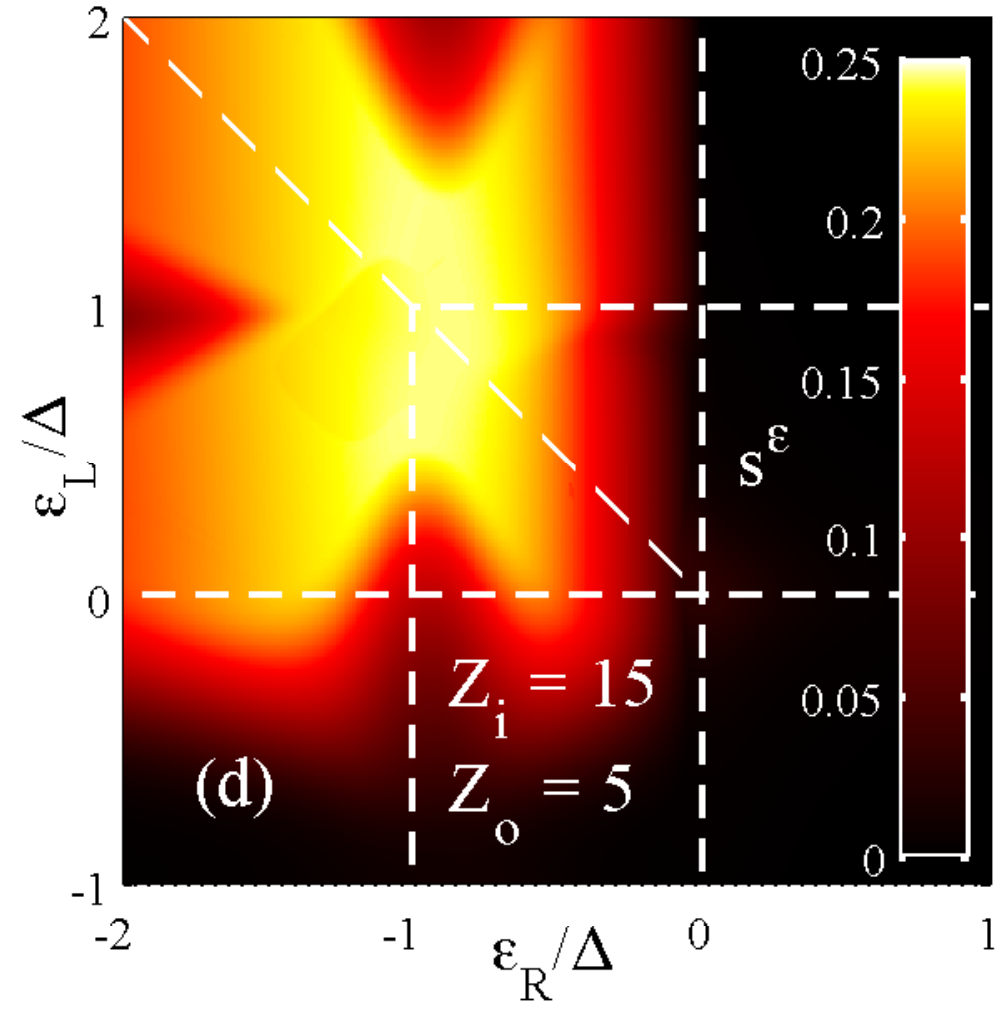}
		} \\ \vspace{-4mm}
		\subfloat{
			\hspace{-3mm} \includegraphics[width=4.3cm]{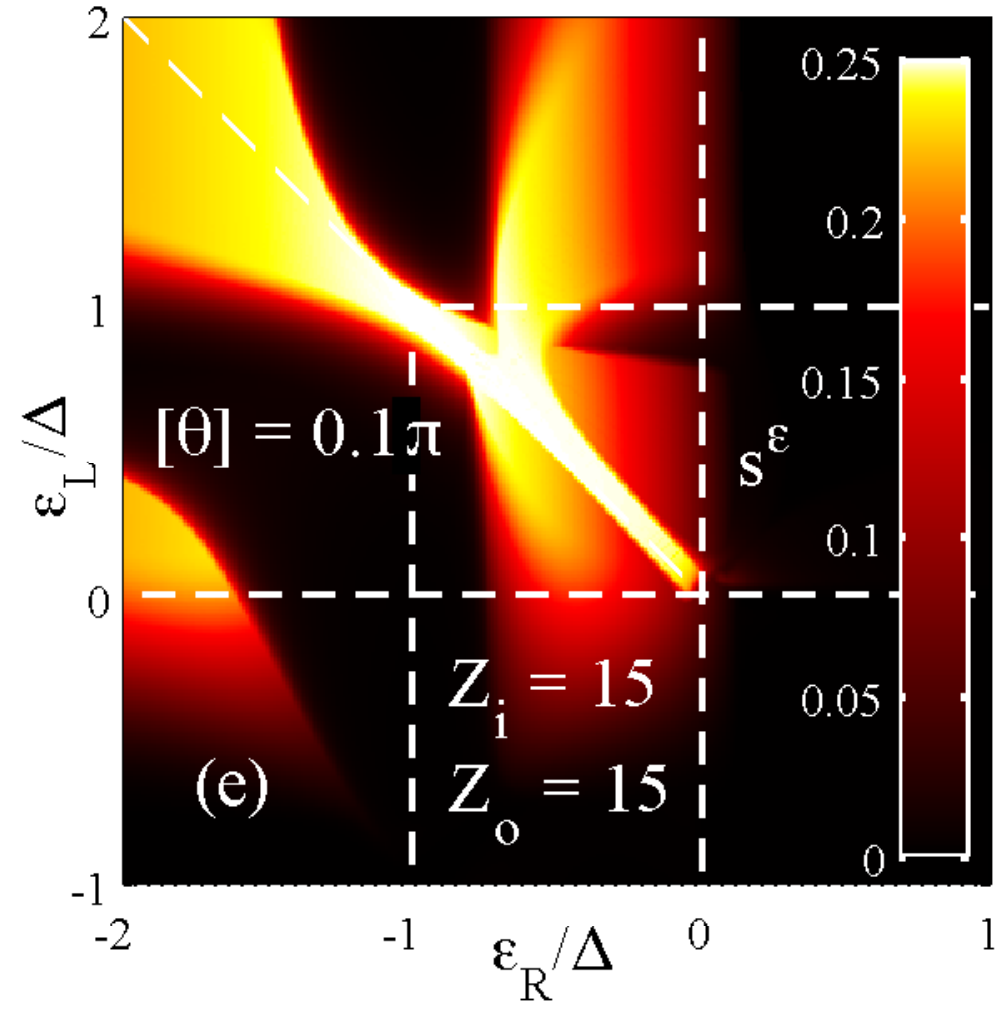}
		}
		\subfloat{
			\hspace{-2mm} \includegraphics[width=4.3cm]{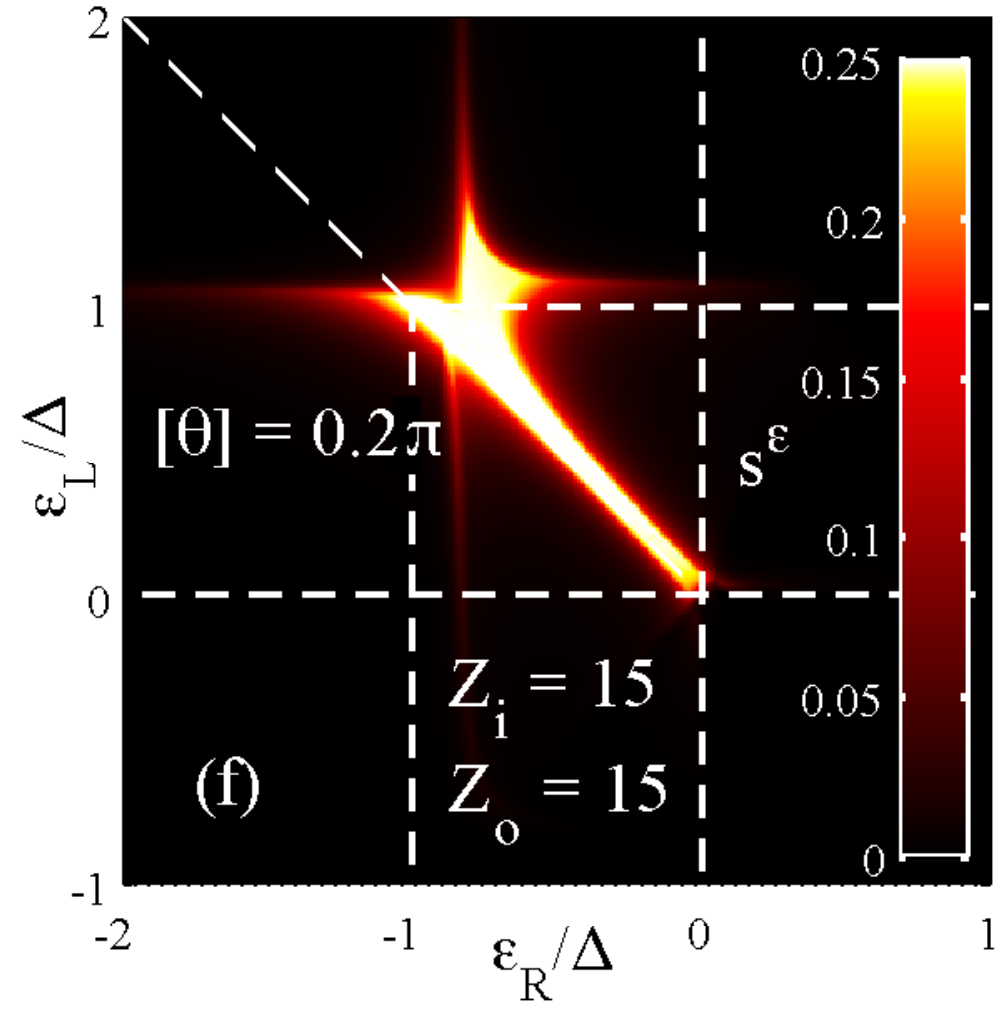}
		}
		\subfloat{
			\hspace{-2mm} \includegraphics[width=4.3cm]{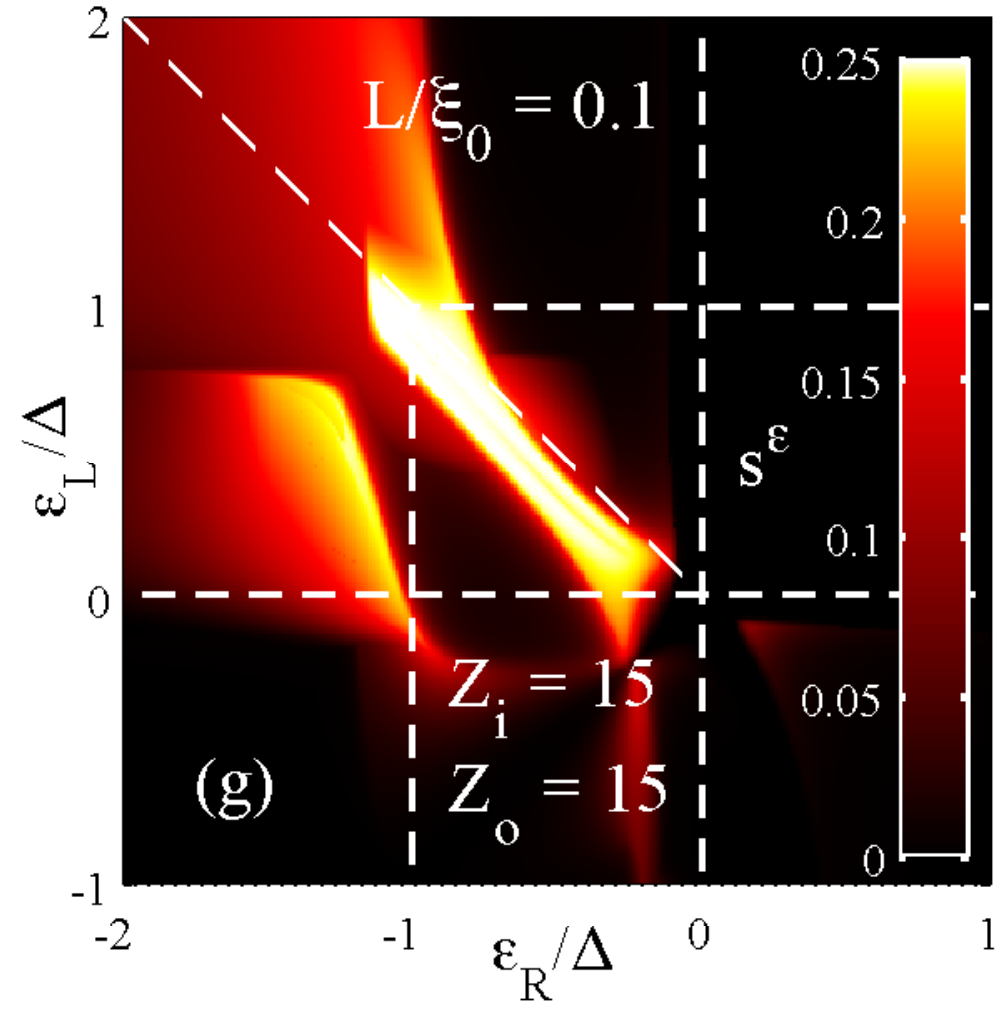}
		}
		\subfloat{
			\hspace{-2mm} \includegraphics[width=4.3cm]{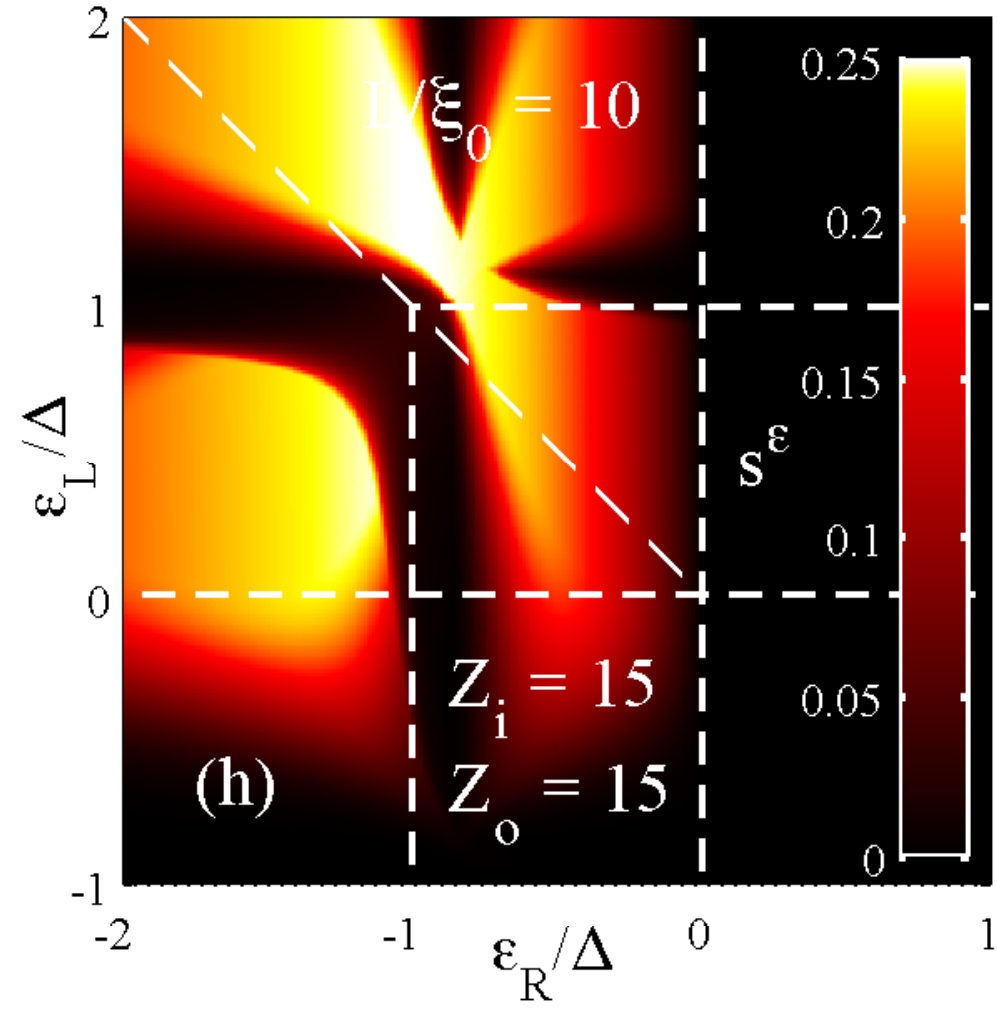}
		}
	\end{center} \vspace{-6mm}
	\caption{
		Color plots of maximal differential noise $s^\varepsilon = 
		\max_\varepsilon\{{\tilde T}_{\rm eh}(1-{\tilde T}_{\rm eh}) - {\tilde T}_{\rm ee}(1-{\tilde T}_{\rm ee})\}$ 
		as functions of~$\varepsilon_\rR$ and $\varepsilon_\rL$ 
		for $L/\xi_0 = 1$, $\delta_\rLR/\Delta = 10$, and $\theta = \pi n$.
		(a)~$Z_{\rm i} = Z_{\rm o} = 5$ ($\Gamma_\sLR / \Delta = 0.39$).
		(b)~$Z_{\rm i} = Z_{\rm o} = 15$ ($\Gamma_\sLR / \Delta = 0.045$).
		(c)~$Z_{\rm i} = 5$ and $Z_{\rm o} = 15$ ($\Gamma_\sLR / \Delta = 0.22$).
		(d)~$Z_{\rm i} = 15$ and $Z_{\rm o} = 5$. 
		(e)~$Z_{\rm i} = Z_{\rm o} = 15$ and $\theta = 0.1\pi + 2\pi n$.
		(f)~$Z_{\rm i} = Z_{\rm o} = 15$ and $\theta = 0.2\pi + 2\pi n$.
		(g)~$Z_{\rm i} = Z_{\rm o} = 15$ and $L/\xi_0 = 0.1$.
		(h)~$Z_{\rm i} = Z_{\rm o} = 15$ and $L/\xi_0 = 10$.
	}
\end{figure}

\subsection{Maximal transparency $\max_\varepsilon\{{\tilde T}_{\rm eh}(\varepsilon)\}$ as a function of $\varepsilon_\rR$ and $\varepsilon_\rL$ for asymmetric dot}

Here we consider different dots and introduce $Z_{\rL{\rm i}}$ and $Z_{\rL{\rm o}}$ for the left dot and $Z_{\rR{\rm i}}$ and $Z_{\rR{\rm o}}$ for the right dot. 

\begin{figure}[H]
	\vspace{-6mm} \begin{center}
		\subfloat{
			\hspace{-3mm} \includegraphics[width=4.4cm]{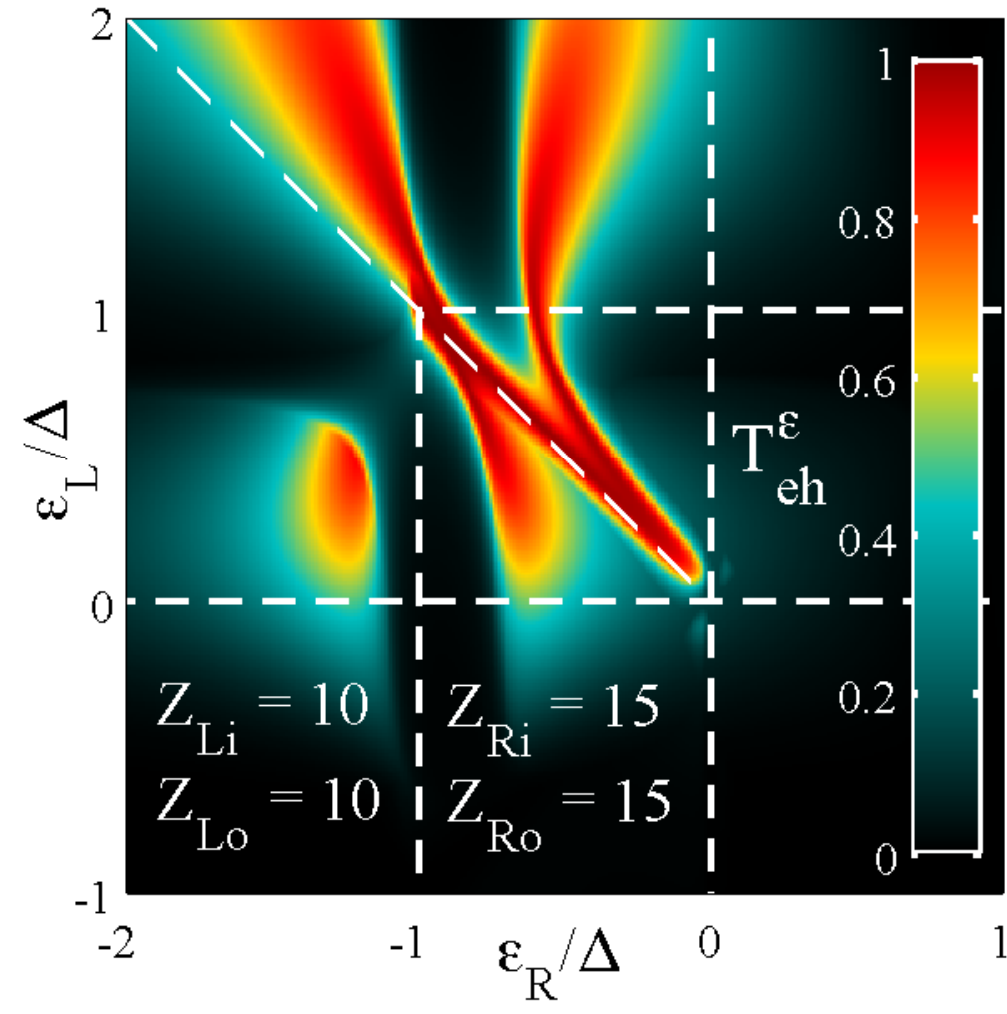}
		}
		\subfloat{
			\hspace{-2mm} \includegraphics[width=4.4cm]{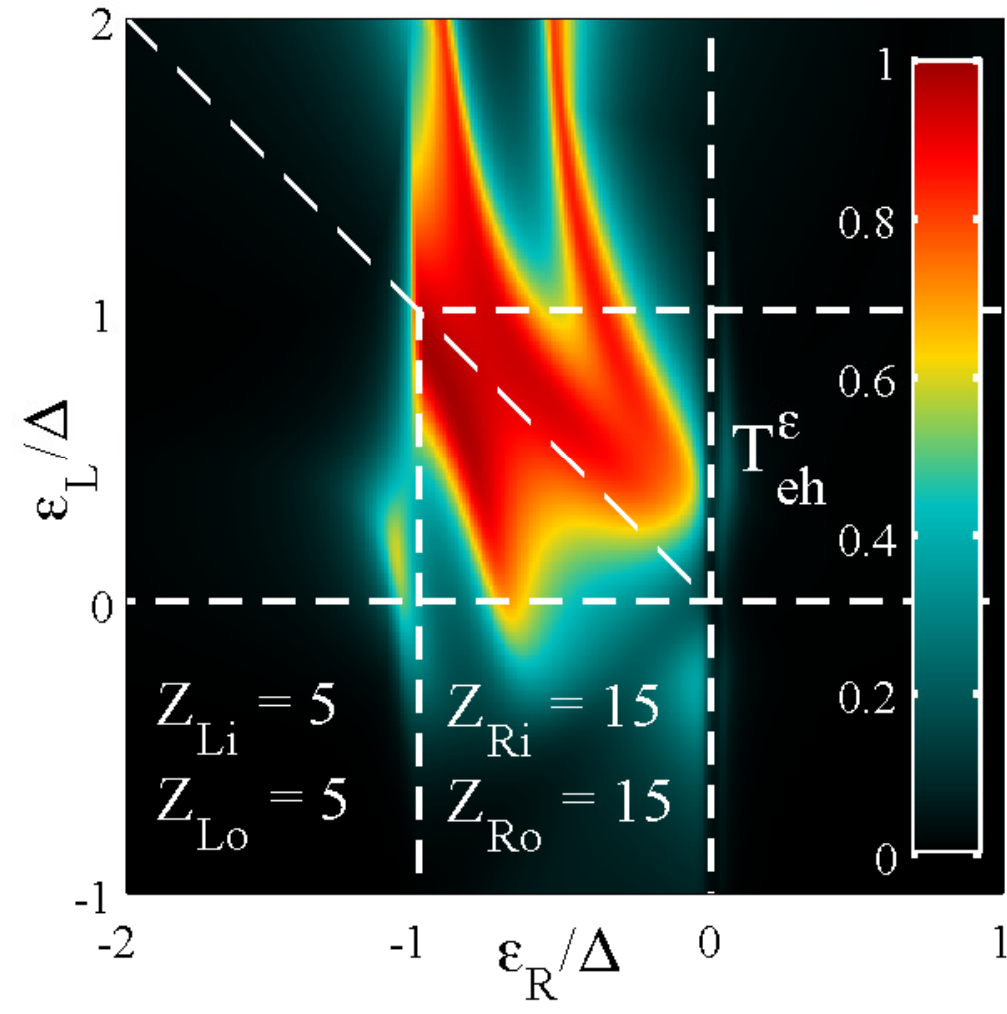}
		}
		\subfloat{
			\hspace{-2mm} \includegraphics[width=4.4cm]{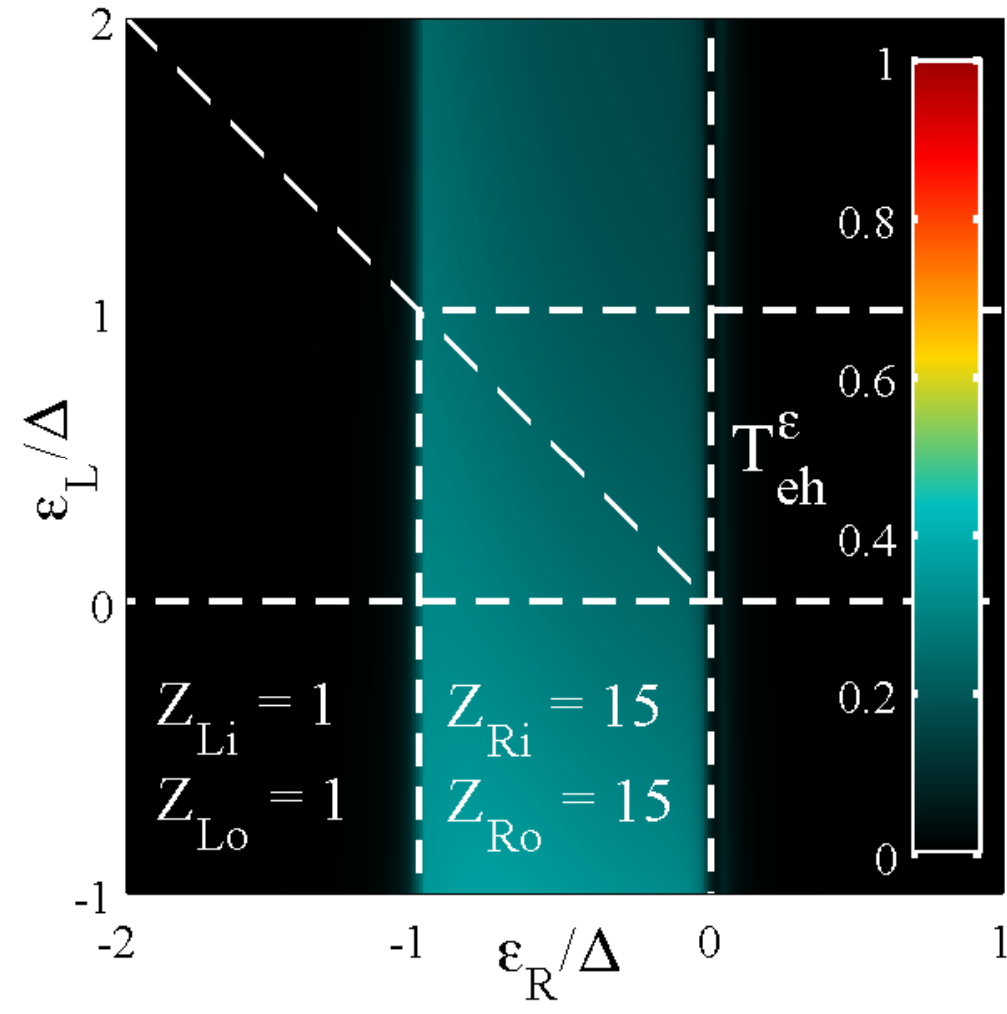}
		} 
		\subfloat{
			\hspace{-2mm} \includegraphics[width=4.4cm]{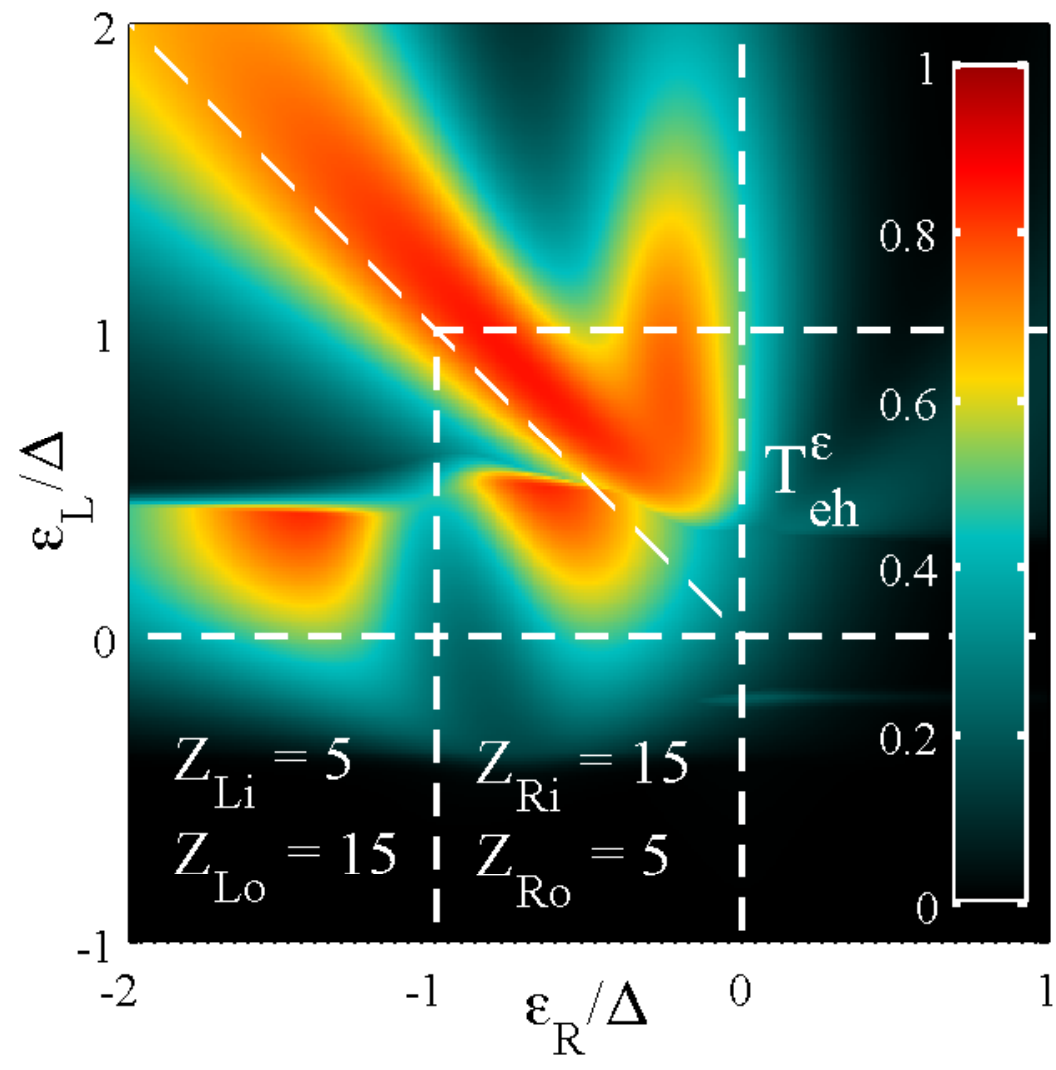}
		} \\ \vspace{-3mm}
		\subfloat{
			\hspace{-3mm} \includegraphics[width=4.4cm]{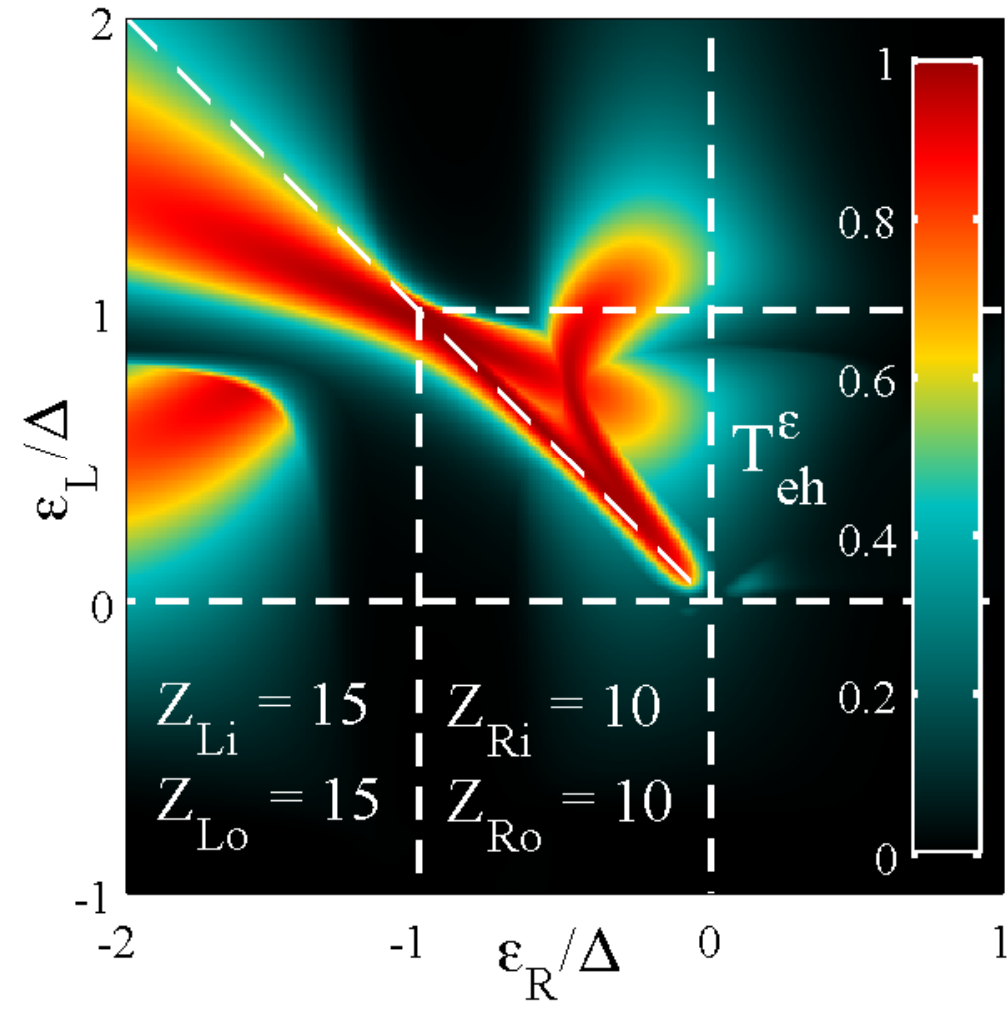}
		}
		\subfloat{
			\hspace{-2mm} \includegraphics[width=4.4cm]{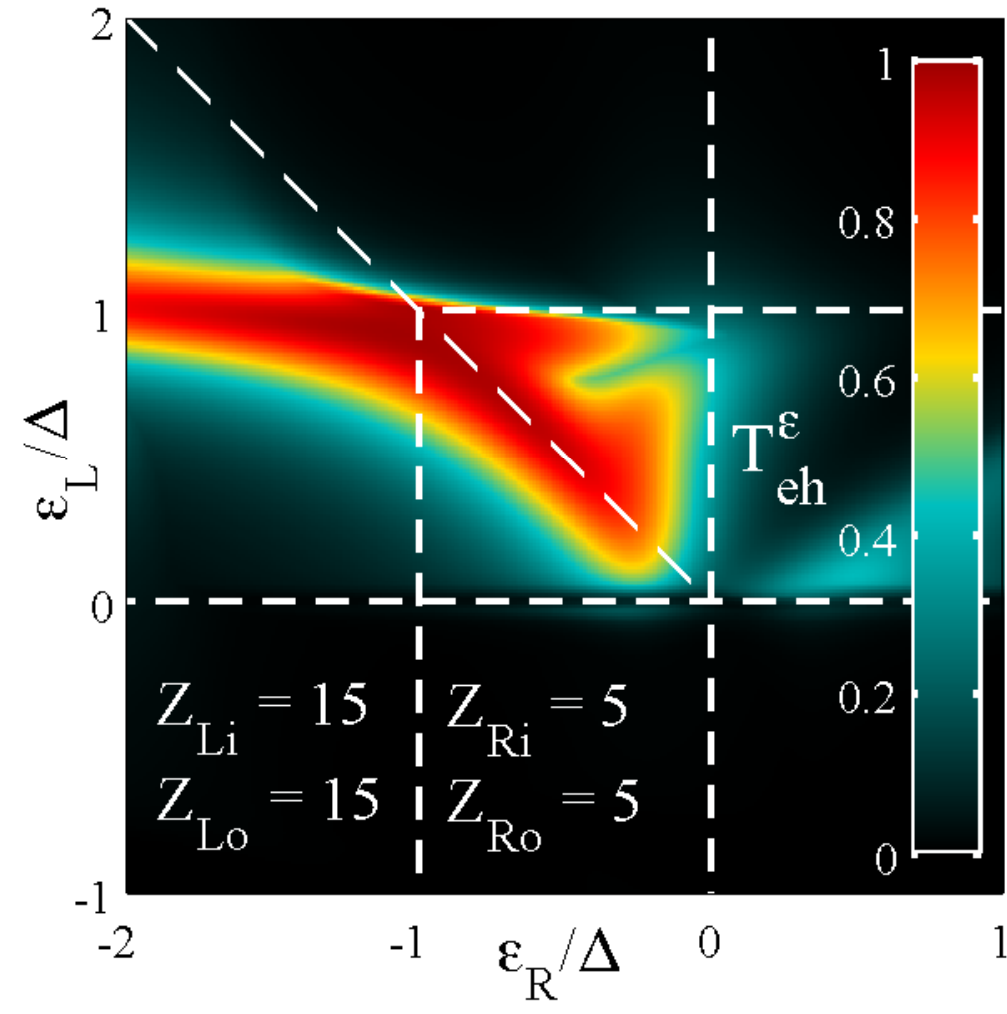}
		}
		\subfloat{
			\hspace{-2mm} \includegraphics[width=4.4cm]{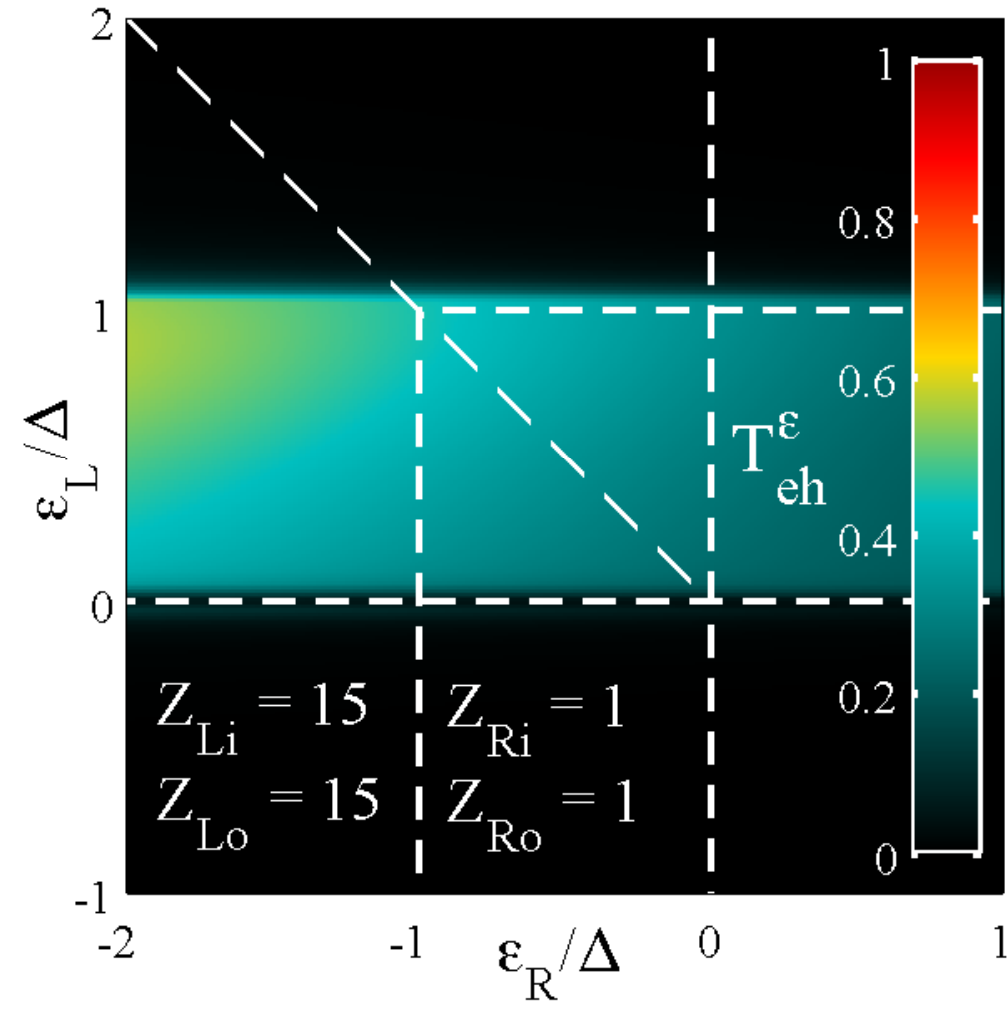}
		}
		\subfloat{
			\hspace{-2mm} \includegraphics[width=4.4cm]{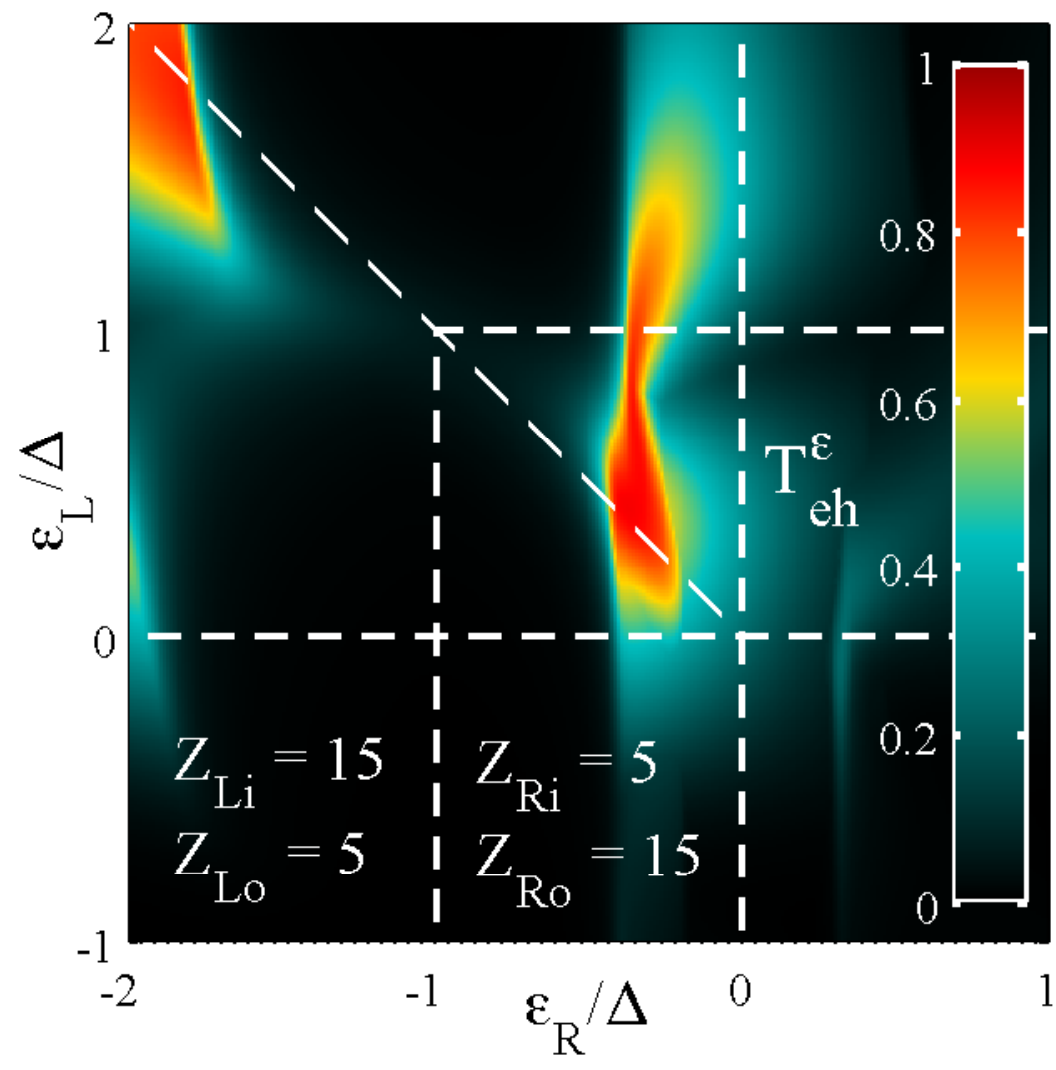}
		}
	\end{center} \vspace{-6mm}
	\caption{
		Maximal transparency $\max_\varepsilon\{{\tilde T}_{\rm eh}(\varepsilon)\}$ 
		as a function of~$\varepsilon_\rR$ and $\varepsilon_\rL$ for $L/\xi_0 = 1$, 
		$\delta_\rLR/\Delta = 1$, 
		and different $Z_{\rL{\rm i}}$, $Z_{\rL{\rm o}}$, $Z_{\rR{\rm i}}$, and $Z_{\rR{\rm o}}$.
	}
\end{figure}

\begin{figure}[H]
	\vspace{-6mm} \begin{center}
		\subfloat{
			\hspace{-3mm} \includegraphics[width=4.4cm]{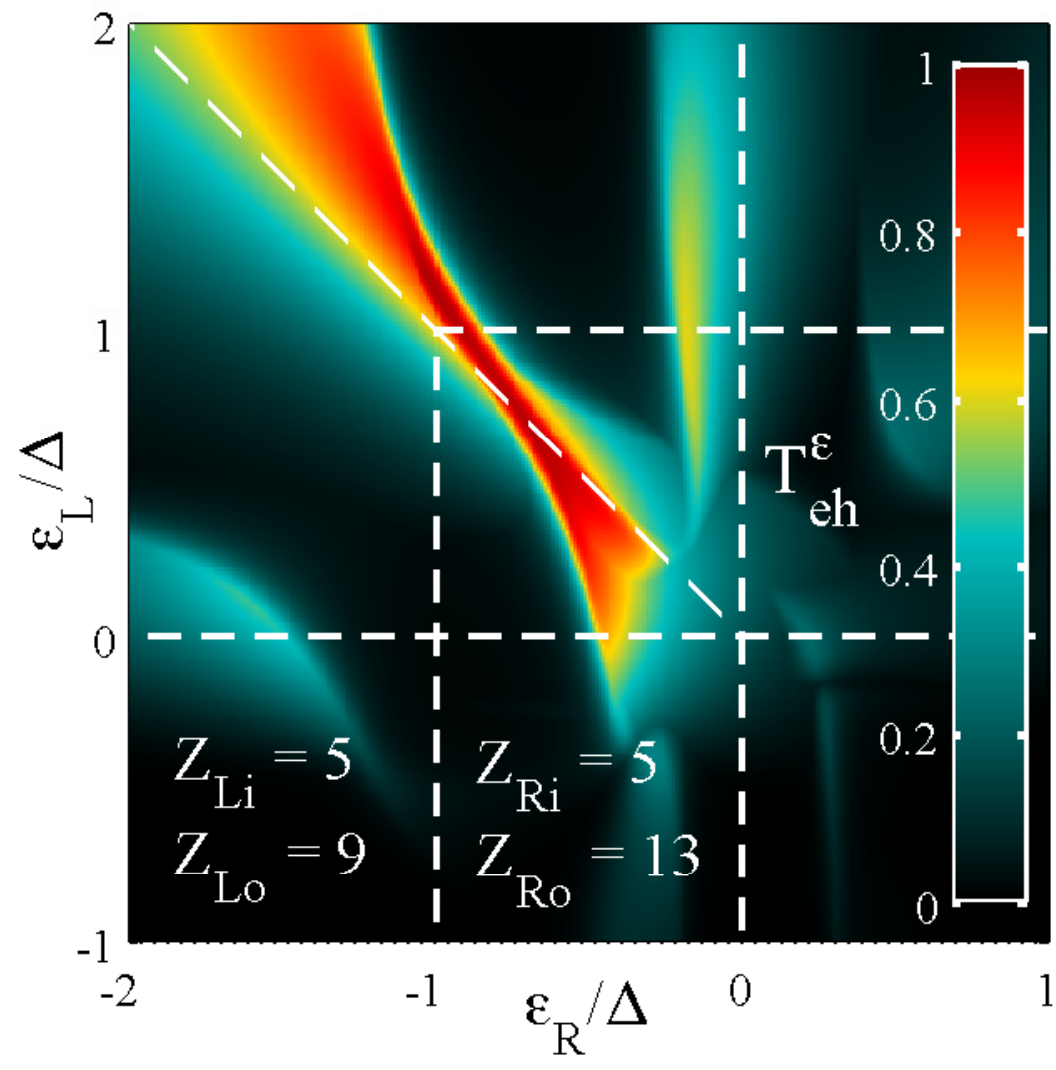}
		}
		\subfloat{
			\hspace{-2mm} \includegraphics[width=4.4cm]{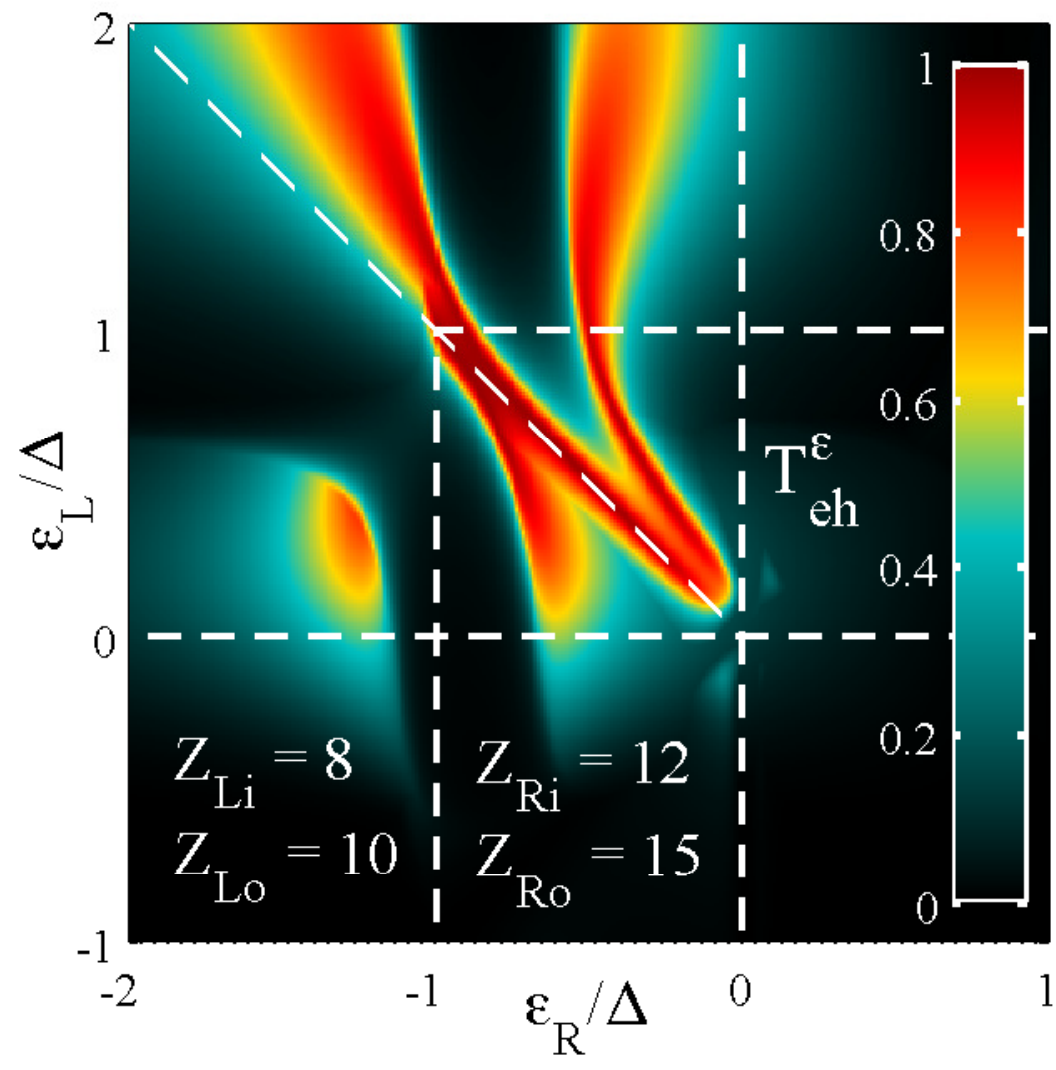}
		}
		\subfloat{
			\hspace{-2mm} \includegraphics[width=4.4cm]{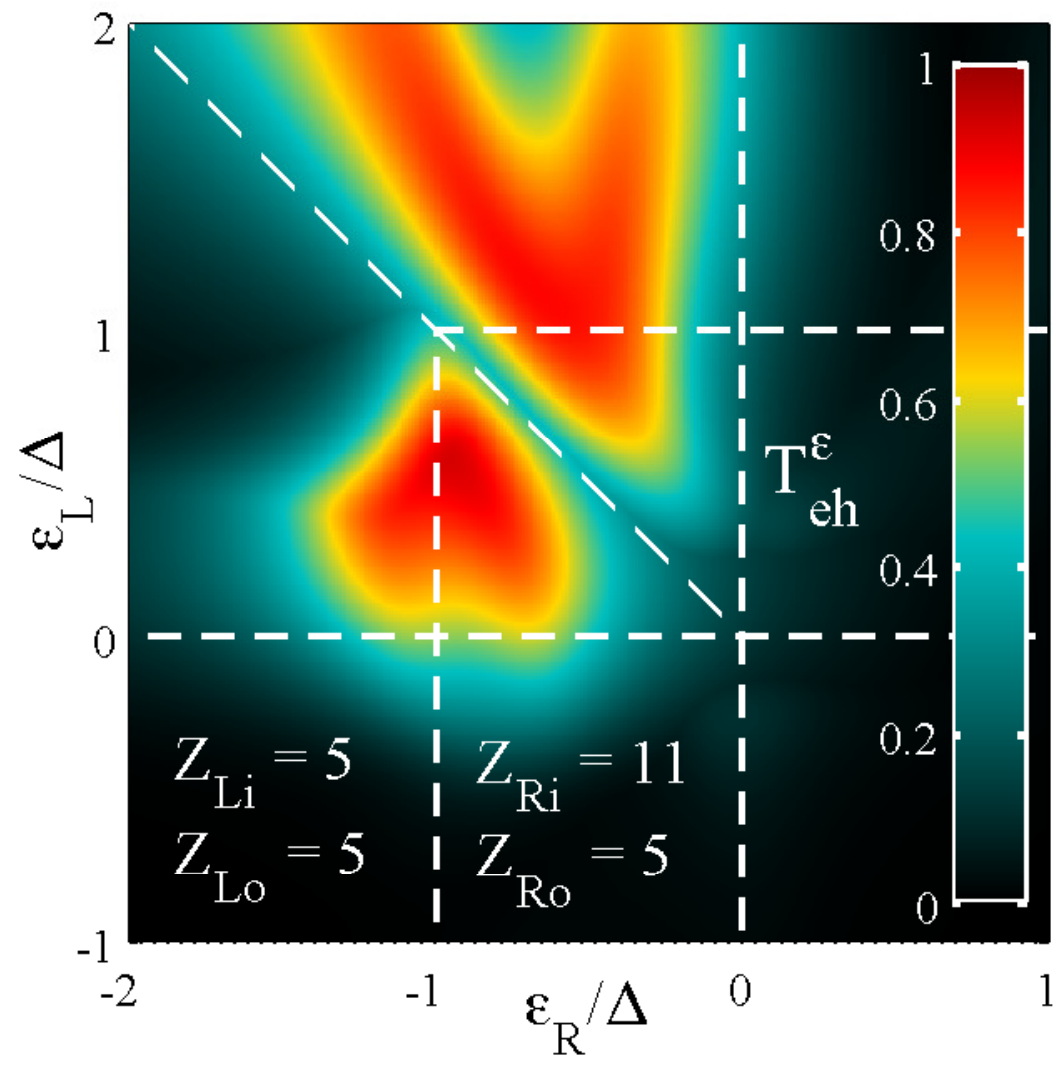}
		}
		\subfloat{
			\hspace{-2mm} \includegraphics[width=4.4cm]{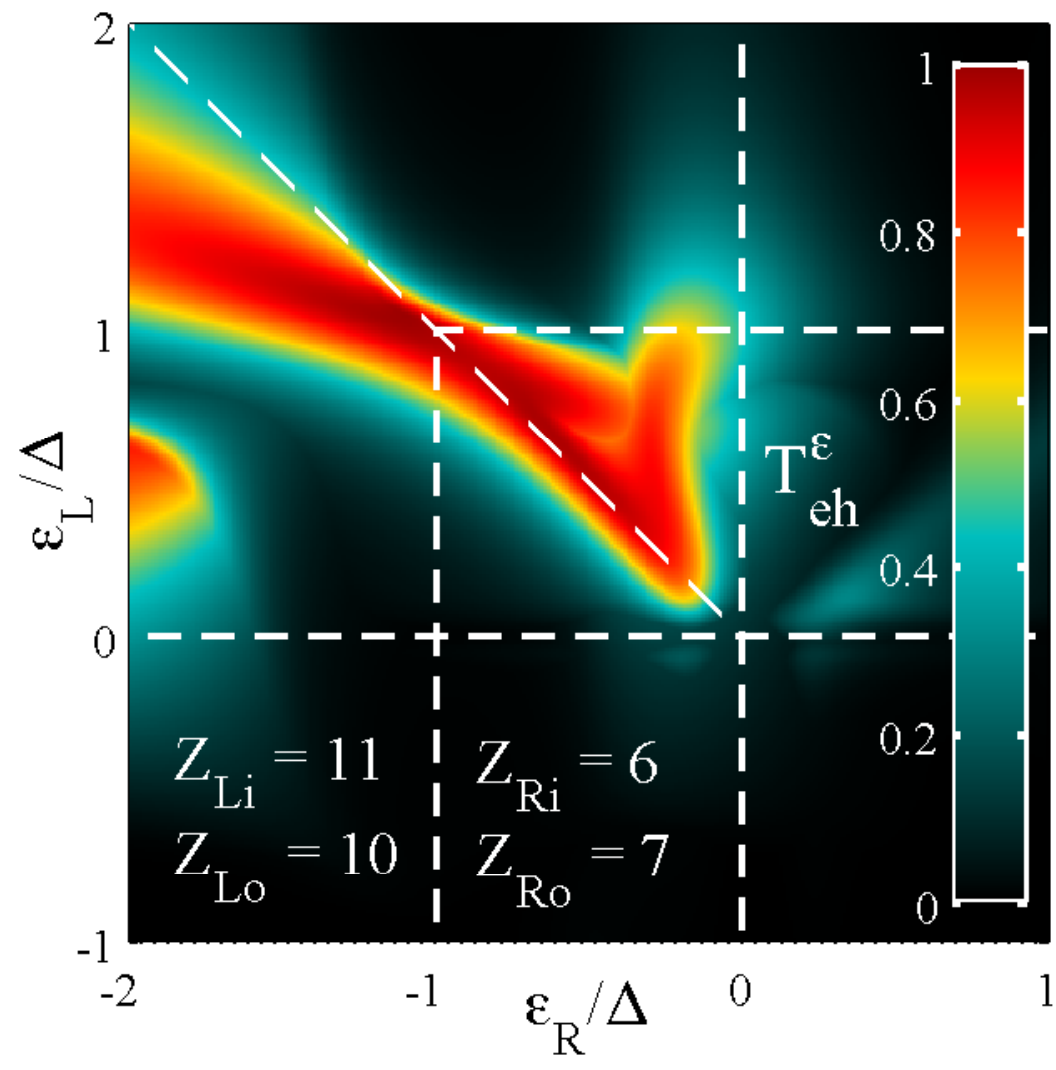}
		} \\ \vspace{-3mm}
		\subfloat{
			\hspace{-3mm} \includegraphics[width=4.4cm]{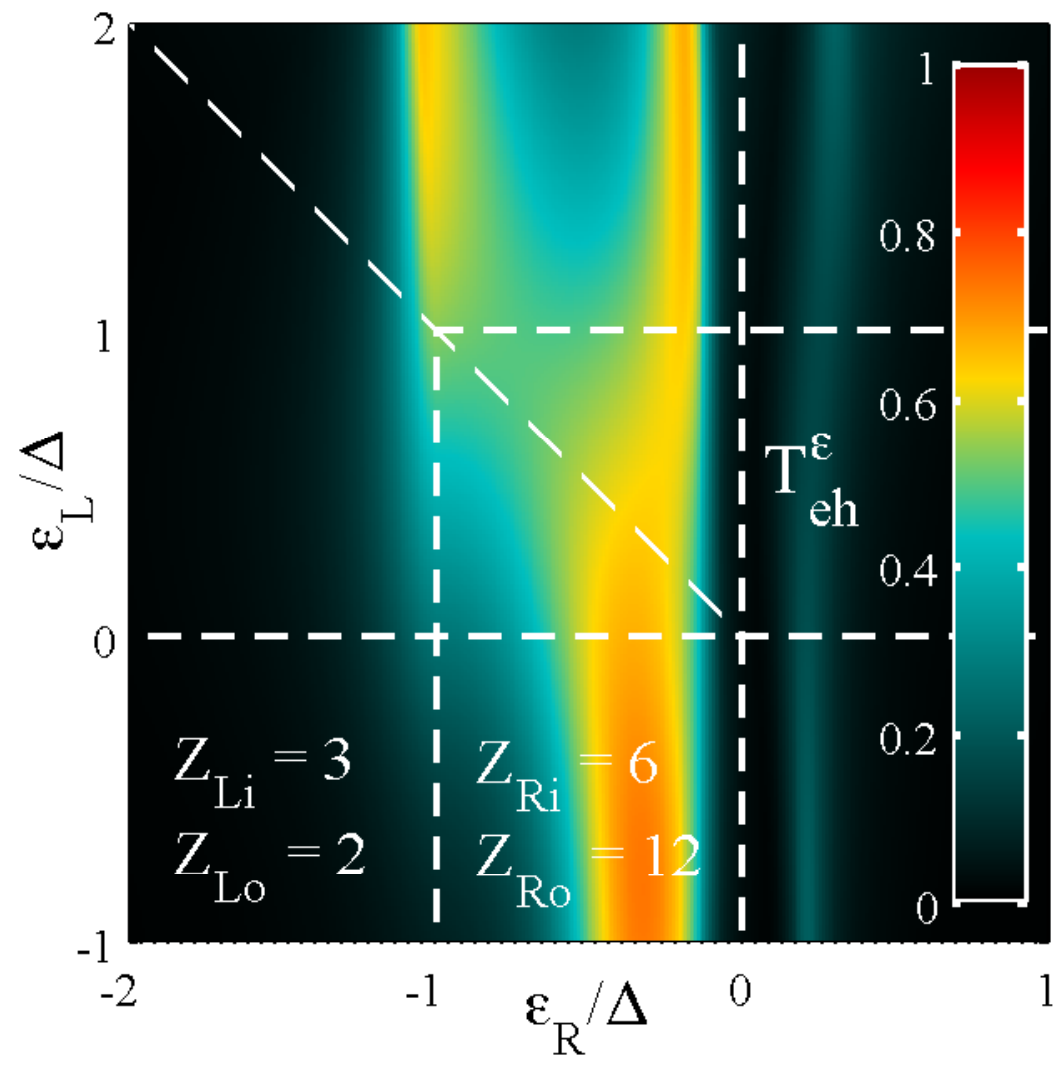}
		}
		\subfloat{
			\hspace{-2mm} \includegraphics[width=4.4cm]{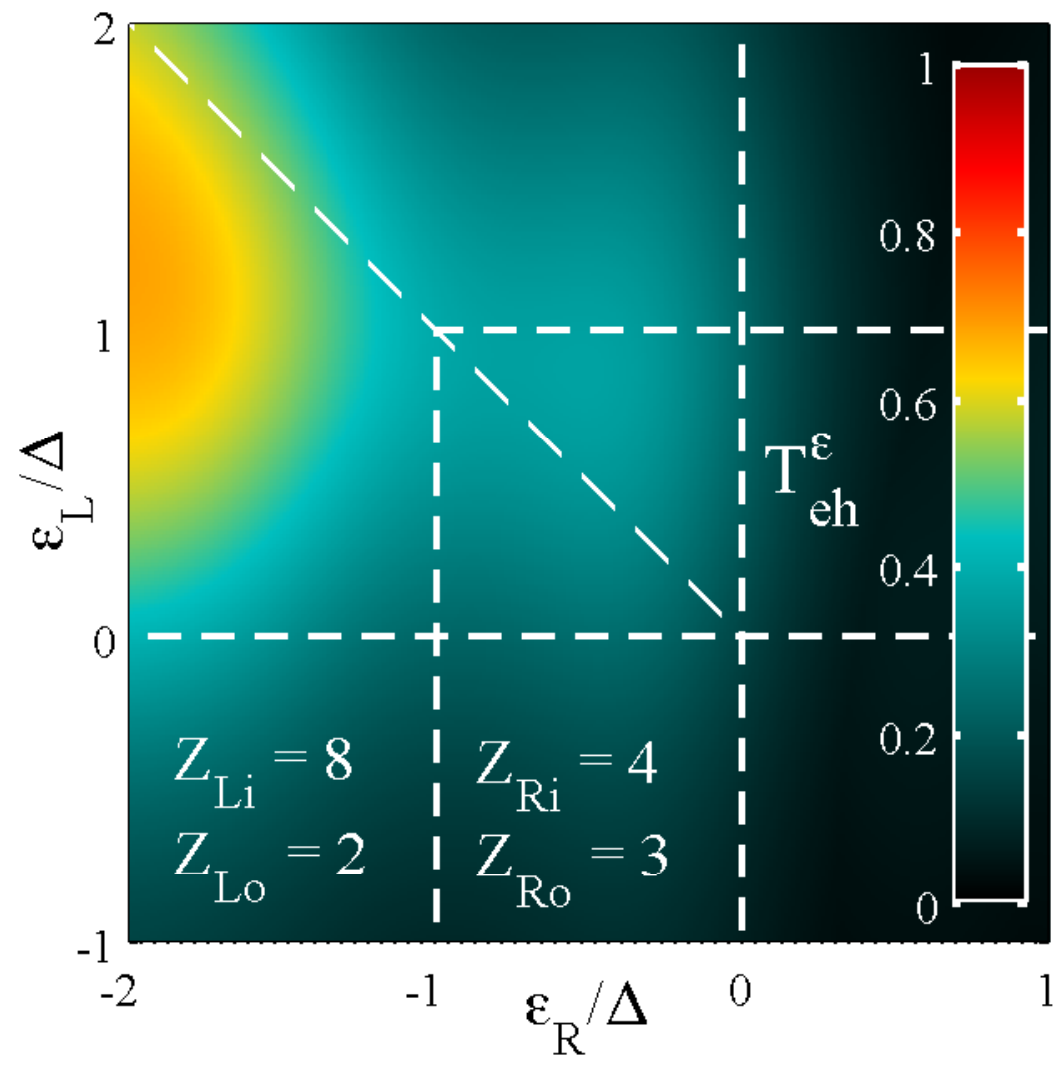}
		}
		\subfloat{
			\hspace{-2mm} \includegraphics[width=4.4cm]{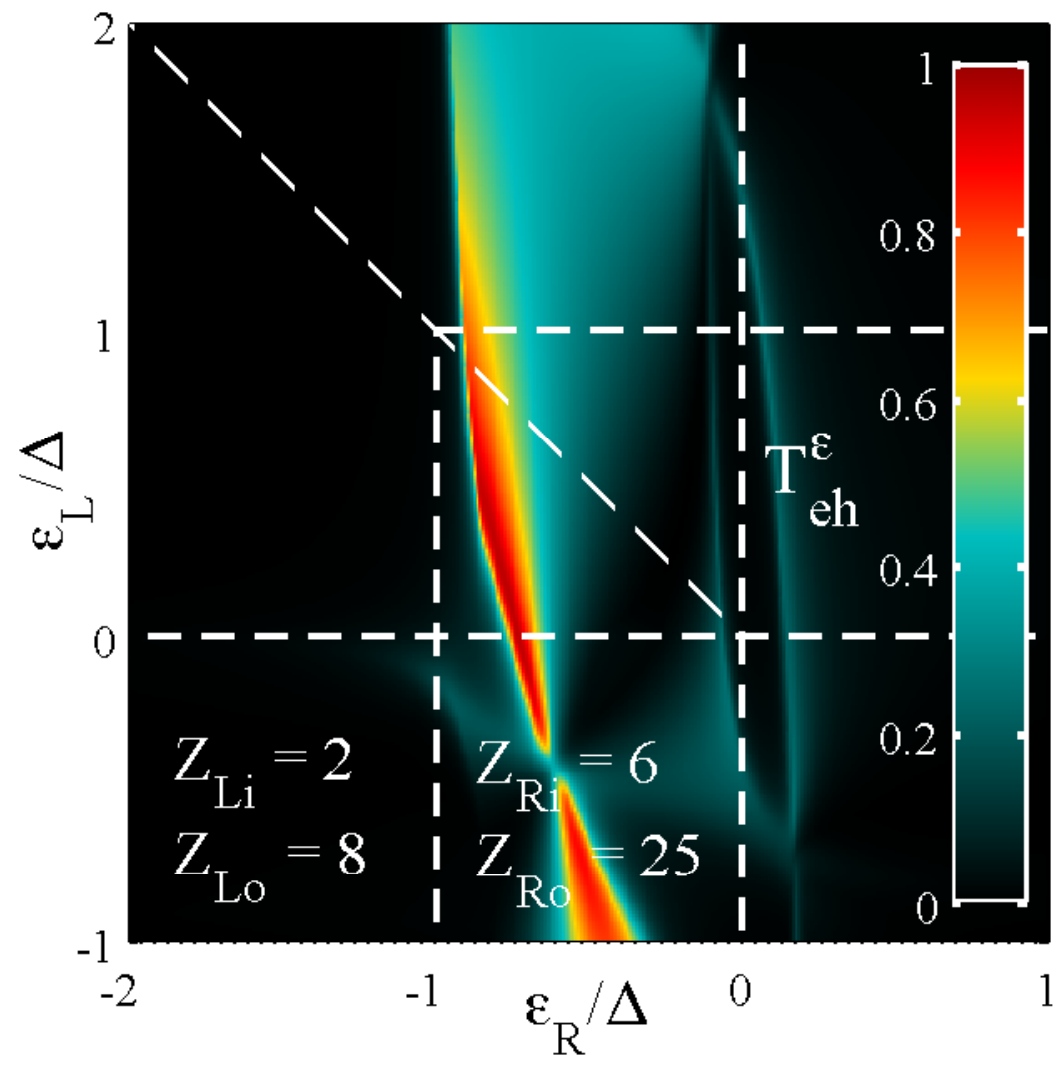}
		}
		\subfloat{
			\hspace{-2mm} \includegraphics[width=4.4cm]{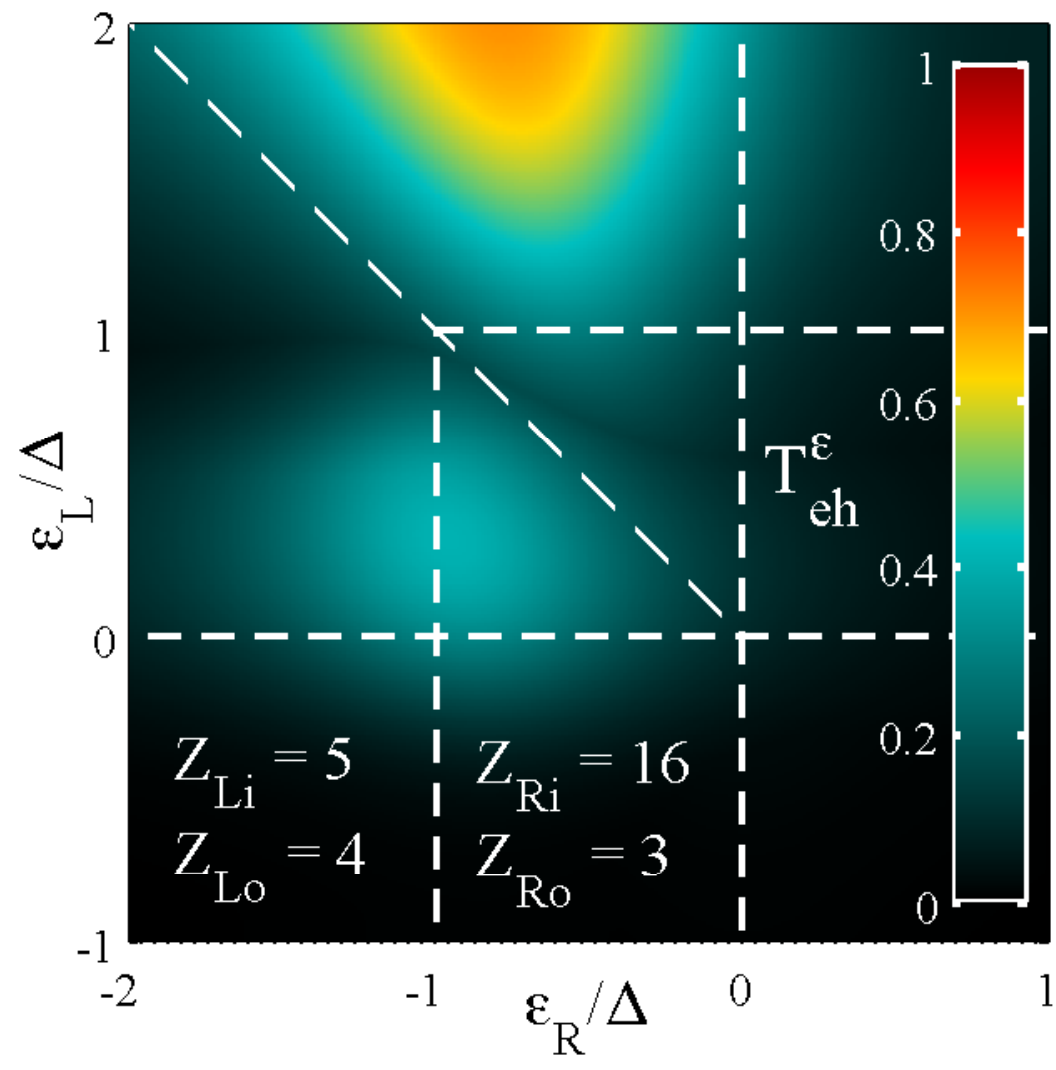}
		}
	\end{center} \vspace{-6mm}
	\caption{
		Maximal transparency $\max_\varepsilon\{{\tilde T}_{\rm eh}(\varepsilon)\}$ 
		as a function of~$\varepsilon_\rR$ and $\varepsilon_\rL$ for $L/\xi_0 = 1$, 
		$\delta_\rLR/\Delta = 1$, 
		and random $Z_{\rL{\rm i}}$, $Z_{\rL{\rm o}}$, $Z_{\rR{\rm i}}$, $Z_{\rR{\rm o}}$ homogeneously distributed in $[5 \ldots 15]$ interval (upper row) and exponentially distributed in $[1 \ldots 25]$ interval (lower row).
	}
\end{figure}

\clearpage \end{widetext}


\begin{thebibliography}{99} 

\bibitem{Einstein:1935}
	A. Einstein, B. Podolsky, and N. Rosen, 
	{\it Can quantum-mechanical description of physical reality be considered complete?}
	Phys. Rev. {\bf 47}, 777 (1935).

\bibitem{Schrodinger:1935}
	E. Schr{\"o}dinger and M. Born,
	{\it Discussion of probability relations between separated systems}, 
	Math. Proc. Cambridge Phil. Soc. {\bf 31}, 555 (1935).
 
\bibitem{Bennett:1993}
	C. H. Bennett, G. Brassard, C. Cr\'epeau, R. Jozsa, A. Peres, and W. K. Wootters,
	{\it Teleporting an unknown quantum state via dual classical and Einstein-Podolsky-Rosen channels},
	Phys. Rev. Lett. {\bf 70}, 1895 (1993).

\bibitem{Steane:1998}
	A. Steane, 
	{\it Quantum computing},
	Rep. Prog. Phys. {\bf 61}, 117 (1998).

\bibitem{Nielsen:2011}
	M. A. Nielsen and I. L. Chuang,
	{\it Quantum computation and quantum information}
	(Cambridge, 2011).

\bibitem{Bohm:2001}
	A. Bohm,
	{\it Quantum mechanics: Foundations and applications}
	(Sprin\-ger, 2001).

\bibitem{Lesovik:2001}
	G. B. Lesovik, T. Martin, and G. Blatter,
	{\it Electronic entanglement in the vicinity of a superconductor},
	Eur. Phys. J. B {\bf 24}, 287 (2001).

\bibitem{Recher:2001}
	P. Recher, E. V. Sukhorukov, and D. Loss, 
	{\it Andreev tunneling, Coulomb blockade, and resonant transport of nonlocal spin-entangled electrons},
	Phys. Rev. B {\bf 63}, 165314 (2001).

\bibitem{Andreev:1964}
	A. F. Andreev, 
	{\it The thermal conductivity of the intermediate state in superconductors},
	ZhETF {\bf 46}, 1823 (1964)
	[Sov. Phys. JETP {\bf 19}, 1228 (1964)].

\bibitem{Bozhko:1982}
	S. I. Bozhko, V. S. Tsoi, and S. E. Yakovlev, 
	{\it Observation of Andreev reflection with the help of transverse electron focusing},
	Pis'ma v ZhETF {\bf 36}, 123 (1982) 
	[JETP Lett. {\bf 36}, 153 (1982)].

\bibitem{Benistant:1983}
	P. A. M. Benistant, H. van Kampen, and P. Wyder, 
	{\it Direct observation of Andreev reflection},
	Phys. Rev. Lett. {\bf 51}, 817 (1983).
	
\bibitem{Byers:1995} 
	J. M. Byers and M. E. Flatt\'e, 
	{Probing spatial correlations with nanoscale two-contact tunneling},
	Phys. Rev. Lett. {\bf 74}, 306 (1995).

\bibitem{Deutscher:2000}
	G. Deutscher and D. Feinberg, 
	{\it Coupling superconducting-ferromagnetic point contacts by Andreev reflections},
	Appl. Phys. Lett. {\bf 76}, 487 (2000).

\bibitem{Falci:2001}
	G. Falci, D. Feinberg, and F. W. J. Hekking, 
	{\it Correlated tunneling into a superconductor in a multiprobe hybrid structure},
	Europhys. Lett. {\bf 54}, 255 (2001).

\bibitem{Beckmann:2004}
	D. Beckmann, H. B. Weber, and H. v. L\"ohneysen, 
	{\it Evidence for crossed Andreev reflection in superconductor-ferromagnet hybrid structures},
	Phys. Rev. Lett. {\bf 93}, 197003 (2004).
	
\bibitem{Russo:2005}
	S. Russo, M. Kroug, T. M. Klapwijk, and A. F. Morpurgo, 
	{\it Experimental observation of bias-dependent nonlocal Andreev reflection},
	Phys. Rev. Lett. {\bf 95}, 027002 (2005).
	
\bibitem{Golubev:2009}
	D. S. Golubev, M. S. Kalenkov, and A. D. Zaikin,
	{\it Crossed Andreev reflection and charge imbalance in diffusive NSN structures},
	Phys. Rev. Lett. {\bf 103}, 067006 (2009).

\bibitem{Choi:2000}
	M. S. Choi, C. Bruder, and D. Loss, 
	{\it Spin-dependent Josephson current through double quantum dots and measurement of entangled electron states},
	Phys. Rev. B {\bf 62}, 13569 (2000).

\bibitem{Leijnse:2013}
	M. Leijnse and K. Flensberg,
	{\it Coupling spin qubits via superconductors},
	Phys. Rev. Lett. {\bf 111}, 060501 (2013).

\bibitem{Feinberg:2003}
	D. Feinberg,
	{\it Andreev scattering and cotunneling between two superconductor-normal metal interfaces: the dirty limit},
	Eur. Phys. J. B {\bf 36}, 419 (2003).

\bibitem{Melin:2008}
	R. M\'elin, C. Benjamin, and T. Martin,
	{\it Positive cross correlations of noise in superconducting hybrid structures: Roles of interfaces and interactions},
	Phys. Rev. B {\bf 77}, 094512 (2008).

\bibitem{Chevallier:2011}
	D. Chevallier, J. Rech, T. Jonckheere, and T. Martin,
	{\it Current and noise correlations in a double-dot Cooper-pair beam splitter},
	Phys. Rev. B {\bf 83}, 125421 (2011).

\bibitem{Floser:2013}
	M. Fl\"oser, D. Feinberg, and R. M\'elin,
	{\it Absence of split pairs in the cross-correlations of a highly transparent normal metal-superconductor-normal metal electron beam splitter},
	Phys. Rev. B {\bf 88}, 094517 (2013).

\bibitem{Hofstetter:2009}
	L. Hofstetter, S. Csonka, J. Nygoard, and C. Sch\"onen\-berger,
	{\it Cooper pair splitter realized in a two-quantum-dot Y-junction}, 
	Nature {\bf 461}, 960 (2009).

\bibitem{Hofstetter:2011}
	L. Hofstetter, S. Csonka, A. Baumgartner, G. F\"ul\"op, S. d'Hollosy, J. Nygard, and C. Sch\"onenberger, 
	{\it Finite-bias Cooper pair splitting},
	Phys. Rev. Lett. {\bf 107}, 136801 (2011).

\bibitem{Herrmann:2010}
	L. G. Herrmann, F. Portier, P. Roche, A. Levy Yeyati, T. Kontos, and C. Strunk, 
	{\it Carbon nanotubes as Cooper-pair beam splitters},
	Phys. Rev. Lett. {\bf 104}, 026801 (2010).

\bibitem{Schindele:2012}
	J. Schindele, A. Baumgartner, and C. Sch\"onenberger, 
	{\it Near-unity Cooper pair splitting efficiency},
	Phys. Rev. Lett. {\bf 109}, 157002 (2012).

\bibitem{Tan:2015}
	Z. B. Tan, D. Cox, T. Nieminen, P. L\"ahteenm\"aki, D. Golubev, G. B. Lesovik, and P. J. Hakonen, 
	{\it Cooper pair splitting by means of graphene quantum dots},
	Phys. Rev. Lett. {\bf 114}, 096602 (2015).

\bibitem{Veldhorst:2010}
	M. Veldhorst and A. Brinkman, 
	{\it Nonlocal Cooper pair splitting in a pSn junction},
	Phys. Rev. Lett. {\bf 105}, 107002 (2010).

\bibitem{Burset:2011}
	P. Burset, W. J. Herrera, and A. Levy Yeyati, 
	{\it Microscopic theory of Cooper pair beam splitters based on carbon nanotubes},
	Phys. Rev. B {\bf 84}, 115448 (2011).

\bibitem{Sadovskyy:2007}
	I. A. Sadovskyy, G. B. Lesovik, and G. Blatter,
	{\it Continuously tunable charge in Andreev quantum dots},
	Phys. Rev. B {\bf 75}, 195334 (2007).

\bibitem{Sadovskyy:2012}
	I. A. Sadovskyy, G. B. Lesovik, G. Blatter, T. Jonckheere, and T. Martin,
	{\it Andreev quantum dot with several conducting channels},
	Phys. Rev. B {\bf 85}, 125442 (2012).

\bibitem{Bogoliubov:1968}
	N. N. Bogoljubov, V. V. Tolmachov, and D. V. Sirkov,
	{\it A new method in the theory of superconductivity}
	(Consultant Bureau, New York, 1959).

\bibitem{GennesBook:1968}
	P. G. de Gennes,
	{\it Superconductivity of metals and alloys} 
	(Benjamin, 1966).

\bibitem{Lesovik:2011}
	G. B. Lesovik and I. A. Sadovskyy,
	{\it Scattering matrix approach to the description of quantum electron transport},
	Usp. Phys. Nauk {\bf 181}, 1041 (2011)
	[Phys.-Usp. {\bf 54}, 1007 (2011)].

\bibitem{Averin:1989}
	D. V. Averin and A. A. Odintsov, 
	{\it Macroscopic quantum tunneling of the electric charge in small tunnel junctions},
	Phys. Lett. A {\bf 140}, 251 (1989).

\bibitem{Averin:1990}
	D. V. Averin and Yu. V. Nazarov, 
	{\it Virtual electron diffusion during quantum tunneling of the electric charge},
	Phys. Rev. Lett. {\bf 65}, 2446 (1990).

\bibitem{Chen:2015}
	W. Chen, D. N. Shi, and D. Y. Xing,
	{\it Long-range Cooper pair splitter with high entanglement production rate},
	Sci. Rep. {\bf 5}, 7607 (2015).

\bibitem{Chtchelkatchev:2002}
	N. M. Chtchelkatchev, G. Blatter, G. B. Lesovik, and T. Martin,
	{\it Bell inequalities and entanglement in solid state devices},
	Phys. Rev. B {\bf 66}, 161320(R) (2002).

\bibitem{Wei:2010}
	J. Wei and V. Chandrasekhar, 
	{\it Positive noise cross-correlation in hybrid superconducting and normal-metal three-terminal devices},
	Nature Phys. {\bf 6}, 494 (2010).

\bibitem{Sadovskyy:2015}
	I. A. Sadovskyy, 
	{\it Reduction of the scattering matrix array},
	Usp. Phys. Nauk {\bf 185}, 941 (2015).

\bibitem{Jarlskog:2005}
	C. Jarlskog,
	{\it A recursive parametrization of unitary matrices},
	J. Math. Phys. {\bf 46}, 103508 (2005).

\end{thebibliography}
\end{document}